Order number: ........

PEOPLE'S DEMOCRATIC REPUBLIC OF ALGERIA
MINISTRY OF HIGHER EDUCATION AND SCIENTIFIC RESEARCH

ECHAHID CHEIKH LARBI TEBESSI UNIVERSITY
FACULTY OF EXACT SCIENCES, NATURAL SCIENCES AND LIFE
DEPARTMENT OF MATHEMATICS AND COMPUTER SCIENCE
LABORATOTY OF MATHEMATICS, INFORMATICS AND SYSTEMS

# DOCTORAL THESIS
# OF THE THIRD CYCLE

Field : Computer science
Specialty : Networks and Multimedia

By

ZAKARIA TOLBA

# CRYPTANALYSIS AND IMPROVEMENT OF MULTIMODAL DATA ENCRYPTION BY MACHINE-LEARNING BASED SYSTEM

Defended on 04 june 2023 in front of the jury :

| | | | |
|---|---|---|---|
| Pr. | HAKIM BENDJENA | University of Tebessa | President |
| Pr. | MOHAMED AMROUNE | University of Tebessa | Examiner |
| Pr. | MOHAMED REDA LAOUAR | University of Tebessa | Examiner |
| Dr. | AHMED AHMIM | University of Souk Ahras | Examiner |
| Dr. | HICHAM TALBI | University of Constantine 2 | Examiner |
| Pr. | MAKHLOUF DERDOUR | University of Oum El Bouaghi | Supervisor |

College year : 2022 - 2023

*I dedicate this thesis to My wife Dr. DEHIMI Nour El Houda and my lovely children Iyad and Baraa,*
*to my parents, brother, sisters, and all colleagues. . . .*

**Abstract**

With the rising popularity of the internet and the widespread use of networks and information systems via the cloud and data centers, the privacy and security of individuals and organizations have become extremely crucial. In this perspective, encryption consolidates effective technologies that can effectively fulfill these requirements by protecting public information exchanges. To achieve these aims, the researchers used a wide assortment of encryption algorithms to accommodate the varied requirements of this field, as well as focusing on complex mathematical issues during their work to substantially complicate the encrypted communication mechanism. as much as possible to preserve personal information while significantly reducing the possibility of attacks. Depending on how complex and distinct the requirements established by these various applications are, the potential of trying to break them continues to occur, and systems for evaluating and verifying the cryptographic algorithms implemented continue to be necessary.

The best approach to analyzing an encryption algorithm is to identify a practical and efficient technique to break it or to learn ways to detect and repair weak aspects in algorithms, which is known as cryptanalysis. Experts in cryptanalysis have discovered several methods for breaking the cipher, such as discovering a critical vulnerability in mathematical equations to derive the secret key or determining the plaintext from the ciphertext.

There are various attacks against secure cryptographic algorithms in the literature, and the strategies and mathematical solutions widely employed empower cryptanalysts to demonstrate their findings, identify weaknesses, and diagnose maintenance failures in algorithms. The establishment of artificial intelligence models and approaches to provide a measure of the relative degree of protection for cryptographic systems represents one of the most recent strategies implemented in this field. This thesis research expanded on work through an automated testing strategy that comprises artificial intelligence methodologies to empower machine learning models to synthesize meaningful patterns and sensitive information from the ciphertext. Unlike classical cryptanalysis, the goal is to accomplish an automated and effective set of experiments without any human intervention.

The added value of our contribution is to enrich the results of previous research which addresses the shortcomings by evaluating P-box permutation algorithms using deep learning. So that we investigate the benefits of deep learning cryptanalysis techniques in the evaluation process, using the features of convolutional neural network in black box attack to detect association between interchangeable entities in order to extract the plaintext effectively and efficiently from the ciphertext without any P-box knowledge. This thesis also deals with the study of the security of lightweight encryption used in the Internet of Things through the realization of several experiments and conclusions.

**Résumé**

Avec la popularité croissante d'Internet et l'utilisation généralisée des réseaux et des systèmes d'information via l'informatique en nuage et les centres de données, la confidentialité et la sécurité des individus et des organisations sont devenues extrêmement cruciales. Dans cette perspective, le chiffrement consolide des technologies performantes qui permettent de répondre efficacement à ces exigences en protégeant les échanges d'informations publiques. Pour atteindre ces objectifs, les chercheurs ont utilisé un large éventail d'algorithmes de cryptage pour répondre aux exigences variées de ce domaine, tout en se concentrant sur des problèmes mathématiques complexes au cours de leur propre travail pour compliquer considérablement le mécanisme de communication cryptée, autant que possible , pour préserver les informations personnelles tout en réduisant considérablement la possibilité de réussite des attaques. En fonction de la complexité et de la distinction des exigences établies par ces différentes applications, le potentiel d'essayer de les briser continue d'exister et des systèmes d'évaluation et de vérification des algorithmes cryptographiques mis en œuvre restent absolument nécessaires. La meilleure approche pour analyser un algorithme de chiffrement consiste à identifier une technique pratique et efficace pour le casser ou à apprendre des moyens de détecter et de réparer les aspects faibles des algorithmes, ce que l'on appelle la cryptanalyse. Les experts en cryptanalyse ont découvert plusieurs méthodes pour casser le chiffrement, comme découvrir une vulnérabilité critique dans les équations mathématiques pour dériver la clé secrète ou déterminer le texte en clair à partir du texte chiffré. Il existe diverses attaques contre les algorithmes cryptographiques sécurisés dans la littérature, et les stratégies et solutions mathématiques largement utilisées permettent aux cryptanalystes de démontrer leurs découvertes, d'identifier les faiblesses et de diagnostiquer les défaillances dans les algorithmes. La mise en place de modèles et d'approches d'intelligence artificielle pour fournir une mesure du degré relatif de protection des systèmes cryptographiques représente l'une des stratégies les plus récentes mises en œuvre dans ce domaine.

Cette thèse a élargi les travaux qui existent, grâce à une stratégie de test automatisée qui comprend des méthodologies d'intelligence artificielle pour permettre aux modèles d'apprentissage automatique de synthétiser des modèles significatifs et des informations sensibles à partir du texte chiffré. Contrairement à la cryptanalyse classique, l'objectif est d'accomplir un ensemble automatisé et efficace d'expériences sans aucune intervention humaine.

La valeur ajoutée de notre contribution est d'enrichir les résultats de recherches antérieures qui comblent les lacunes en évaluant des algorithmes de permutation de P-box en utilisant l'apprentissage en profondeur. Afin que nous étudiions les avantages des techniques de cryptanalyse d'apprentissage en profondeur dans le processus d'évaluation, en utilisant les fonctionnalités du réseau neuronal convolutif dans l'attaque de la boîte noire pour détecter la corrélation entre des entités permutables afin d'extraire le texte en clair de manière efficace et efficiente du texte chiffré sans aucune P-box connaissance. En outre , cette thèse porte également sur l'étude de la sécurité du chiffrement léger utilisé dans l'Internet des Objets à travers la réalisation de plusieurs expérimentations et conclusions.

[illegible]

[illegible]

# ACKNOWLEDGMENTS

I WANT to thank my adviser, Pr. Makhlouf DERDOUR, for guiding me over these four years.
You were extremely helpful for guidance, offered excellent research instructions, and had wonderful patience when it came to teaching me how to write, verify things, and gain an understanding of what we were doing.
Working beside you taught me rigor, tenacity, and hard work, and I aim to emulate you in the future.
I would like to express my gratitude to Pr. Mohamed reda Laouar and Pr. Mohamed Amroune, Dr. Ahmed Ahmim, and Dr. Hicham Talbi for agreeing to review this thesis, as well as Pr Hakim Bendjena for agreeing to serve as the president of my defense jury.
My deep gratitude goes to the members of the LAMIS Laboratory, and the head of the laboratory, Pr. Hakim Bendjenna, without forgetting, of course, Pr. Amroune Mohamed and Dr. Laimeche Lakhdar. I never forget your help, your availability, and your kindness. It is a pleasure for me to collaborate with these talented and loving experts on this team.
I would like to deeply thank Dr. Mohamed Amine Ferrag from the University of Guelma, Prof. S. M. Muyeen from Qatar University in Doha, Qatar, and Prof. Mohamed Benbouzid from the University of Brest in France. Thank you for all your encouragement, advice, and support.
Finally, I want to thank my friends from all around the country and the Computer Science Department of Tebessa, without whom none of this would have been possible.

Tebessa, in 2023.

# CONTENTS

# List of Figures

# LIST OF TABLES

# GENERAL INTRODUCTION

NOWADAYS, the world has seen widespread internet use, a massive exchange of information, and an enormous amount of big data that circulates in all directions, whether in social networks or centralized information systems or in services provided by public and private companies via remote access to cloud services. Furthermore, the current revolution has seen widespread celebration and acceptance of technology such as the Internet of Things and cyber physical systems. To accomplish their purpose, the designers developed a wide security policy that comprises a large number of secure protocols, complex authentication procedures, access permissions, and a variety of additional approaches used to safeguard all of these systems and the privacy of organizations and individuals in which cryptography is one of the most important techniques for securing publicly available digital information and summarizes the effective techniques to realistically achieve these needs.

To address these issues, the designers completed several such cryptographic techniques that corresponded to the various requirements of the area of interest, as well as specifically based on their research on mathematical techniques to significantly affect the encrypted communication procedure to be as complex as possible to safeguard confidential information while minimizing the possibility of breaking cryptographic primitives.

Despite the availability of these diverse methodologies' beneficial high-end characteristics, the possibility of breaking them persists. The above tends to indicate that systems for analyzing and approving the cryptographic algorithms employed are required. Typically, the best various methods against cryptograms appear to break it, which is known as cryptanalysis.

Experts in the field of cryptanalysis have discovered numerous methods for breaking encryption, such as finding a vulnerability in the mathematical expression to deduce the secret key or the plain text from the cipher text or attacking direct implementations of algorithms on a well-specified computer system. In the literature on this subject, various methods against cryptograms appear secure. where mathematical, heuristic, and meta-heuristic approaches aid cryptanalysts in proving their work.

Cryptanalysis is a critical phase in the approach and process of developing new cryptograms, mostly because it provides for the detection of drawbacks in algorithms as well as the identification of major weaknesses in cryptographic algorithms.

According to the Kirchhoff principle, cryptanalysis is based on a variety of attacks that vary depending on the attacker's circumstances and have the primary goal of determining the key used during encryption or the plain text from the encrypted information. This is the study of theoretical or technological mechanisms aiming at breaking (trying to break) an encryption scheme, i.e. discovering the message $M$ from $C$ without prior knowledge of the key $K$. In some circumstances, it will also be necessary to locate this key $K$.
A cipher is used to an original message or plaintext to create a confused message or ciphertext through encryption and decryption is the inverse of this procedure, which converts ciphertext back into plaintext.

## 0.1 Problematic

A simple internet search on cryptanalysis will yield a large amount of work on various fields of cryptography, with multiple types of techniques and methods. The classic methods used by experts in the field of cryptanalysis are generally, in the pure case, attacks with brute force to break the encryption by trying all the possible keys, which takes a long time and requires considerable computing capacity, and the experiments are based on the evaluation of the decryption results, or else they are based on the search for weaknesses in the mathematical expressions to deduce sensitive information from the parameters of the algorithm, such as the number of rounds used, part of the private key, or a relationship between the key and the ciphertext, in which most of the works that we can found in the literature of this field, are based on the use of heuristic and meta-heuristic optimization methods to reduce the time of attacks and improve the results. Human participation in this entire process is extremely evident, especially during the analyzing phase of the attack outcomes. In this situation, the experts analyze the findings using their experience, either by attempting to conclude the private key employed and the text cipher in the textual data or by analyzing the visual results in the case of cipher picture analysis. This intervention's goal is to discover certain features and extract patterns from a large amount of data manually. The major problem is that human intervention in the evaluation, which is essential, may ignore important criteria or not be able to detect digital features in the tests carried out, despite the quality and performance of the attack. which directly implies the results of the tests and the possibility of detecting the algorithm's flaws and defining recommendations for correcting them.

## 0.2 Motivations

The emergence of artificial intelligence in general and machine learning approaches specifically has accelerated the growth of this discipline alongside the development of AI. Machine learning is a data analysis technique that automates computers in learning certain qualities of data which can then be used to predict outputs based on a set of inputs. Neural networks, which connect a series of nodes that each act to determine correlations, are a type of machine learning used to reveal associations.
In the field of application, for example; computer vision, the ability to recognize objects through machine learning has exceeded the abilities of human beings. In automated driving and cyber-security, this field has proven its power, moreover, in medical imaging, the emergence of AI techniques has consolidated very deeply the detection of concerted tumors, as well as the prediction of disease symptoms and the ability to propose automated treatment protocols. All these advantages have inspired experts in the field of cryptography and cryptanalysis to exploit AI techniques to improve the field. If you wish to view some instances, in cryptography, designers have developed numerous encryption algorithms by utilizing the architecture of neural networks to automate the encryption and decryption procedures in a lightweight manner and other works have used hybrid methods between classical encryption techniques and the combination of deep learning models to implement more reliable and more tolerant systems. in the exploitation of AI techniques in the investigation phase of encryption security, the state of the art is full of scientific articles on this subject in general. but lately, the exploitation of machine learning has experienced a period of research explosion. more particularly, the use of these techniques to minimize the time of expertise and to reinforce the detection of the faults and the weaknesses of the cryptograms with profitability more than that of the traditional methods used previously.

## 0.3 Objectives and Methodology

In this subject, we wish to exploit the techniques of artificial intelligence to overcome the problems encountered by the experts in the phase of evaluation of the results of the attacks. Deep learning models can provide automated testing and evaluation methods, easing the complex task of cryptanalysis and simplifying audits. Furthermore, these models are very easily reusable, increasing the accessibility, reuse, and development of verification and validation tools and, as a result, providing a better foundation for the emergence of new research works to improve and support the phase. verification of encryption. Inside the situation that artificial intelligence (AI) can automatically monitor several more weak cryptograms that conventional techniques do not reach, this same potential of deep learning models to critically analyze a large amount of data from an arbitrary database and extract essential features without human assistance could indeed open up a great port for research in the field of verification and validation of cryptographic systems by machine learning approaches. While deep learning has recently emphasized the importance of cryptography professionals, especially those who take an interest in block cipher cryptanalysis, the vast majority of research has decided to focus on deep learning-based black-box cryptosystem attacks. The early implementations of machine learning models throughout cryptanalysis concentrated attention on conducting experiments to try to emulate cipher behavioral patterns under that same assumption of a currently offered encryption key.

## 0.4 Organization of the thesis

This thesis is written according to the following organization:

- **Part I. Basic Concepts and State-of-the-Art:**

– **Chapter 1:** in his first part, it briefly describes the methods and techniques of modern cryptography to secure digital data in public use, in this chapter we will make an overview of public key and private key systems, or otherwise: asymmetric and symmetric cryptography respectively. also, we will see some basic notions in this field and take a deeper idea of light cryptography.
in this second part, we see the main axes of the work of cryptanalysis by the classical methods and we detail the most famous known attacks.

– **Chapter 2:** This chapter is divided into three parts, the first explains in detail the principle of machine learning and the methods used, as well as the large families of models that exist in the literature; the second part, an explanation that well describes the relationship between machine learning and deep learning, as well as details of deep learning techniques, the famous models and techniques of optimization and function activation, and the most well-known practical problems of implementation. In the third part, a detailed description of similar work on the use of techniques in cryptanalysis is provided. A significant effort has been made to classify these works using appropriate criteria, allowing for explanatory comparisons and revealing the strengths and weaknesses of each family of these approaches.

- **Part II. Contribution and Validation:**

– **Chapter 3:** This chapter explains our first contribution to the body of knowledge with a well-defined study on the problem of testing permutation primitives and the weak points of the classical work carried out in the cryptanalysis of these primitives. Where we approach this subject by adding value to the state of the art through experiments with a different view of the concept treated previously. The study is based on the exploitation of black box tests to extract the clear image from its clear counterpart by training a DL model on different databases with a diversification of the parameters and patterns of the primitives of permutation to prove the effectiveness and generality of the proposed approach, as well as the possibility of their re-use simply and effectively in the procedures of the tests of these primitives.

– **Chapter 4:** This last chapter in its first and second parts summarizes a series of lightweight algorithm security verification tests that are widely used on the internet of things. A set of experiments on some implementations of these ciphers have been briefly explained and detailed using deep learning models, and the results obtained also demonstrate the effectiveness and objectivity of the use of these types of investigations to detect encryption weaknesses and offer recommendations for the best settings that can increase security. In this third part, an architecture of a platform has been proposed for the detection of the regions of interest in the phase of the analysis of the results of the decryption of the images. This platform can be used to evaluate results more accurately and quickly than the naked eye of an expert. It uses the genetic algorithm in its operation to improve the results after each iteration and exploits pre-developed models such as the RCNN and Fast RCNN to detect objects in the results.

- **General Conclusion and Research Perspective:**
  At the end of this thesis, a general conclusion will be drawn on the state of the art of the use of DL models to verify and inspect the security of ciphers, as well as on the open problems in this subject, the challenges that exist, and the perspectives and future directions in this field.

# Part I

# Fundamental Concepts and the State of the Art

# Cryptography and Cryptanalysis 1

## Contents

## 1.1 Introduction

HUMANS' urge to keep secrets is perhaps as ancient as their capacity to speak. Although our forefathers may have found whispering to be an acceptable answer, the emergence of the written word created a new problem: how to safeguard conversations when they take place through letters or other written messages that may be intercepted.

Cryptography is the study of secure communications, specifically how to safeguard data from untrustworthy parties. Cryptography includes design, cryptography, and analysis, or cryptanalysis.
Several strategies used to modify or encrypt communications or plaintexts into seemingly nonsensical ciphertexts that can be converted back into the original message if a secret decryption procedure is understood are described in ancient history and folklore.
The majority of these solutions rely on either transposition or rearrangement of the message's symbols or substitution of the symbols by other symbols in the same or a different alphabet. However, the first cryptanalysts would eventually undermine these early cryptographers' approaches. Simple substitution ciphers, for example, can be cracked by analyzing the relative frequencies of each symbol, which should resemble the relative distributions of the letters of the alphabet. The history of cryptography is the back and forth between cryptography and cryptanalysis. One of the essential principles of cryptography was expressed in the late nineteenth century by the Dutch scholar Auguste Kerchkoffs [1] and has come to be known as Kerchkoffs' principle or desideratum. It asserts that the security of a cryptographic system must never rely on the scheme's obfuscation, but must instead rely on a randomly generated key. An encryption technique of this type includes two inputs: the plaintext and a key, which is a random sequence of letters, and a single output, which is the ciphertext. It is impossible to extract the plaintext from the ciphertext without the key. This assures that the scheme remains safe even if an eavesdropper is fully aware of how it was built. In general, a cryptographer should never underestimate a prospective adversary's skills. The usage of keys also has practical benefits: for example, keys may be changed regularly so that if one of them falls into the hands of the enemy, they will only be able to view the section of the communication in which that specific key was used. The Second World War was a watershed point in cryptography history. The Germans designed the Enigma and Lorentz machines for use in their respective fields and high-level communications. Both devices used electrical signals to flow through revolving rotors and a telephone-like switchboard to conduct a sophisticated substitution cipher. These systems were cracked owing to the perseverance of Polish and British cryptanalysts, whose accomplishments would tragically stay classified and largely unrecognized by the public for decades after the war's conclusion. The emergence of the first electro-mechanical and completely electronic computers enabled these cryptanalysis operations. Computers would become indispensable instruments for cryptography and cryptanalysis from this point forward. Mathematics was another discipline that would prove to have a tight as-sociation with cryptography. The advancement of information theory would lay the theoretical basis of cryptography. Shannon [2] demonstrated in particular that it is possible to design an unconditionally secure cipher, in the sense that any adversary is provably unable to derive anything about the plaintext from the ciphertext without previous knowledge of the key. This encryption is known as the one-time pad, and Miller and Vernam [3, 4] described similar variants of it. The one-time pad encrypts plaintext by replacing each symbol in the alphabet with another according to a key as long as the message itself. If the message and the key are both bit strings, the ciphertext is derived by XORing the key and the plaintext. Although it has excellent theoretical qualities, it has several practical issues. It needs the safe exchange of keys that are the same length as the message, and information about the plaintext can leak if the key is not created properly [5].

## 1.2 CRYPTOGRAPHY

Modern cryptography considers security in terms of computing cost: most, if not all, existing cryptosystems may be cracked by employing a general attack strategy. For example, if the key is a 128-bit string, an adversary may simply try to decode a message using all conceivable keys until they obtain a plaintext that makes sense. This would need the attacker doing test decryptions, which are usually thought to be beyond the capabilities of today and near-future computers. Of course, there is still the potential that some undisclosed shortcut exists that would allow an test to be carried out at a significantly lower cost. It is the responsibility of cryptographers to make their designs known to the community, and it is the responsibility of cryptanalysts to evaluate them to find such weaknesses. A construction's confidence should be only as deep as the community's inspection of it.

– **Confusion and Diffusion:** Shannon [2] was the first to solve this question and establish the concepts of Confusion and Diffusion, which would determine whether a cipher fits security criteria. In reality, the majority of block ciphers used today are product ciphers, which are a collection of following operations for dispersion and confusion. The confusion property is usually attained by the non-linear layer operation in a cipher, which is primarily the substitution layer.
The linear layer, which is commonly depicted as permutation or "mixing," usually achieves the diffusion feature. However, it is difficult to distinguish between the cipher components that accomplish confusion or dispersion on their own. Using confusion and dispersion methods can increase the power consumption and execution time of the cipher. Confusion often aims to confound as much as possible the link between the ciphertext and the key. A small change in a Sbox's input, for example, results in a complex change in the output.
In reality, a complicated relationship between each bit/byte of the cipher text and the encryption key is required for a solid crypto-system. As a result, a cryptanalyst will not be concerned about the system's security. A diffusion layer, however, must be introduced to diffuse this alteration across the state. The most common method of implementing diffusion is to use a bit permutation layer, which is easily achieved on the hardware level. In this situation, the adversary will have a tough time attacking the system since the redundancy has been distributed across a large number of ciphertext bits/bytes. A cryptographic system may seek to achieve one of three major security objectives:

- Confidentiality: Interlocutors must be able to exchange encrypted communications that, even if intercepted by a third party, convey no information about their contents.
- Authenticity: When exchanging communications, the interlocutors must be able to validate each other's identities so that a third party cannot impersonate them.
- Integrity: A third party must be unable to change the content, substitute, or fake communications between the interlocutors without being detected.

The introduction of computers, and subsequently computer networks and the Internet, sparked a surge in interest in cryptography, initially from the commercial sector and then from public research institutes.
This heightened interest eventually resulted in the emergence of a community dedicated to the advancement (and sometimes rediscovery) of concepts and techniques in cryptography and cryptanalysis for the benefit of the general public. This ever-expanding collection of public knowledge would become indispensable in the age of computer-based communication. Using a symmetric or an asymmetric algorithm entails significant changes in their design. Asymmetric cryptography is based on tough mathematical problems such as enormous number factorization, discrete logarithms on finite fields or elliptic curves, and lattice challenges, among others. Symmetric cryptography, on the other hand, is based on Shannon's principles of confusion and diffusion and uses fundamental arithmetic, logical processes, and/or look-up tables. Indeed, both cryptographic families collaborate to secure massive amounts of data: asymmetric cryptography ensures that a secret key is safely shared, while symmetric algorithms use this shared key to efficiently safeguard massive amounts of data.

### 1.2.1 Asymmetric cryptography

In the late 1970s, another significant advance was the advent of public-key or asymmetric encryption. Traditional cryptography is private-key or symmetric, meaning that all (pairs of) interlocutors use the same key and have the same encryption and decryption capabilities. Diffie and Hellman [6] defined the essential notion of asymmetric cryptography, and Rivest, Shamir, and Adleman [7] established the first public-key cryptosystem. Asymmetric cryptography employs pairs of related keys that are connected by a difficult-to-solve mathematical problem.
For example, RSA employs a private key comprised of two big prime integers, which are multiplied to get the public key. Factorization, which is considered a difficult task, is used to reconstruct the private key from the public key. As the name indicates, the public key can be released to the general public, whilst the private key is held by the person who developed it. When it comes to encryption, anybody may use the public key to send messages, but only the intended recipient has the private key to decrypt them. In the case of digital signatures, only the private key can be used to produce signatures, but the public key may be used to validate them.

Beyond their core design intent, some significant distinctions between public-key and symmetric cryptography strongly impact their implementation in the real world. The security of asymmetric structures is typically based on mathematical issues with significant computing complexity, such as factorization or the discrete logarithm.

Asymmetric constructions, on the other hand, frequently rely more heavily on the lack of known attacks as a security argument: if no weaknesses have been discovered despite years of rigorous analysis by cryptanalysts, then a design may be regarded as trustworthy.

However, this is a hazy line: asymmetric Cryptography also employs cryptanalysis attempts as a security justification, and several symmetric constructs include security proofs. Another notable distinction is the computational cost: public-key cryptography sometimes necessitates greater processing resources to implement. Many applications employ public-key cryptography as a key setup tool, allowing private-key cryptography to be used for communication.

Perhaps the most significant movement in Cryptography in recent decades has been the progressive trend toward public research and wider application. Until the mid-twentieth century, cryptography was a discipline strongly connected with military and governmental concerns, with most studies conducted behind closed doors.

## 1.2.2 Symmetric cryptography

This section introduces the fundamentals of block cipher creation and use, using the DES [8] standard as an example. We also go through several other typical symmetric cryptography structures, such as stream ciphers and hash functions. As previously stated, symmetric encryption is based on a single secret key that is used for both encryption and decryption. An encryption technique Enc takes as input a secret key of length k bits represented by K and a plaintext string P of p bits to generate symmetric encryption.

### 1.2.2.1 Block Ciphers

Block ciphers are one of the most often used symmetric cryptographic constructs. A block cipher encrypts plaintexts or blocks of a defined number of bits, often 64, 128, or 256, using a secret key of a certain length.

**– Definition 1.1 (Block cipher):**

A map E is a block cipher E with block length n and key size: $\mathbb{F}_2^n \cdot \mathbb{F}_2^k \to \mathbb{F}_2^n$ for which: $E_k = E(., k)$ is a bijective map (also known as permutation in Cryptography) for any key $k \in \mathbb{F}_2^n$ This map accepts a plaintext $x$ and produces a ciphertext $y$ given a fixed key K: $y = E_k(x)$.

The encryption map is the name given to the map $E_k$. Meanwhile, $E_k^{-1}$ an inverse decryption map exists for each key K.

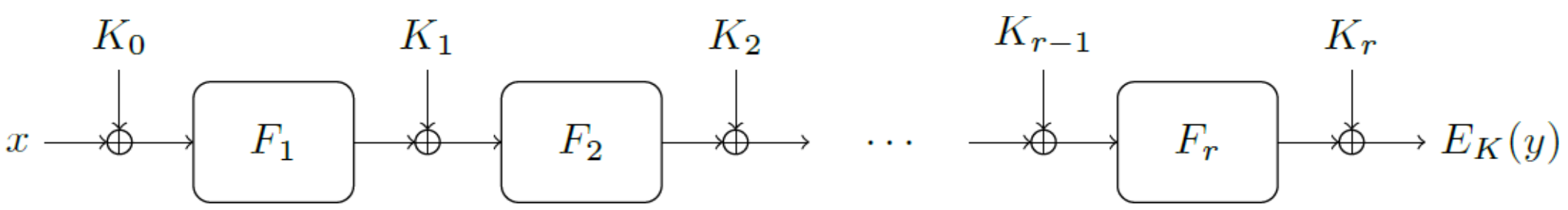

Figure 1.1 – *An r-round key-alternating block cipher diagram*

Of course, not every map E that meets the previous criteria would be a usable block cipher, as numerous performance and security requirements must be met. When the key K is known, the encryption and decryption procedures should be able to be performed at a very minimal computational cost.
This guarantees that vast amounts of data may be encrypted and decrypted quickly in a variety of application scenarios. However, we also want the block cipher to be secure. For example, obtaining any information about the key K by comparing pairings of plaintexts and ciphertexts or querying a black box that computes $E_k$ should be computationally challenging. In an ideal world, the permutation $E_k$ would be essentially indistinguishable from a randomly selected permutation of Fn2, a concept that will be discussed in further depth in the Subsection (Scenarios of Cryptographic Attacks).

#### 1.2.2.1.1 Architecture of Block Ciphers

Many block cipher designs make use of numerous common constructs. The majority of block ciphers are iterative or product ciphers in the sense that they are composed of the sequential application of simpler rounds. Each round performs some (potentially key-dependent) change that is simple to calculate, intending to obtain a safe block cipher after a sufficient number of rounds. The round sub-key is the key material that is utilized in each round. A key schedule is used to create round sub-keys from the master key K. The state refers to each of the intermediate outcomes or registers to which the round transformations are performed. Many iterative block ciphers are key-alternating [9], which implies they alternate between bitwise key additions and an independent round transformation:

**– Definition 1.2 (Key-changing block cipher ):**

$E : \mathbb{F}_2^n \cdot \mathbb{F}_2^k \to \mathbb{F}_2^n$ is an iterative key-alternating block cipher if it can be written as the following composition:

$$E_k = AL_{k_r} o F_r \;\; o \;\; AL_{k_{r-1}} o F_{r-1} \;\; o \;\; AL_{k_{r-2}} o F_{r-2} \text{.......} o \;\; AL_{k_1} o F_1 \;\; o \;\; AL_{k_0} \quad (1.1)$$

where: $F_i : \mathbb{F}_2^n \to \mathbb{F}_2^n$ are the (bijective) round functions and $AL_{k_i} : \mathbb{F}_2^n \to \mathbb{F}_2^n$ are the round key additions.

The following is the decryption map for a key-alternating block cipher:

$$E_k^{-1} = AL_{k_0} o F_1^{-1} \;\; o \;\; AL_{k_1} o F_2^{-1} \;\; o \;\; AL_{k_2} o F_3^{-1} \text{.......} o \;\; AL_{k_{r-1}} o F_r^{-1} \;\; o \;\; AL_{k_r} \quad (1.2)$$

Please keep in mind that while a block cipher may meet Definition 2.1, the specification may differ. For example, the round and sub-key numbering may change, or

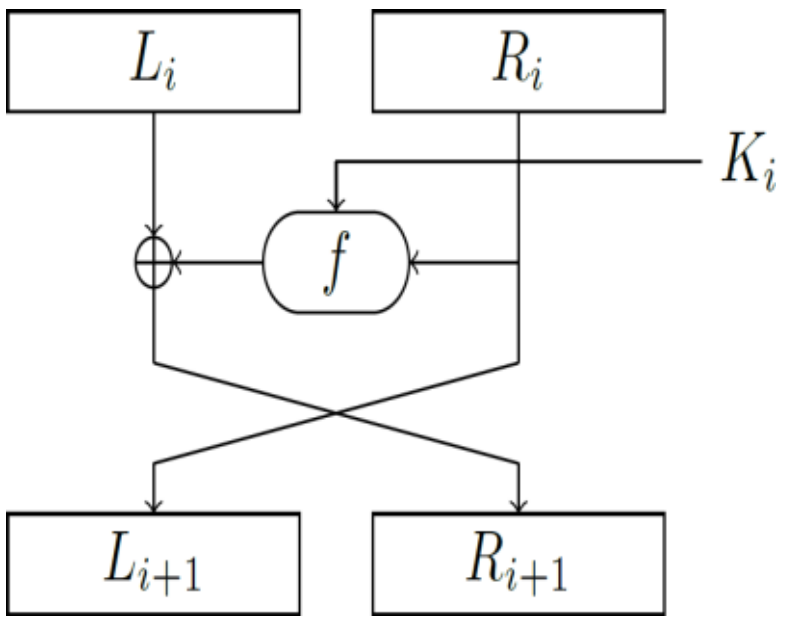

Figure 1.2 – *A Feistel network's round functions*

key addition may be regarded as the beginning rather than the last step of the round. In rare circumstances, the key-alternating structure may be obscured (for example, when the round sub-key is only XOR-ed to part of the state). We shall follow the round numbering and name rules provided in the original specification whenever we address a specific block cipher in this article.
More specifically, five common iterative constructs arise often in proposed block ciphers and have received an extensive analysis:

**a- Feistel networks:** This structure was popularized by Feistel and Coppersmith when inventing ciphers at IBM in the early 1970s, and it served as the foundation for the Data Encryption Standard, or DES [8]. GOST [10], Blowfish [11], Twofish [12], Camellia [13], CLEFIA [14], Simon [15], and Speck [15] are some more instances. A Feistel cipher's state is made up of two equal-length components, $Xi = (Li, Ri)$.
Every round, the transformation $(L_{i+1}, R_{i+1}) = (Ri, Li \oplus f(Ri, Ki))$ is used, where f can be any (non-bijective) function between $\mathbb{F}_2^{\frac{n}{2}} x \mathbb{F}_2^{k}$ to $\mathbb{F}_2^{\frac{n}{2}}$. The security aspects of this architecture have been extensively examined in publications such as those of Luby and Rackoff [33], and numerous generalizations, such as the Lay-Massey construction [34] and generalized Feistel networks, have been suggested.

**b- Substitution permutation networks (SPNs):** This is also a long-standing structure (it was already described by Feistel in [16]). An SPN's round function consists of two operations: the parallel application of tiny non-linear mappings known as S-boxes (also known as the S-box layer or substitution layer) and a linear transformation on the entire state (often called the linear layer or permutation layer). The concept is that the S-boxes produce a very complicated but local state modification (sometimes referred to as confusion), whereas the linear layer provides a less complex but global alteration (often called diffusion).
SPN block ciphers frequently include the round key by XORing it to the state, resulting in key-alternating block ciphers. SPN is frequently reserved for ciphers in which the linear layer is a permutation of the state's bits or words, such as PRESENT [17], RECTANGLE [18], and GIFT [19]. Other ciphers, such as the Advanced Encryption Standard or AES [20], Serpent [21], and Noekeon [22], employ linear layers that do not bit permutations but are nonetheless referred to be SPN in the literature. A popular way to build linear layers [23] is to divide them into a mixing layer and a transposition layer. For example, the 128-bit AES state is made up of 16 8-bit words that are seen as $F_{2^8}$ elements and are organized in a 4x4 grid. Each of the 8-bit registers is assigned a nonlinear map by the S-box layer Sub-Bytes. The mixing layer Mix-Columns performs a linear transformation on each grid column, which is specified by a GL matrix ($F^4_{2^8}$, $F^4_{2^8}$ ). Shift-Rows' transposition layer performs a permutation on the text in each grid row.

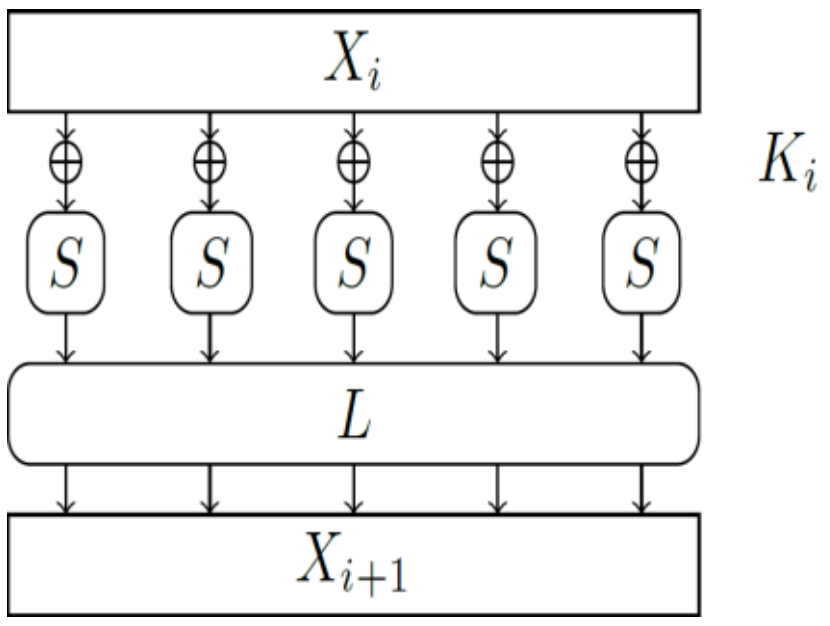

Figure 1.3 – *A SPN's round functions*

**c- Add-rotate-XOR (ARX) constructions:** ARX structures divide the state into many registers of the same amount of bits l and perform three fundamental operations on them: addition modulo 2l, circular bit rotations, and bitwise XOR. Because all of these operations are contained in the instruction sets of most processors, the goal is to obtain primitives with very excellent software performance. The stream ciphers Salsa20 and ChaCha [24] and the block cipher family SPECK [15] are well-known examples.
These constructs are not mutually exclusive and can be used in primitives other than block ciphers, but most block ciphers employ at least one of them. Key scheduling is another significant aspect of cipher design. Key schedules that are poorly constructed can lead to a variety of attacks, including meet-in-the-middle [25] and slide [26, 27] tests.

**d- NLFSR-Based**: These are ciphers that use stream cipher building blocks. They use nonlinear-feedback shift registers as their foundation. They are efficient in terms of hardware implementation, and their inner components' security is based on stream cipher analysis.

**e- Hybrid**: This one incorporates many forms of the aforementioned block ciphers. It often seeks to improve certain cipher metrics like throughput, number of rounds, and execution time.

#### 1.2.2.1.2 Block Cipher Operational modes

A block cipher E may only encrypt plaintexts with a set length equal to the block size n, which is typically 64, 128, or 256 bits. For messages that do not fit inside this specified length, a mode of operation must be defined that explains how larger inputs will be handled. The message must be broken into n-bit blocks, and in some circumstances, a padding scheme must be defined to increase the message length to a multiple of n. We also need to decide how to feed these n-bit blocks through the

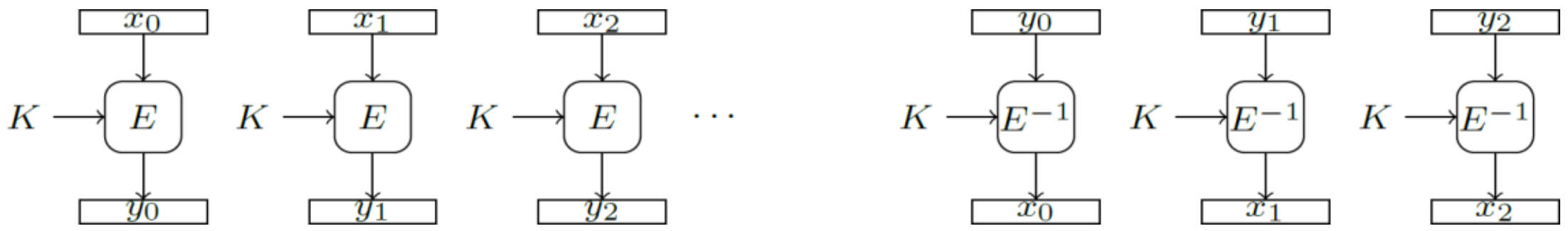

Figure 1.4 – *The (ECB) mode encryption and decryption*

block cipher. [28] The NIST recognizes five modes of operation that enable message confidentiality:

- **Electronic codebook (ECB):** Each block in ECB mode (Figure 2.4) is encrypted separately using E. Although it has several advantages, such as the ability to decode each portion of the message independently, it has poor security. An adversary, for example, may be able to recognize recurring blocks in the message because similar plaintext blocks result in identical ciphertext blocks.

- **Cipher block chaining (CBC):** In the commonly used CBC mode [29] (Figure 2.5), the result of each block's encryption is utilized to whiten the input of the following block. This has the effect of removing any patterns in the plaintext from the ciphertext. Any modification in a plaintext block is propagated to all following blocks. The first block is whitened with a random initialization vector IV. In general, the initialization vector is regarded as a nonce, that is, a non-secret string that is only used once. This also has the advantage of producing distinct ciphertexts each time a message is encrypted.

- **Cipher Feedback (CFB):** In CFB mode (Figure 2.6), each encryption output block is sent through the block cipher before being XORed to the next input block. This option will also remove any plaintext patterns. It offers various advantages over CBC mode, including the elimination of the need to construct block cipher decryption because E is used for both encryption and decryption. It also eliminates the requirement for plaintext padding because truncating the block cipher's final output to the right size suffices.

- **Output feedback (OFB):** OFB mode (Figure 2.7) constructs a stream cipher using the block cipher (see the next section). A keystream is created by passing an initialization vector through many iterations of E, which may then be XORed to the plaintext to obtain the ciphertext, in a manner similar to the one-time pad.

- **Counter mode (CTR):** CTR mode (Figure 2.8) is also a stream cipher. Each plaintext block is XORed to the block cipher's output, which is given an input consisting of a nonce and a counter indicating the location of the block in the plaintext. This model has the benefit of allowing you to encrypt and decode individual message blocks separately.

Additional modes for authenticated encryption exist in addition to these confidentiality-only ones. Along with the ciphertext, these modes generate a tag or message authentication code (MAC), which may be used to verify the message's authenticity. Some authenticated encryption techniques combine an encryption mode with a tag calculation, whereas others do both at the same time. In recent years, there has been a focus on authenticated encryption with associated data (AEAD), which allows for the binding of certain unencrypted associated data to the ciphertext while simultaneously authenticating the associated data.

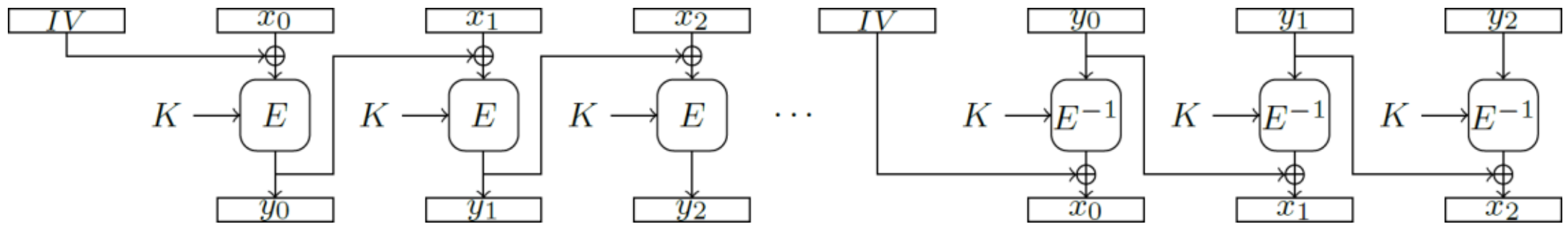

Figure 1.5 – *The (CBC) mode Encryption and decryption*

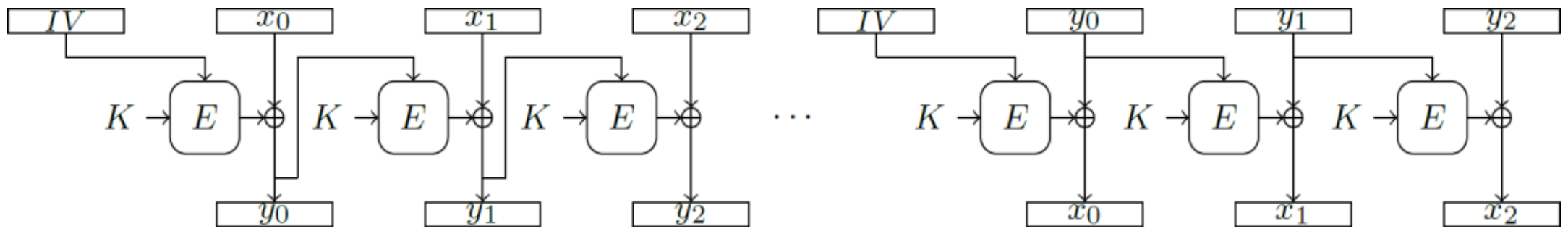

Figure 1.6 – *The(CFB) mode encryption and decryption*

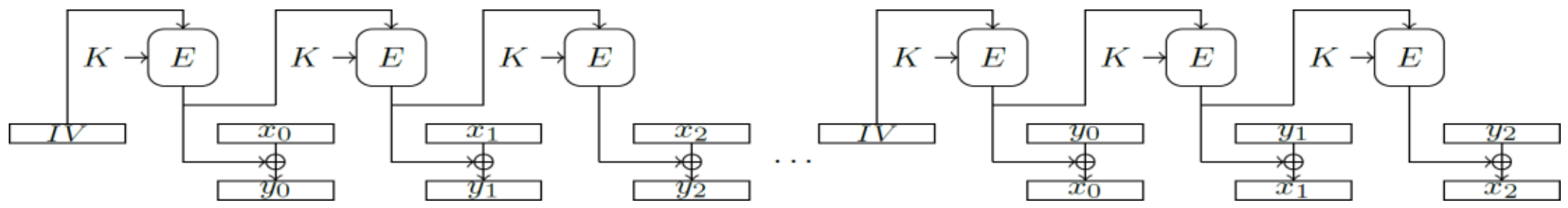

Figure 1.7 – *The (OFB) mode encryption and decryption*

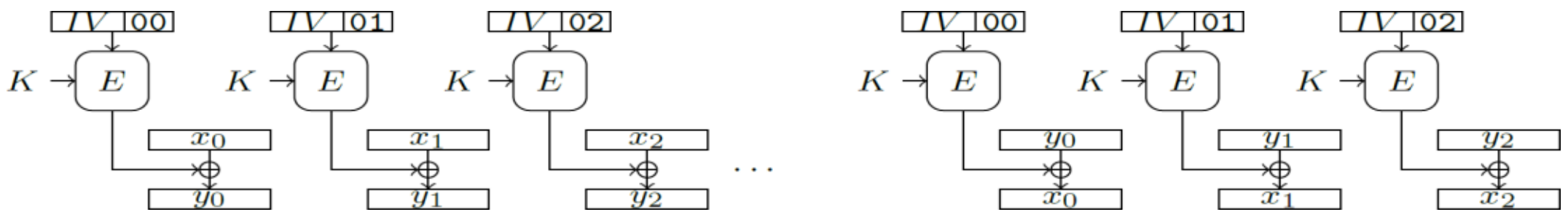

Figure 1.8 – *The (CTR) mode encryption and decryption*

#### 1.2.2.1.3 Other Symmetric architectures

Other frequent fundamental primitives and constructions in symmetric cryptography, in addition to block ciphers, can be employed in situations where a block cipher would be impracticable (stream ciphers) or if an aim other than encryption is sought (hash functions). We will not adopt a hard classification because the goal of this section is only to familiarize the reader with certain typical cryptographic algorithms that are not block ciphers: for example, a block cipher in OFB or CTR mode can be called a stream cipher.

### 1.2.2.2 Stream Ciphers

A stream cipher is an encryption method that is based on an internal state that is created using a secret key and an Initial Value (IV). To produce a key stream, this state is iteratively updated. To compute the ciphertext, the plaintext is XORed with the produced key stream. The need to employ a more practical version of the one-time pad drives the development of stream ciphers. Instead of creating a random key as long as the plaintext, the interlocutors share a shorter master or seed key, and an expansion function is applied to it to produce a pseudo-random bit sequence termed the keystream, which may be XORed to the plaintext. Stream ciphers are typically easier to build and are commonly employed when the duration of the plaintext cannot be established in advance, as in phone conversations.
Many stream cipher designs are made up of three separate parts:

- An initialization function that employs the master key and maybe a nonce to populate an initial internal state.
- An update function that is applied repeatedly and alters the internal state. Linear feedback shift register (LFSR) and nonlinear feedback shift register (NLFSR)-based update functions are widely used.
- An extraction function that creates a key stream segment from the state between update function operations.

Other techniques to stream cipher design exist. ChaCha [24], one of the most popular stream ciphers, is based on an ARX-based pseudo-random permutation that is utilized in counter mode. Along with the nonce and the counter, the 256-bit seed key is included in the permutation's input.
The security required from a permutation is difficult to define and is dependent on how the permutation is used. A good permutation should, heuristically, avoid any structural feature that may be met for given input and output pairings.

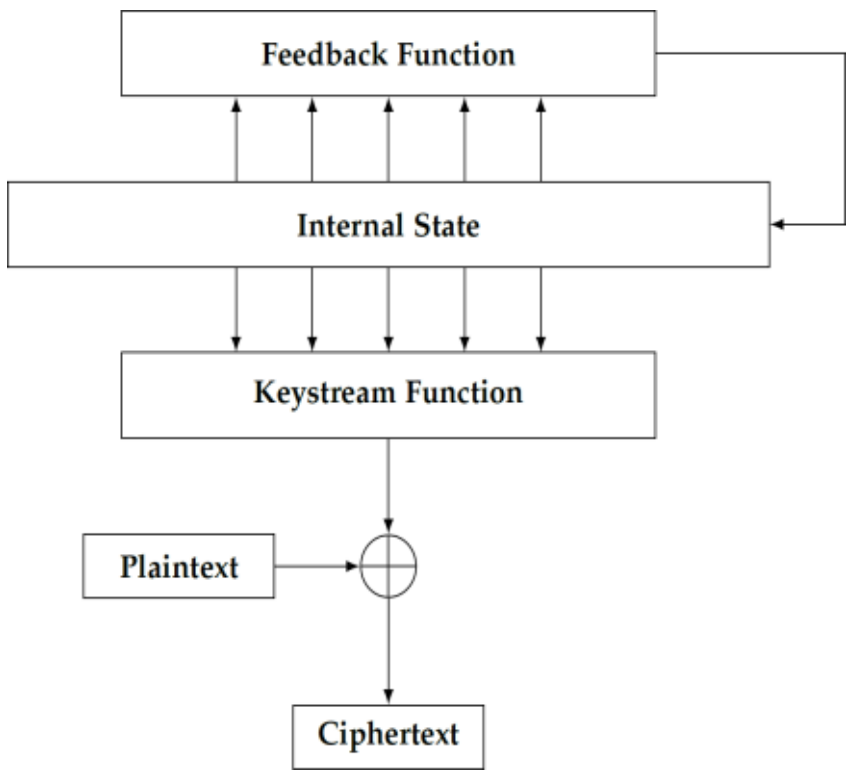

Figure 1.9 – *Stream cipher architecture*

### 1.2.3 Permutation based cryptography

Cryptographic permutations can be employed as a building block in symmetric primitives like the Even-Mansour construction [30] or the sponge, duplex, and variations [31]. Cryptographic permutations are therefore a component of many primitives, ranging from block and stream ciphers to hash functions and even AEAD primitives. Informally, a cryptographic permutation is a reasonably simple bijective map from Fn2 to itself that behaves like a randomly selected permutation, or a block cipher without a key. When employed as a symmetric primitive, we want the resultant construction to be as secure as if a real random permutation had been used, and there are no attacks that take advantage of the cryptographic permutation's special structure. In the case of sponges, for example, this approach is known as the hermetic sponge strategy: there should be no particular features or structural distinguishers specific to the permutation that an attacker may use, such that the only conceivable attacks are those generic to any permutation.
Cryptologists are also interested in pseudo-random functions, which are constructed similarly but without the bijection condition. However, from a cryptographic standpoint, pseudo-random functions are entirely distinct things, as the occurrence of collisions of the kind f(x) = f(y) can have significant consequences on security. Several alternative structures, such as the Feistel network, can be used to derive a permutation from a pseudorandom function.

### 1.2.4 Hash Functions based cryptography

Although hash functions have no key and so theoretically do not belong under either symmetric or asymmetric cryptography, they are frequently believed to be part of the former due to similarities in both design and cryptanalysis. A hash function is a simple map H that accepts any length bit string as input and returns a hash with a fixed length. Hash functions have several applications in computer science, both inside and outside of cryptography. Cryptographic hash functions, in particular, have a variety of uses, including digital signatures and password management. A hash function must have the following criteria to be deemed cryptographically secure:

- Preimage resistance: Hash functions should only have one direction. Given a goal hash y, finding a preimage x that hashes to y, that is, y = H(x), must be challenging.
- Second preimage resistance: Given an input x, finding a second message $x'$ with $x'$ = x so that both messages have the same hash, H(x) = H($x'$), should be tough.
- Collision resistance: Collisions of the hash function, that is, pairings x, $x'$ for which H(x) = H($x'$), should be difficult to discover.

The most recent NIST standard for cryptographic hash functions is the SHA-3 family of hash functions [32], which was chosen in a selection procedure and includes functions that generate hashes of 224, 256, 384, and 512 bits using the Keccak sponge architecture. The same standard offers the SHAKE128 and SHAKE256 extensible output functions (XOFs), which may generate outputs of any length d.

### 1.2.5 Lightweight cryptography

Because of the critical requirement for innovative algorithms that can fit in today's gadgets, lightweight cryptography has recently evolved. Indeed, the embedded devices that are frequently employed in many systems have their unique requirements. The linking of various embedded devices results in Mark Weiser's renowned idea of ubiquitous computing [210]. The emerging paradigm in information technology is widely acknowledged to be ubiquitous computing.
Today, we live in the era of everything being connected; the phrase being used is the well-known IoT Internet of Things. RFID tags, wireless sensors, embedded sensors and devices, and so on are all devices having processing power, battery life, and memory limits.
Lightweight cryptography has been proposed to address the limits that traditional ciphers cannot, particularly in IoT. These algorithms must adhere to hardware/software constraints while maintaining a high level of security.
Block ciphers have been widely utilized in lightweight cryptography. The majority of the block ciphers proposed and implemented for restricted devices are table below:

- **BLOCK LIGHTWEIGHT CIPHERS:** Block ciphers have been widely utilized to implement lightweight cryptography. The majority of block ciphers proposed and implemented for constrained devices are listed here: 3-Way (1994) [211], RC5 (1995) [212], Misty1 (1997) [213], BKSQ (1998) [214], Khazad (2000) [215], PRESENT (2007) [216], RECTANGLE (2015) [217], SKINNY (2016) [218], SPARX (2017) [219].
- **STREAM LIGHTWEIGHT CIPHERS:** The most well-known lightweight ciphers suggested for limited devices are listed below: Trivium (2006) [220],

Enocoro-80 (2008) [221], MICKEY v2 (2008) [222], A2U2 (2011) [223], Sprout (2015) [224].

- **LIGHTWEIGHT HASH FUNCTIONS:**It is more difficult to develop and implement a lightweight hash function than it is to create and implement a lightweight cipher. True, they generally demand a bigger internal state, which is relevant for desktop computers, but this would be pricey on a constrained device. However, we include some of the lightweight hash functions below: Armadillo (2010) [225], Spongent (2011) [226], Blake2s/b (2013) [227], Quark (2013) [228].

## 1.3 CRYPTANALYSIS

Cryptanalysis is an audit process that drives designers to construct more resilient cryptographic algorithms and evaluates the overall performance of the methods.
Cryptanalysis is the study of cryptographic algorithms by cryptanalysts from the perspective of an adversary attempting to breach the security of cryptographic systems. Informally, the longer an algorithm has been studied, the more trustworthy it is regarded by the cryptography community. The new ideas might be inspired by prior cryptographic systems that have earned the confidence of the cryptographic community and benefit from earlier security analyses, or they can be constructed from the ground up, offering novel design techniques that increase performance, security, or both. In any scenario, they must be thoroughly investigated. Only after confidence is gained via solid design justification and cryptanalysis and communicated throughout the cryptographic community can a cryptographic technique be employed in real-world settings.
Even if an algorithm has no significant faults, the discovery of undesired qualities might warrant distrust. When faults emerge after extensive deployment, the effects can be disastrous. It can lead to dishonest persons exploiting these vulnerabilities, necessitating swift updates to avoid these possible attacks.

### 1.3.1 Cryptanalysis Attacks

This part will go over some fundamental features of cryptanalysis of block ciphers, as this is a topic addressed in multiple chapters of this paper. We shall begin by presenting a general taxonomy of block cipher attacks. The section continues with a very quick description of differential cryptanalysis, one of the most thoroughly researched families of attacks against block ciphers and other symmetric primitives in general. We also go through a variation of the differential approach using boomerang and rectangular distinguishers.

**- Complexity and Strategies of Cryptanalysis**

The tactics used to target symmetric encryption algorithms may be divided into three categories: statistical, algebraic, and structural attacks.

- Statistical attacks are carried out by exploiting statistical biases in cipher output or certain intermediate variables, as well as unusual correlations with the inputs. With the appropriate input, the odd statistical behavior may be detected and used to recover secret information about the original plaintext or the secret key. In order to exploit this undesired statistical tendency, these statistical attacks sometimes require a significant degree of data complexity. Differential and linear cryptanalysis of block ciphers such as DES is the most well-known statistical attack.

- Algorithmic attacks recover hidden information stored in the secret key or plaintext by generating and solving multivariate algebraic systems over a finite field. The internal architecture of a cipher dictates the structure and availability of the system. Designers strive to boost non-linearity and improve linear dispersion across the cipher to prevent algebraic attacks.

- The last category, structural attacks, depend on faults in system design to reveal them. They disregard intrinsic elements of the fundamental components, such as statistical or algebraic properties, and instead focus on the structure itself. The square attack, for example, which was designed to target the Square cipher but is now applicable to a round-reduced version of AES, makes use of the structural peculiarities of the ciphers.

Following the stat of the art, the following three quantities are measured to determine an attack's efficiency:

- Data Complexity: The total number of queries queried to Oracle is utilized to assess an attack's data complexity. It is frequently measured in terms of the number of encryption/decryption requests made in a certain attack scenario. It is sometimes referred to as online computational cost.

- Time Complexity: The time complexity of an attack quantifies its offline computing cost. It is frequently measured in terms of a single encryption/decryption process run. It should be noted that the time complexity does not include the computational cost of Oracle related to online queries.

- Difficulty with memory. It determines how much memory is required for an opponent to launch an test. It is commonly described in terms of the cipher's block length on which the attack is undertaken.

### 1.3.2 Scenarios of Cryptographic Attacks

When the earliest cryptanalysts sought to crack early kinds of encryption, their primary goal was to extract information from ciphertext intercepted in the real world. However, modern symmetric cryptography relies on the efforts of cryptanalysts to appropriately assess the security of most primitives. This is a significant shift in viewpoint on the goal of cryptanalysis: although in the past, most attacks had to be practicable, modern cryptanalysts frequently explore methods that would be entirely impossible with current technology. In general, we prefer to think about attacks in terms of scenarios that are far larger than what is actually achievable in the real world.
This seeks to build a sufficiently significant gap between real-world attackers' greatest capabilities and the capabilities required to attack the structure. In block cipher attacks, we frequently assume that the key K is fixed and unknown to the opponent. We can categorize these attacks based on the type of data available:

- **Known ciphertext attack:** The attacker only has access to a set of ciphertexts and certain keys. There is relatively little information regarding the linked plaintexts (for example, that they consist of ASCII characters, or that they are formatted in some other specific way). The number of accessible ciphertexts determines the data complexity.

- **Known plaintext attack:** The attacker has got a (supposedly random) sample of plaintext/ciphertext pairings that they may examine. We differentiate a subset of separate known plaintext tests by assuming that no two pairings are similar. The number of accessible plaintext-ciphertext pairings determines data complexity.

- **Chosen plaintext attack:** An encryption oracle is available to the attacker. They may query this oracle with the plaintext x, and it will return the ciphertext Ek(x) using the key K. Non-adaptive and adaptively selected plaintext attacks are distinguished. In the former, the adversary may only query a list of plaintexts and is unable to perform further questions after the results are received. In the latter case, the adversary can query plaintexts individually and create further requests based on the replies received. The number of queries to the oracle represents the data complexity.

- **Chosen ciphertext attack:** The attacker has access to decryption (and, more often than not, encryption) oracle. We also differentiate between adaptive and non-adaptive selected ciphertext attacks. The total number of queries to the oracle is also the data complexity.

- **Related-key attack:** The oracle supports related-key queries in this form of the previous attack scenarios. Typically, the attacker can query encryption or decryption using the key $K \oplus \delta$, where K is the secret key and is a difference of the attacker's choosing.
  We can also examine the attacker's distinct goals:

- **Distinguishing attack:** The opponent possesses either a data list or an encryption or decryption oracle. These might be a keyed instance of E for some unknown key K or a randomly selected permutation P: $\mathbb{F}_2^n \rightarrow \mathbb{F}_2^k$. The attacker attempts to figure out which of the two possibilities is accurate.

- **Target encryption/decryption:** The attacker is given either plaintext or ciphertext and is charged with cracking it. It is encrypted or decrypted using the mysterious key K. The attacker has access to a data set or an oracle of *Ek* or $Ek^{-1}$ that may be searched for any input other than the target.

- **Key recovery:** The attacker has access to either a list of data or an encryption or decryption oracle under the unknown key K and is trying to figure out what this secret key is.

Each of these goals is more stringent than the one before it, in the sense that a key recovery attack, for example, may differentiate a block cipher from a random permutation. There may be some intermediary attacks as well. An adversary, for example, may create an alternate version of the maps $Ek$ or $Ek^{-1}$ that can encrypt or decode any plaintext or ciphertext without explicitly retrieving K. We also need to specify a baseline as to what makes an effective attack or "break". The majority of current ciphers include precise security claims that define what constitutes an attack. The exhaustive search or brute-force tests is a general key recovery technique that may be used for any block ciphers. It uses a single plaintext-ciphertext pair to find the key in no more than two s, where is the number of bits in the key.
The attack encrypts the plaintext with all possible key values until the properly related ciphertext is created. As a result, it is usual to equate the temporal complexity of block cipher tests to that of exhaustive search. This comparison is typically aided by defining the temporal complexity in terms of analogous encryptions, i.e. comparing the cost of the attack in binary operations to the cost of conducting two complete block cipher encryptions. A typical security claim is that the block cipher does not suffer from any effective differentiating attacks in an adaptively selected ciphertext scenario with a temporal complexity less than that of exhaustive search, i.e., two equal encryptions. If such a shortcut attack exists, we call the cipher broken since its true security does not match its parameters. Please keep in mind that a cipher might be declared cracked even if no practical attack exists.
Because most block ciphers are based on round iteration, we may consider a variation of the block cipher with fewer rounds than the entire specification as an attack target. We anticipate that a reduced number of rounds will be less secure and hence more vulnerable to attack. The security margin is the difference between the maximum number of rounds on which a shortcut attack is known and the entire number of rounds. Furthermore, resilience against related-key attacks is anticipated in many applications. When it comes to attacks on block ciphers, there is one more factor to consider: attempting to quantify their efficacy. This is critical when comparing tests since many attacks have a trade-off between their efficacy and their data and time complications. In the case of key-recovery tests, the efficacy may be determined by calculating the probability of success PS, or the likelihood that the attack would recover the proper key K. In the case of differentiating tests, the chance of success is not a relevant metric because a coin flip that chooses between both alternatives already succeeds with a probability of 50 %. As a result, a hypothesis-based approach is employed:

### 1.3.3 Differential cryptanalysis

Biham and Shamir originally presented differential cryptanalysis [35] to the public in 1990, and it would become the first of a lengthy series of statistical attacks on block ciphers. The fundamental principle behind differential cryptanalysis is to take advan-tage of XOR discrepancies between pairs of states (or other fixed shifts by an element of a group) that propagate with high probability across the internal components of the block cipher. This section seeks to offer the essential information required to describe the differential-linear and rectangular attacks that will be discussed later in this article.
Differential cryptanalysis is a very active family of attacks on various block ciphers and other structures. There are other varieties of the differential attack, including impossible differential attacks [36], truncated and higher-order differentials [37], and boomerang [38] and rectangle [39, 40] attacks, which are explored in Subsection 1.3.2.3. There are several instances of impossible, truncated, and traditional differential attacks, including attacks on FEAL [41, 42], DES [8], SAFER [43], CRYPTON [44], Camellia [45, 46, 47], TEA and XTEA [48], Skipjack [49], and CLEFIA [50].

#### 1.3.3.1 Differentials attack Characteristics

**– Definition 1.4 (Differential):** Let f: $\mathbb{F}_2^n \rightarrow \mathbb{F}_2^k$ A differential is any difference between input and output : $\Delta, \Delta' \in \mathbb{F}_2^n$ often denoted by $\Delta \rightarrow \Delta'$
if no confusio n is possible. The differential probability of the differential $\Delta \rightarrow \Delta'$ is:

$$DP \quad \Delta \xrightarrow{f} \Delta' \quad = \frac{1}{2^n} |x \in \mathbb{F}_2^k : f(x) + f(x + \Delta) = \Delta'|.$$

*Thisde f initionappliestoanykeyedblockcipherinstance.Becausetheprobabilityo f thedi f f erentialmaybede*

$\Delta$ and ending at $\Delta'$.
To compute the differential probability of the corresponding differential features, examine all potential intermediate differences on each round of the cipher. Because the number of differential features is frequently exceedingly enormous, it is typical to exclude any characteristics with differential probability less than a specific threshold and regard their contribution as unimportant. This can get relatively accurate answers depending on the distribution of the differential probabilities of the features. In certain circumstances, a single differential feature dominates the overall likelihood of the difference.

#### 1.3.3.2 Differential Key Recovery Attacks

Differential cryptanalysis is frequently used on block ciphers as a key recovery attack, such as Biham and Shamir's differential attack on DES [8].
It extends a differential $\Delta \rightarrow \Delta'$ for a few rounds by adding certain important guesses. The attacker conducts encryption queries for pairs of plaintexts carefully chosen to display the input difference at the start of the differential for each key guess in the first few rounds of the cipher.
Similarly, the key guess at the bottom is used to verify the output difference $\Delta'$. If enough pairs are formed for each key guess, then excellent differential pairings should arise with a considerably higher probability for the key guess corresponding to the secret key K than for the other guesses. An extensive search will yield the remainder of the key. From an algorithmic standpoint, we must deal with key guesses on the plaintext side, which affect pair production, and key guesses on the ciphertext side, which affect pair checking.
The differential's output difference propagates through the final rounds of the cipher, producing an indeterminate difference in certain sections of the ciphertext while leaving the remainder of the ciphertext unchanged. Given a ciphertext pair, we must estimate the key k sections that interact to partly invert the past few rounds and determine if the output difference holds. If there is a discrepancy between the ciphertexts in, we can reject the pair for all key guesses. After we've eliminated all of the rejected pairings, we can identify the key estimates that result in the output difference for each of the remaining pairs. We count the number of excellent pairings for each key guesses.
The key guessing process may be made more time efficient by utilizing the fact that any ciphertext pairings that are not equal will be discarded (this is often called early rejection, sieving, or filtering). Indeed, by sorting all the plaintexts and ciphertexts in a structure by value, we may quickly locate all pairs of plaintexts that have the same value. We can see which key guesses in the top portion $k_{up}$ will provide the proper input difference and which guesses in the bottom part k will produce the output difference for each of these potential pairings. Each guesses ($k_{up}$, $k_{down}$) that results in a good pair is recorded for the final exhaustive search phase.

#### 1.3.3.3 Boomerang and Rectangle Attacks

The boomerang attack [38], which was invented by Wagner, is one of the most prominent variations of differential cryptanalysis. The boomerang attack is a selected ciphertext attack that divides a block cipher E into two halves.
The calculation of the likelihood of the boomerang returning is rather wrong. For example, the fact that the middle differences are not fixed may increase the likelihood, implying that the total probability is the consequence of combining the probabilities for all conceivable middle differences. Furthermore, the assumption that both differentials act independently is frequently incorrect.
Several tools, the most well-known of which is the Boomerang Connectivity Table or BCT [51], have been introduced in recent years to better understand the behavior during the transition between both differentials.
The amplified boomerang or rectangle attack, invented by Kelsey et al. [52] and Biham et al. [53], is a frequent version of the boomerang attack. The rectangle attack attempts to convert the boomerang distinguisher, which is a selected ciphertext attack, into a chosen plaintext attack by focusing on the output difference$\nabla$ rather than the input difference $\Delta$.
The literature has several instances of boomerang and rectangle attacks on block ciphers and reduced-round versions, including COCONUT98 [38, 54], Serpent [52, 53, 55], IDEA [53], and KASUMI [53].
Later, many cryptanalytic techniques are developed based on differential cryptanalysis, like, impossible differential [56], mixture differential cryptanalysis [57], yoyo cryptanalysis [58], retracing boomerang cryptanalysis [59], exchange attack [60], higher-order differentials [61], truncated differential attack [61, 62], extended truncated differential attack [63], related key attack [64, 65, 66], meet-in-the-middle (MITM) attack [67], etc. Except for the secret key, the basic design of the cipher is deemed publicly known in such attacks. There are, however, attacks in which the underlying s-box is regarded as secret [68, 69, 70].

### 1.3.4 Linear cryptanalysis

Any iterated encryption scheme, such as substitution-permutation networks, can be subjected to linear cryptanalysis. This is an established plaintext attack. As a result, we have a huge number of pairings of plaintexts and ciphertexts. Linear cryptanalysis, in its most basic and wide sense, can be defined as a class of attacks on symmetric structures that employ probabilistically imbalanced linear combinations of bits. This concept may be traced back to known-plaintext attacks on the block cipher FEAL [71] in works like [72, 73], as well as (rapid) correlation attacks on stream ciphers [74, 75]. Both techniques in Matsui's linear cryptanalysis employ a linear approximation with a high correlation in absolute value because the data complexity of a linear attack is inversely proportional to the attack's complexity. square of the bias. This brings up the following issues:

- How can we estimate the correlation of a block cipher given a linear approximation?
- How can we find linear approximations that have a high absolute correlation?

These concerns have sparked a significant amount of research, which is still ongoing. In general, we wish to utilize the cipher's structure to determine the behavior of its linear approximations from the attributes of its components.

## 1.4 CONCLUSION

A broad cryptographic notion has been provided in this chapter. First, we discussed the fundamentals and broad ideas of cryptography. The basic security services are presented, with explanations for each. The fundamental premise of cryptography The principle of Kirchhoff's is explained. Following that, we distinguished between symmetric and asymmetric algorithms, as well as their methods of implementation. We discussed block ciphers and their modes of operation in the symmetric cryptography section. Then we went over stream ciphers. The hash functions are also shown. The limits that hash functions encounter are also addressed. Each of these cipher primitives can perform one or more security functions. We discovered that hybrid systems offer the highest probability of obtaining a completely secured crypto-system. Symmetric algorithms, as well as asymmetric crypto-systems, are extensively employed today, and both are being implemented to combat new and hazardous threats and enable secure communication. This type of system is well-known and is known as a hybrid system. To summarize, the purpose of this chapter is to provide the groundwork for the subsequent chapters, which will deal with more in-depth elements of cryptography.

# ARTIFICIAL INTELLIGENCE AND CRYPTOGRAPHY

# 2

CONTENTS

## 2.1 Introduction

ARTIFICIAL intelligence (AI) is a relatively recent branch of science and engineering. After WWII, work began in earnest, and the term was coined in 1956. Frequently recognized as an "area I wish I was in" by scientists from other fields. Throughout history, four approaches to IA have been pursued, each by a different group of people using different methods. A human-centered approach must include empirical research, including observations and hypotheses about human behavior. A rational approach entails a combination of mathematics and engineering. The various groups are defined and mutually assisted; here are the four approaches:

- **Acting humanly :**The Turing test is defined as "the art of designing machines that do activities that need intellect when performed by humans." [115]
- **Thinking humanly :** Cognitive modeling "The exciting new challenge of building computers that think, machines with consciousnesses, literally and figuratively" [116]
- **Thinking rationally :** Laws of Thought are defined as "the study of the computations that enable perception, thinking, and action"[117].
- **Acting rationally :** the rational agents "IA is the study of intelligent agent conception." [118]

AI is divided into numerous subdomains, ranging from the most basic (learning and perception) to the most specialized (playing chess, showing mathematical theorems, creating poetry, driving a car, or detecting illnesses). AI is proven to be beneficial in all cognitive activities. It is genuinely a multi-disciplinary and global area.

**– Objectives of artificial intelligence:**
The objective of AI research is to develop technology that enables computers and machines to operate intelligently. The overarching challenge of developing intelligence has been subdivided into various sub-problems. These are the abilities that scientists believe an intelligent machine will be able to do.

- **Reasoning and problem solving:** For tough tasks, an algorithm demands a high number of computational resources, and the quantity of memory and execution time required might become astronomical. A primary focus is a hunt for more efficient problem-solving algorithms.
- **Knowledge representation:** In AI, knowledge representation is critical. Several of the challenges that the machines will have to answer will need extensive knowledge of the world. Items, attributes, categories, relationships between objects, situations, events, causes and consequences, knowledge about knowledge, and many other things must be represented by AI.

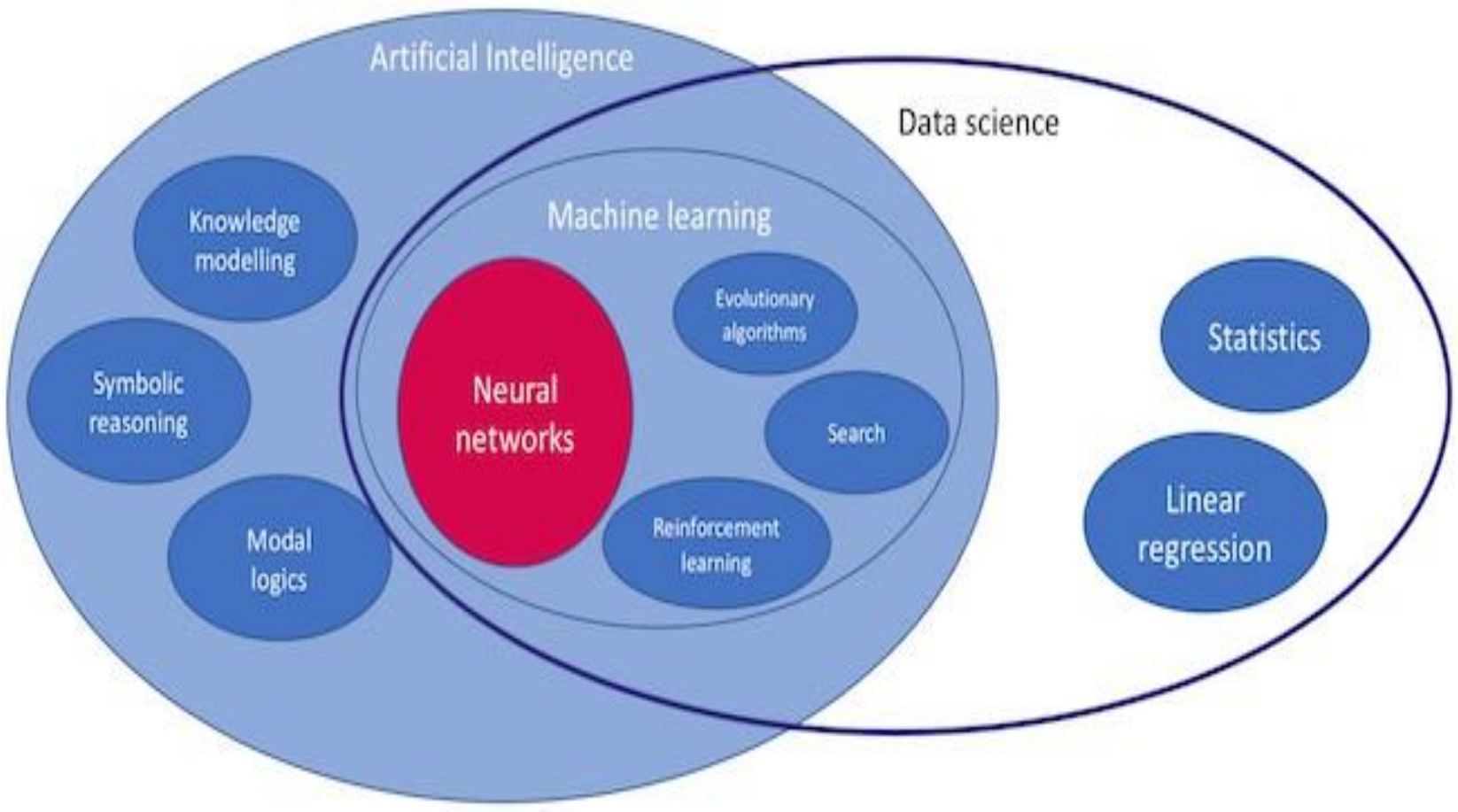

Figure 2.1 – *Relation between Machine Learning and Artificial Intelligence .* [287]

- **Planning:** Intelligent agents must be able to select and achieve goals. They need a mechanism to depict the status of the world and forecast how their actions will affect it.
- **Learning:** ML, or machine learning, is the study of algorithms that have been refined via trial and error.
- **Creativity:** Artificial creativity is at the crossroads of AI, cognitive psychology, philosophy, and the arts. The goal is to use computers to mimic or reproduce creativity to develop robots endowed with human creativity, to better understand human creativity, and to enhance its level without having to be creative yourself.
- **Perception and manipulation:** Biological systems have the ability to adapt to new settings, but not always successfully. Human programs are currently unstable; if they are compiled for one architecture, they cannot execute on another. Can we create software that installs itself on an unknown architecture?

## 2.2 Fundamental concepts

Machine learning is an AI subject that allows computers to learn from a set of observations known as a training set. Each observation, such as “I ate such and such foods at such and such a time of day during such a period, which caused such a disease” is described using two types of variables:

- The first are known as predictive factors (or attributes or characteristics), such as my age, medical history, and medical antecedents. These are the variables from which predictions are hoped to be made. The n predictive variables associated with observation will be represented as a vector $x = (x_1, ......, x_n)$ with n components. An ensemble of M observations will be made up of M of these vectors $x^1,.... ,x^M$.

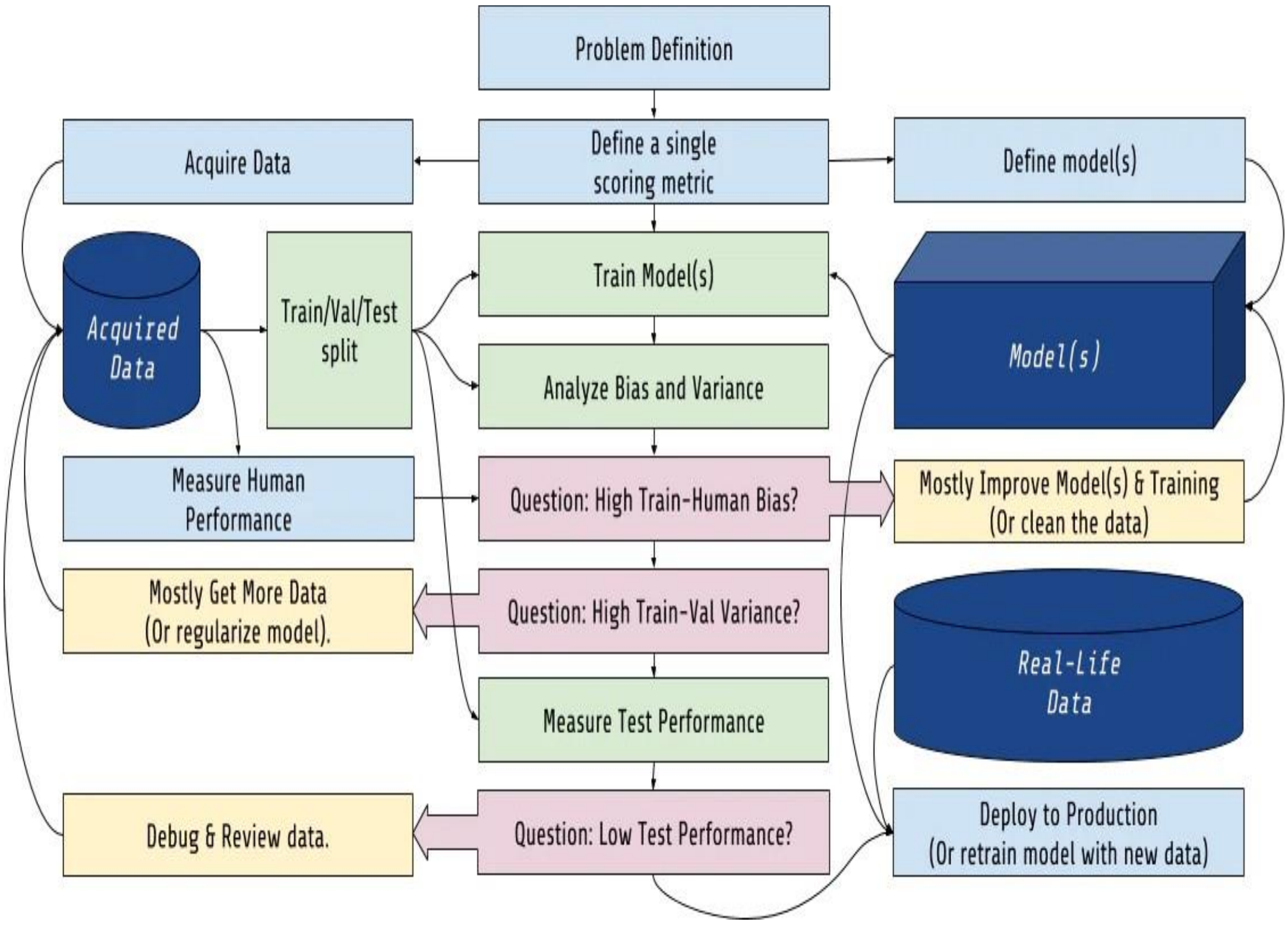

Figure 2.2 – *Machine learning process* [288]

- A target variable whose value we wish to anticipate for future events. In our case, this would be the sickness that was contracted. This variable will be referred to as y.

Learning is the process of acquiring new information or abilities, and it is one of the most basic aspects of intelligence (Simon, 1983). An agent's learning capacity, whether biological or artificial, allows it to accomplish a job more efficiently than the rest of the population (Simon, 1983). Since Hebb's postulate was presented in the late 1940s, different descriptions have been developed to suit various artificial intelligence (AI) topologies (Judd, 1990). In 1959, Arthur Samuel defined AI as a "field of study concerned with the ability of computers to learn without being explicitly programmed" (Samuel, 1959). Despite its novelty, the notion has acquired significant popularity in the scientific community in recent years.

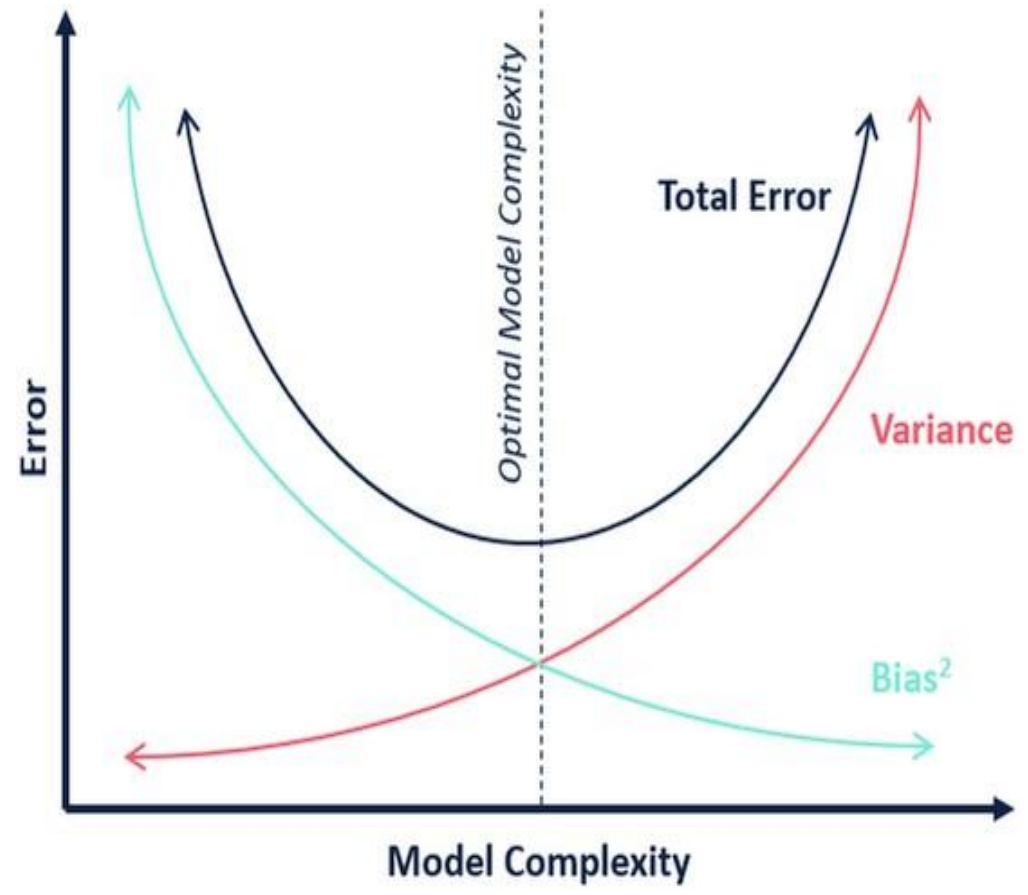

Figure 2.3 – *Model Overfitting and Underfitting in Machine Learning* [289]

### 2.2.1 Machine learning model performance measurement

We can think that the performance of a model will be a function of the correct predictions made on the observation set used for learning, the higher it is, the better.
However, this is absolutely incorrect; what we desire from ML is not to properly forecast the values of known target variables that have been used for learning, but to anticipate those that have not yet been seen. As a result, the quality of an ML algorithm is determined by its ability to correctly predict future observations based on the features learned during the training phase. As a result, we must avoid having an ML model that is so over-trained that it can precisely predict the training data but cannot generalize to the test data. This is referred to as over-learning.
Over-fitting occurs when the model is excessively complicated concerning the function F that we intend to learn.
Over-learning: the graph shows the evolution of the error committed on the test set compared to that committed on the training set, the two errors decrease but as soon as we enter a phase of over-training, the training error continues to decrease while that of the test increases.
Figure 2.1 describes the learning process employing the pre-processed database as well as the overall process from model subsequent generations to deployment. We divide the supplied data into two separate categories to overcome this challenge. The first will be the training set, followed by the test set. Cross-validation is used to provide a decent separation of data into training and test data. The objective is to randomly divide the data we have into k equal-sized chunks. One of these k parts will serve as a test set, while the rest will comprise the learning set. After each sample has served as a test set once. The prediction error is estimated by taking the average of the k mean errors.

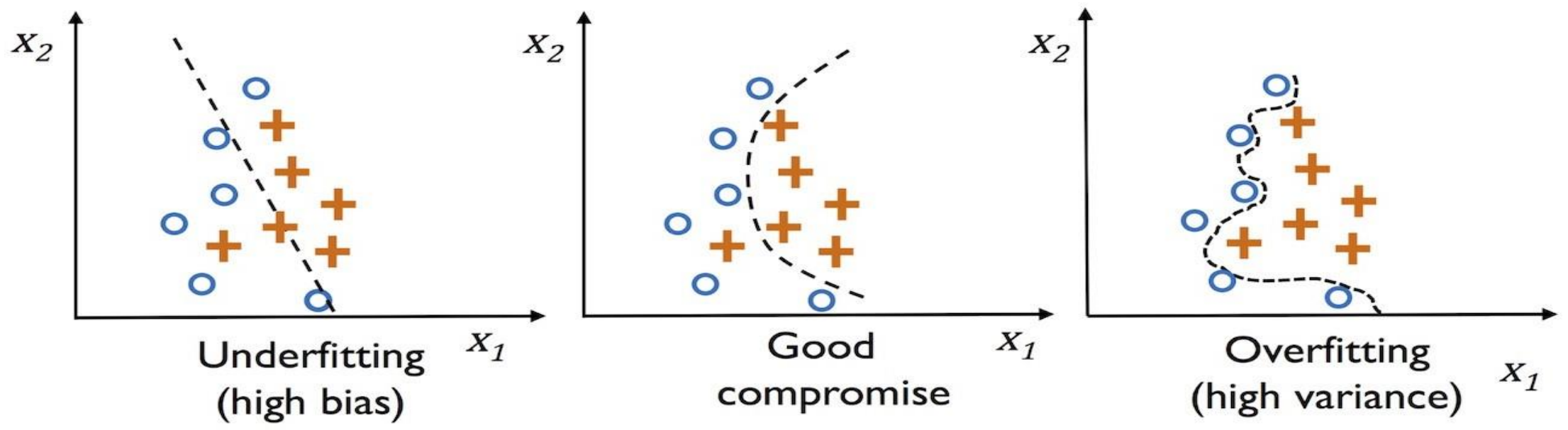

Figure 2.4 – *Model Performance, Underfitting and Overfitting*
[290]

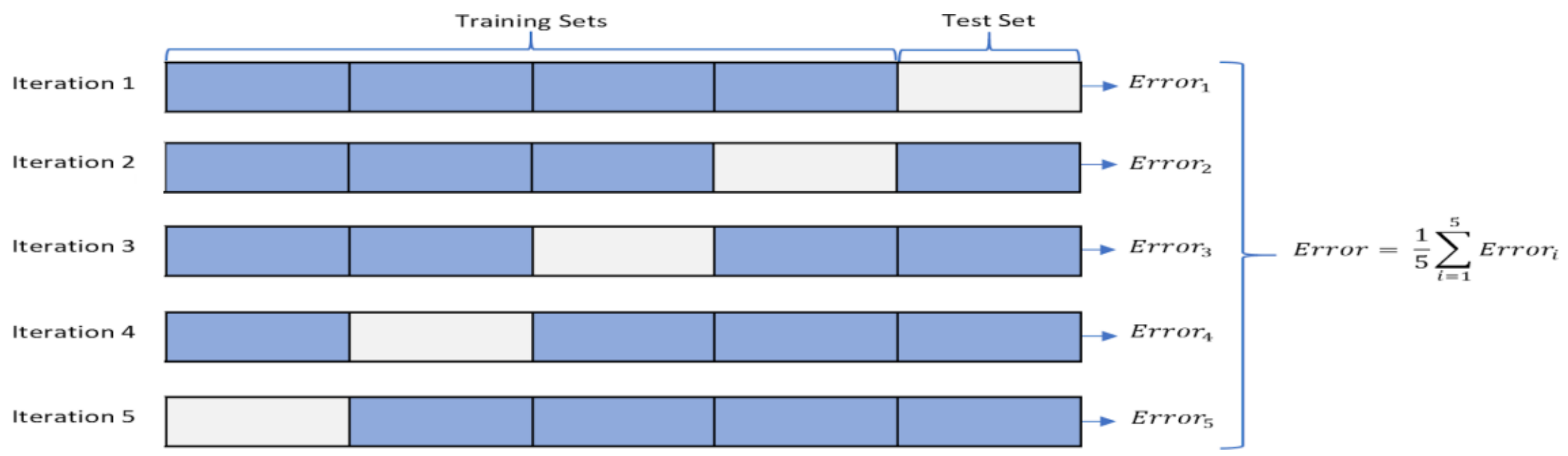

Figure 2.5 – *Cross-validation*
[289]

### 2.2.2 The types of machine learning

Machine learning is classified into three types: supervised, unsupervised, and semi-supervised. While supervised learning is a well-established area of machine learning, some consider unsupervised and semi-supervised learning to be newborns. Although novel learning paradigms are gaining popularity, supervised machine learning algorithms continue to dominate research.

#### a- Supervised learning

The task of supervised learning is:
let the training set be composed of N examples of input-output pairs:
$(x^1, y^1), (x^2, y^2), \ldots\ldots\ldots, (x^M, y^M)$
each $y^i$ was generated by an unknown function $F(x) = y$, The primary purpose of supervised learning is to discover the pattern that connects inputs and outputs and find the function $f$ that approximates $F$.
Simply said, each input value represents an integer dimension vector of data that the machine learning system must learn. During training, the dimensional vector, also known as the feature vector, conveys the algorithm's comprehension of X. From Y's perspective, the output can be anything, but it is assumed that it correlates to a category or nominal variable in the training dataset.

When the result Y is categorical, the machine learning algorithm is said to be doing a classification problem; when the result Y is a real number, the approach is said to be conducting a regression task.

**– Regression and classification:**

- A classification model is an ML model whose outputs fall into a certain range of values (example: good, average, bad)
- A regression model is an ML model with numbers as outputs (for example, tomorrow's temperature).

**b- Unsupervised learning**

Unsupervised learning or clustering does not require the data to be labeled beforehand. The aim is for the model to be successful at categorizing the available observations on its own.

**c- Semi-supervised learning**

Semi-supervised learning falls somewhere in the middle of these two approaches. The model includes some labeled instances, but the majority of the data is unlabeled. There are several applications where getting data is simple but labeling it involves work, time, or money, such as:

- It costs nothing to record a significant volume of speech in speech recognition, but identifying it requires individuals who listen to it.
- There are billions of online pages available, but to rate them, you must read them.

## 2.2.3 The types of machine learning algorithms

### 2.2.3.1 Supervised machine learning model

**a- Linear regression:**
Linear regression is a supervised ML model that has the form $y = w_1x + w_0$ with $x$ as input and $y$ as output, where $w_0$ and $w_1$ are real values to learn.
We define w as the vector $[w_0 , w_1]$ and conclude:$f(x) = w_1x + w_0$.

- Benefits: The advantage of this model is that learning is focused on solving the equation $W^* = Argmin_w Loss_f$ that seems to be right:$w^* = (X^TX)^{-1}X^Ty$
- Drawbacks: Sensitive to noise and Interactions between predictor variables are ignored.

**b- K-Nearest Neighbors (KNN):**
A supervised classification technique, the k- Nearest Neighbors (KNN) algorithm. Each training set observation is represented as a point in an n-dimensional space, where n is the number of predictor variables. We seek the k points closest to this example to forecast the class of an observation. The class of the target variable is the one with the most representation among the k nearest neighbors. There are versions of the technique in which we weight the k observations based on their distance from the example we wish to categorize [119], with the observations further away from our example being regarded as less essential.

- Benefits: Easy to design.
- Drawbacks: Sensitive to noise and the calculation of the distance becomes prohibitively costly when there are a significant number of predictive factors.

**c- Naive Bayes classifiers:**
The naive Bayes classifier is a probabilistic supervised method that believes the existence of a feature for a class is independent of the existence of other features (thus the word "naive"). A guy is defined as someone who weighs a specific amount and stands a given height. Even if these features are connected in reality, a naïve Bayesian classifier will decide that a person is a man based on height and weight alone.
Despite incredibly simple underlying assumptions, this classifier produces very good results in a wide range of difficult real-world circumstances. Research published in 2004 demonstrated that there are theoretical grounds for this surprising efficacy [120]. Another research from 2006, however, found that more current techniques (reinforced trees, random forests) provide superior results[121].

- Benefits: The algorithm offers good performance.
- Drawbacks: The prediction becomes erroneous if the Conditional Independence Assumption is invalid.

**d- Decision Trees:**
Decision trees are supervised machine-learning models that may be used for both classification and regression. A decision tree is a function that accepts a vector of characteristics as input and returns a single-value decision. Discrete or continuous inputs and outputs are possible. A decision tree produces choices by conducting a test sequence; each internal node of the tree corresponds to a test of an attribute's value, and the branches that emerge from the node are the potential values of the attribute. The leaf in which the observation comes after the test sequence then determines the class of the target variable.
The learning step involves determining the best test sequence. We must first pick which characteristics to maintain. A useful attribute splits the instances into homogeneous sets, which only contain cases from the same class, but a worthless feature leaves the examples with almost the same proportion of value for the target variable. A formal definition of "good" and "useless" is required. There are standard homogenized metrics for measuring the homogeneity of a set for this purpose. The Gini diversity index and entropy [122] are the most well-known.
The entropy of a random variable V with values $vk$ each with probability $P(vk)$ is defined in general as

$$H(V)=-\sum_K P(v_k)Log_2P(v_k). \quad (2.1)$$

- Benefits:
  - It is a white box model, simple to understand and interpret.
  - Little data preparation.
  - Input predictor variables can be both qualitative and quantitative.
  - Efficient on large datasets.

- Drawbacks: The existence of a risk of overfitting if the tree becomes very complex. Pruning procedures are used to circumvent this problem.

**e- Random Forests:**

Combining several independent weak classifiers to build a single classification system, known as Classifier Sets, has piqued the scientific community's attention. The effectiveness of classifier combinations is primarily determined by their capacity to use the complementary of individual classifiers to improve total generalization performance as much as feasible. The random Forests method [123], formed of a series of elementary classifiers of the decision tree type, is one way for generating sets of classifiers. The purpose of the algorithm is to maintain the advantages of decision trees. decision while minimizing their shortcomings, notably their susceptibility to overfitting. It is a method that may be used for both classification and regression.
The algorithm is built around three key concepts:

- By drawing with replacement, we construct B new samples of the same size M for M observations from the training set, each defined by n predictive factors. This is referred to as bootstrap. A decision tree will be trained using each of the B samples.

- For each n feature, a number k n is taken at random so that at each node of the tree, a subset of k features is drawn at random, and the best one is then chosen for partitioning.

- We use a majority vote to categorize a fresh observation. This observation is run through the B trees, and its class is the most common among the B predictions.

Some research projects have concentrated on the number of decision trees that should be planted within a forest. When Breiman introduced random exercises, he also proved that beyond a certain number of trees, adding more did not always increase overall performance. This finding suggests that the number of trees in a random forest does not have to be as enormous as feasible to generate a high-performing classifier.

- Benefits:
  - In terms of accuracy, it is one of the top algorithms.

– Incorporation of cross-validation.
–Even when there is missing data, it retains strong predictive power.
– It does not suffer from over-fitting.

- Drawbacks: A difficult implementation.

**f- Support Vector Machine (SVM):**

Support Vector Machines (SVMs) [124] are very effective nonlinear binary classification techniques.
SVMs work by generating a nonlinear separation band of maximum width that divides two sets of observations and then utilizing it to create predictions. SVMs employ a nonlinear $T$ transform to transport the points $x^1$, ......$x^M$ from the original n-dimensional space (n is the number of predictive variables) to new points $T(x^{(1)})$, , $T(x^{(M)})$ in a space greater than n or they will be easier to separate.
SVMs are classifiers that are based on two essential concepts:
The initial concept is to find a linear divider with the greatest possible width; this is the concept of maximum margin. The separation border and the nearest samples are separated by the margin. These are known as support vectors. The difficulty is determining the best separation border.
The choice of the separating hyperplane is not evident if the issue is linearly separable. There are an infinite number of separating hyperplanes whose performances in the learning phase are identical but whose performances in the test phase can differ greatly. It has been demonstrated [125] that there is a single optimum hyperplane, defined as the hyperplane that maximizes the gap between the samples and the separating hyperplane, to solve this problem. This decision was made on theoretical grounds. According to Vapnik [125], the capacity of distinguishing hyperplane classes reduces as their margin rises.
To cope with scenarios when the data is not linearly separable, the second important notion of SVMs is to change the representation space of the input data into a higher-dimensional space where a linear separation is likely. This is accomplished by the use of a kernel function, which must obey the constraints of Mercer's theorem [126] and has the advantage of not needing explicit knowledge of the transformation to be used for space change. Kernel functions enable the costly transformation of a dot product in a high-dimensional space into a simple point evaluation of a function. The kernel trick is the name given to this procedure.
The polynomial kernel and the Gaussian kernel are the two most commonly utilized kernel functions.

- Benefits:
  – It allows dealing with complex nonlinear classification problems.
  – SVMs are an alternative to neural networks because they are easier to train.

- Drawbacks: SVMs are often less efficient than random Forests.

**g- Multi-Layer Perceptron (MLP):**

The Multi-Layer Perceptron (MLP) [127] is a neural network-like classifier composed of numerous layers, each built by one or more formal neurons. It employs the gradient back propagation technique in supervised learning [128].

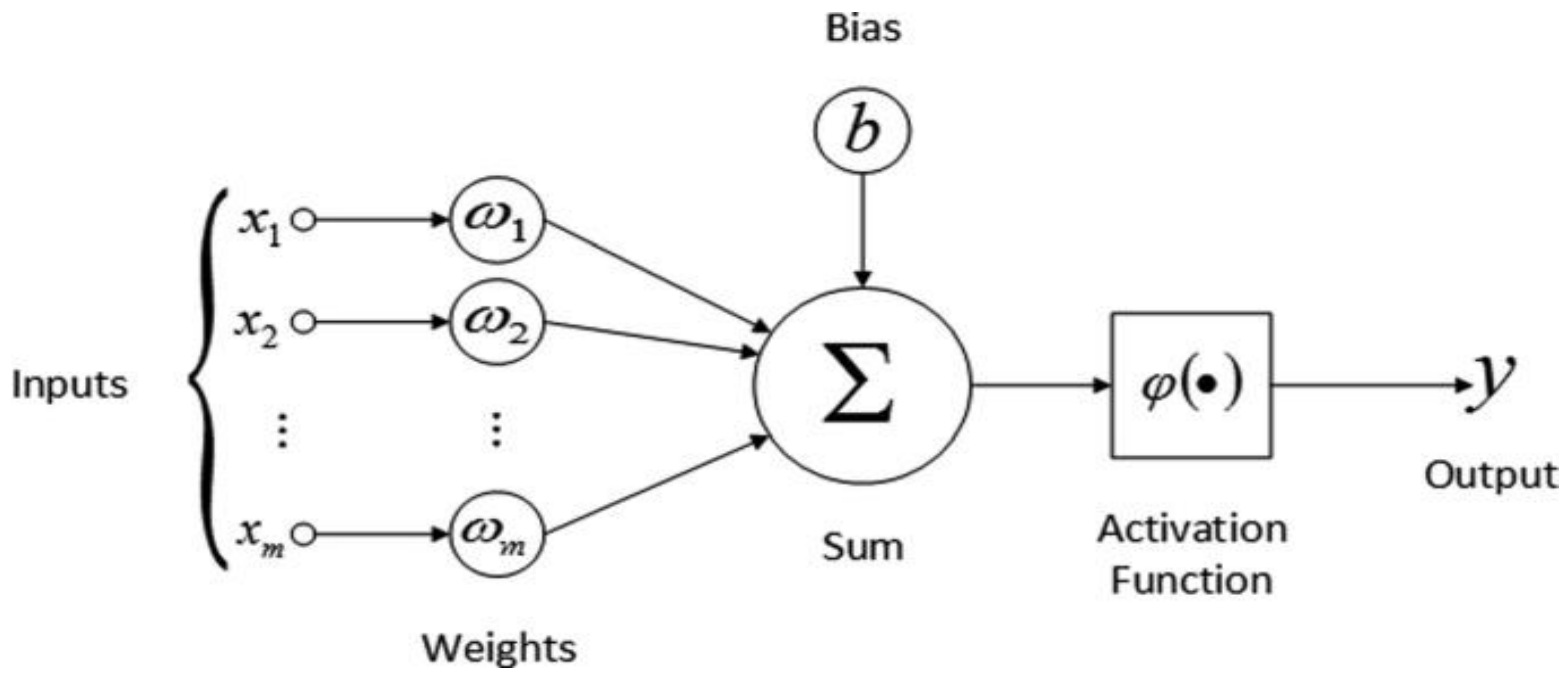

Figure 2.6 – *The figure depicts the operation of a formal neuron. It is a calculating component that computes the weighted sum of the signals received at the input and then applies an activation function to them to Y .*
[292]

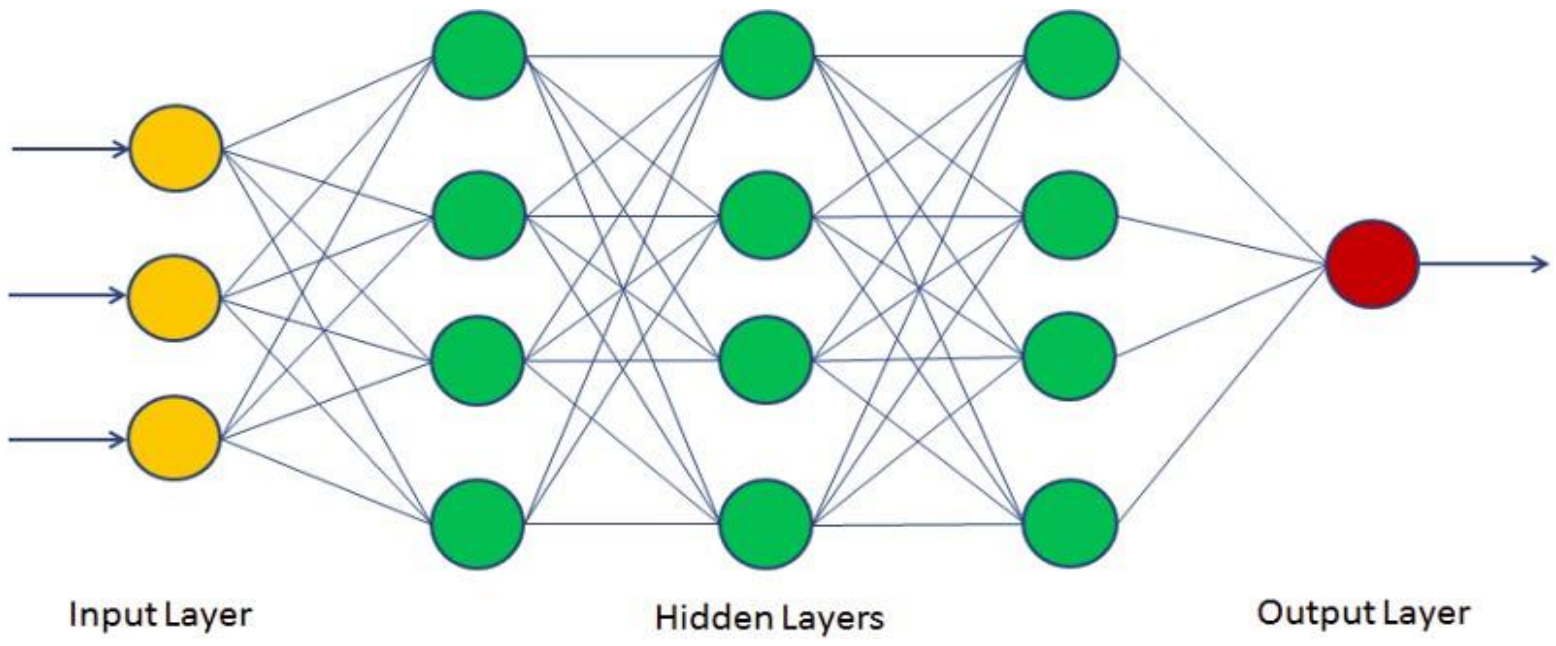

Figure 2.7 – *An MLP is made up of an input layer, three hidden layers, and an output layer .*
[291]

The learning algorithm involves feeding the network inputs and instructing it to change its weight to discover the associated output. We first propagate the inputs forward until we have an output computed by the network, then we compare this output to the intended output, and lastly, we alter the weights so that the error between the actual output and the desired output is reduced at the next iteration. This method is repeated until the output error is insignificant.

- Benefits:
  – Capability to identify dependencies on their own.
  – Noise resistance.

- Drawbacks: It is a black box model with no explanation for its judgments.

ML is a huge and ever-changing field. His algorithms are inspired by a variety of sources, including probability theory, geometric intuitions, and heuristic techniques.

#### 2.2.3.2 Unsupervised machine learning model

**a- k-means** Unsupervised algorithms are used in the k-means method. Each observation is represented by a point in an n-dimensional space, where n is the number of descriptive variables. This algorithm will split a training set of $M$ observations $x^1$, ......$x^M$ into $k$ clusters so that the Euclidean distance between the points in the center of gravity and the group to which they are allocated is as little as possible. The algorithm's stages are as follows:

- Choose k points to indicate the clusters' average location.
- repeat until stabilization of the central points:
  – assign each of the M points to the closest to the k center points.
  – update the center points by calculating the centrists of the k clusters.

- Benefits: Suitable for big amounts of data.
- Drawbacks:
  – The choice of the parameter k is not discovered but chosen by the user.
  – The solution depends on the k center of gravity chosen during the initialization.

## 2.3 Deep Learning

Deep Learning is a new ML study topic that was established to bring ML closer to its core goal: artificial intelligence. It is about algorithms that are inspired by the structure and function of the brain. They may learn many layers of representation to simulate complicated data connections.

Figure 2.8 – *The relationship between artificial intelligence, ML and deep learning*

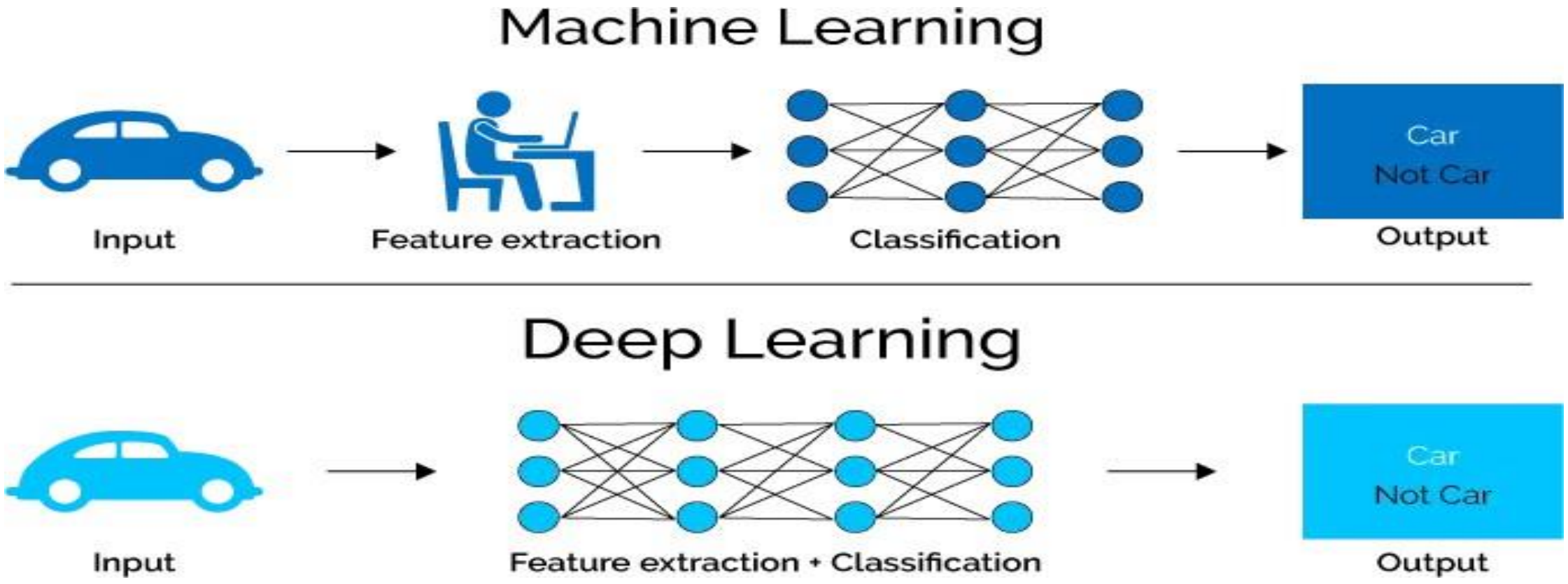

Figure 2.9 – *The process of classical ML compared to that of Deep Learning* [293]

Deep Learning is an artificial neural network-based technique that is designed to handle massive quantities of data by adding layers to the network. A deep learning model may extract features from raw data using several layers of processing that include various linear and nonlinear transformations, and then learn about those characteristics layer by layer with minimum human interaction [129],[130],[131],[132],[133],[134],[135].
Deep learning has progressed from a niche sector with just a few academics interested in it to the field most valued by researchers in the last five years. Deep learning research is currently being published in prestigious publications such as Science [136], Nature [137], and Nature Methods [138], to mention a few. Deep learning has persuaded the google [139], taught a vehicle to drive [140], detected cancer [141] and autism [142], and even helped a person become an artist [143]. Dechter (1986) [144] coined the term "Deep Learning" for machine learning, and Aizenberg et al (2000) [145] for artificial neural networks.

The machine learning methods discussed in the first section are effective for a wide range of tasks. They did not, however, address certain fundamental AI challenges, such as speech recognition and object identification.
The failure of standard algorithms in such an AI assignment prompted the creation of deep learning. But it wasn't until bigger amounts of data were available, courtesy of Big Data and linked things, and calculating engines got more powerful that we realized the true potential of Deep Learning.

One of the key differences between Deep Learning and other ML algorithms is that it scales effectively; the more data supplied, the better the Deep Learning algorithm's performance. Unlike many standard ML algorithms, which have a limit on the amount of data they can receive, known as the "performance board" Deep Learning models have no such restrictions (theoretically) and have even outperformed human performance in areas such as image processing.

The feature extraction stage is another distinction between regular ML algorithms and Deep Learning methods. Traditional ML algorithms do feature extraction manually, which is a complex and time-consuming phase that needs a professional in the subject, whereas Deep Learning methods accomplish this step automatically.

### 2.3.1 The types of deep learning algorithms

Deep architectures come in a variety of flavors. The majority of them are derived from original parent architectures. Because not all designs are examined on the same data sets, it is not always viable to compare their performance. Deep Learning is a constantly evolving discipline, with new structures, versions, and algorithms appearing every week.

#### 2.3.1.1 Convolutional Neural Network (CNN)

Convolutional Neural Networks (CNN) is a sort of network of specialized neurons with a grid-like architecture used for data processing. Time series data, which can be thought of as a 1D grid by capturing samples at regular time intervals, and picture data, which can be thought of as a 2D grid of pixels, are two examples. Convolutional networks have had a lot of success in real-world applications. The term "convolutional neural network" refers to the network's use of the mathematical process of convolution. Convolution is a particular type of linear operation. Convolutional networks are simple neural networks with at least one layer that employs convolution rather than matrix multiplication.

They have a wide range of applications, including image and video recognition [146], recommendation systems, and natural language processing [147].

**– Inspiration:**

A convolutional network is a sort of feed-forward neural network in machine learning that was inspired by biological processes [148]. The structure of the animal visual cortex influenced the notion of the connection between neurons in a CNN. Individual cortical neurons respond to stimuli in a limited space known as the receptive field. The receptive regions of the several neurons partially overlap, covering the visual field. A convolution process can approximate the response of an individual neuron to stimuli in its receptive area quantitatively.

**– The convolution operation:**

Convolution, in its most generic form, is an operation on two real argument functions. To explain the reasoning behind convolution, consider two functions that may be used. Assume we're using a laser sensor to track the location of a spaceship. Our laser sensor outputs x(t), which represents the spacecraft's location at time t. Because x and t are real numbers, we can acquire a different reading from the laser sensor at any moment.

Assume our laser sensor is relatively noisy. We'd want to integrate numerous readings to get a less noisy estimate of the spacecraft's location. Of course, more recent measures are more significant, therefore we want to weigh these measurements and give greater weight to recent measurements. This is possible using a weighting function w(a), where an is the age of the measurement.

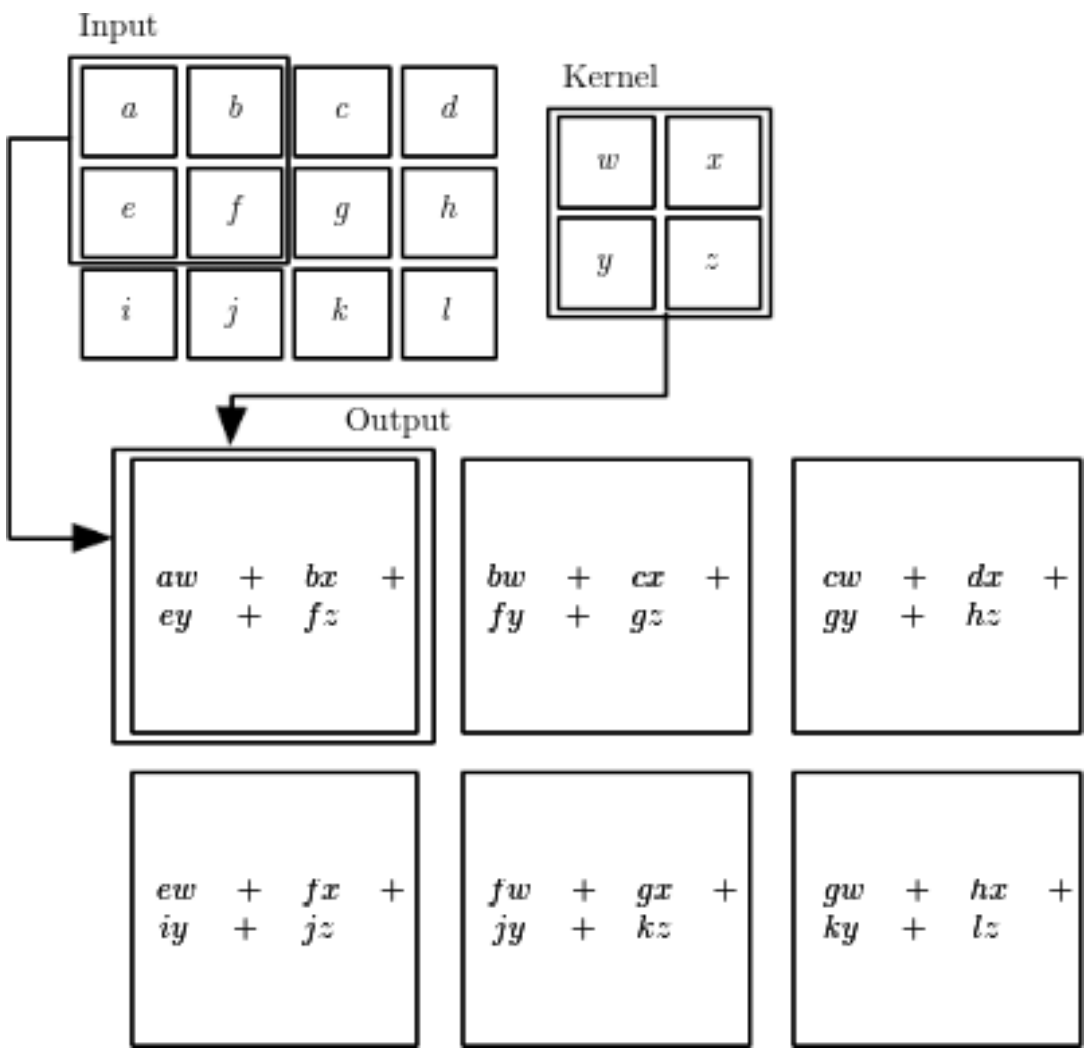

Figure 2.10 – *2D convolution example* [294]

We get a new function that offers a smooth estimate of the spacecraft's position if we use such a weighted average operation at each instant:

$$s(t) = \int x(a)w(t - a)da \tag{2.2}$$

Convolution is the name given to this procedure. An asterisk generally denotes the convolution operation:

$$s(t) = (x*w)t \tag{2.3}$$

In this scenario, the concept of a laser sensor that can deliver measurements at any moment is unrealistic. When we interact with data on a computer, time is often discrete (digitalized), and our sensor provides data at regular intervals. In our scenario, it is more reasonable to believe that our laser measures once every second. The time index t can only take integer values at that point. Assuming that x and w are integers, we may define the discrete convolution as follows:

$$s(t) = (x*w)(t) = \sum_{a=-\infty}^{\infty} x(a)w(t - a) \tag{2.4}$$

Finally, we frequently apply convolutions on several axes at the same time. Image 3.10 As an example, use the following two-dimensional picture x as input:

$$s(i; j) = (x * w)(i; j) = \sum_n \sum_m x(m; n)w(i - m; j - n) \tag{2.5}$$

The first parameter (in this case, the x function) of the convolution is generally referred to as the input and the second argument (in this case, the w function) is the kernel in convolutional network nomenclature. The result is commonly referred to as a feature map.

**– Convolutional layer:**

Convolution is based on three key notions that might help an ML system improve: sparse interactions, parameter sharing, and equivariant representations.
**– Pooling layer:**
An unusual convolutional network design comprises three different sorts of layers. First, a convolutional layer is used to create a collection of linear activations, which are then passed via a nonlinear activation layer such as Rectified Linear Unit (ReLu), before being pooled [129].

- It enables a steady reduction of the size of the representations in order to limit the number of parameters and computations in the network and, therefore, control overfitting.
- It enables tiny translations to be invariant.
- Useful when you want to know if a characteristic exists rather than where it exists.
- There are several forms of pooling (MAX pooling (extremely famous), AVG pooling, etc.).

**– Perceptron:**
After determining the properties of the inputs, we add a perceptron or an MLP to the network's end. The perceptron takes the extracted characteristics as input and outputs a vector of N dimensions, where N is the class number and each element represents the likelihood of belonging to a class. In the scenario when the classes are purely mutual, each probability is computed using the softmax function.

$$\text{softmax}(z)_i = \frac{e^{z_i}}{\sum_j e^{z_j}} \tag{2.6}$$

We can name a few well-known convolutional networks:

- LeNet [149] : Yann LeCun pioneered the use of convolutional networks in the 1990s.The most well-known of them is the LeNet architecture, which is used to read zip codes, numerals, and so on.
- AlexNet[150]: AlexNet, developed by Alex Krizhevsky, Ilya Sutskever, and Geoff Hinton, was the first study to popularize convolutional networks in computer vision. In 2012, AlexNet entered into the ImageNet ILSVRC competition [151] and surpassed its opponents. The network's design was quite similar to LeNet's, but it was deeper, bigger, and featured convolutional layers layered on top of each other (previously it was common to have only one convolutional layer always immediately followed by a pooling layer).
- ZFnet[152]: The winner of the 2013 ILSVRC competition was a convolutional network created by Matthew Zeiler and Rob Fergus. It was renamed ZFNet (short for Zeiler and Fergus Net). It was an upgrade to AlexNet by modifying the architecture's hyper-parameters, namely increasing the size of the convolutional layers and decreasing the kernel size on the first layer.

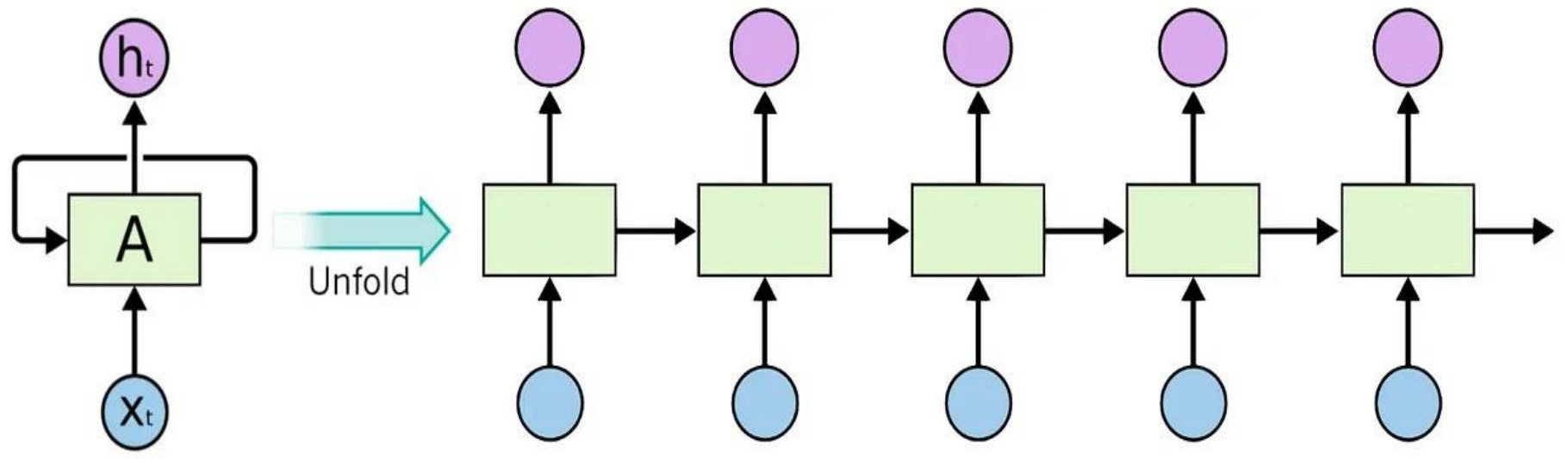

Figure 2.11 – *(left) An RNN. (right) Its unrolled version* [295]

- GoogLeNet[153]: A convolutional network by Szegedy et al. from Google won the ILSVRC competition in 2014. His main contribution was the development of an inception module, which significantly reduced the number of network parameters (compared to AlexNet's 60M). Furthermore, this module employs global AVG pooling rather than PMC at the network's end, which removes several settings. There are also other GoogLeNet variants, including Inception-v4 [154].
- ResNet[155]: Kaiming He et alresidual .'s network was the ILSVRC 2015 winner. It has connection hops and makes extensive use of batch normalization. It also employs global AVG pooling rather than PMC at the end.

#### 2.3.1.2 Recurrent Neural Network (RNN)

Humans do not start their thinking at zero every time. When we read a book, we understand each word concerning the preceding words. We don't forget everything, so let's get back to thinking. Our thoughts have tenacity. Traditional neural networks cannot accomplish this, which is a significant disadvantage. Assume we want to categorize the type of event that happens at each segment of the film. It's unclear how a typical neural network might utilize prior events in the movie to guide subsequent ones.
This issue is addressed by RNN (Recurrent Neural Networks). They are looped networks that allow information to persist. RNNs are designed to use sequential information. In a typical neural network, all inputs (and outputs) are assumed to be independent of one another. However, for many tasks, this is a terrible notion. We need to know the words that came before to predict the following word in a phrase. Recurrent neural networks (RNNs) are so-called because they complete the same job for each element of a sequence, with the outcome dependent on past computations. RNNs may also be thought of as having a "memory" that stores information about what has been calculated thus far. In principle, RNNs can utilize information in infinitely long sequences, but in practice, they can only go back a few stages. This is what a typical RNN looks like in figure 3.11. The illustration above depicts an unrolled RNN. Scrolling simply means that we are seeing the network for the whole sequence.

If the sequence of interest is a 5-word phrase, for example, the network would be unrolled into a 5-layer neural network, one layer for each word. The following formulae regulate the computations in an RNN:

- $x_t$ is the input at time t.
- The parameters that the network will learn from the training data are U, V, and W.
- The hidden state at time t is denoted by st. The network's "memory" is what it is. st is determined it using the previous hidden state and the current step's input:

$$st = f(Ux_t + Ws_{t-1}) \tag{2.7}$$

  Where f is a nonlinear function such as ReLu or Hyperbolic tangent (tanh).

- $o_t$ represents the output at time t. If we wish to anticipate the next word in a phrase, for example, it would be a vector of probabilities in a vocabulary.

$$o_t = softmax(Vs_t) \tag{2.8}$$

- The network's memory may be thought of as the hidden state $s_t$. $s_t$ records information about what occurred in the preceding phases. The output $o_t$ is determined only by the memory at time t. As briefly explained above, it's a little more problematic in practice because you can't normally record information for lengthy periods of time.
- Unlike a standard neural network, which utilizes distinct parameters at each layer, an RNN uses the same parameters (here: U; V; W) throughout all stages (parameter sharing). This reflects the fact that we are executing the same job with different inputs at each phase. This decreases the overall number of parameters that the network must learn.
- The figure above shows outputs at all times, however, this may not be essential depending on the work. For example, while estimating the mood given by a phrase, we must consider the end output as well as the sensation after each word. Similarly, we do not require inputs at every stage.

**– RNN learning process:**

Backpropagation Through Time (BBTT) [156],[157] is a significantly modified variant of backpropagation that we employ for RNN training. Because the parameters in the network are shared at all times, the gradient at each exit is determined not only by the computations of the present instant but also by the preceding steps. Following that, the chain rule is implemented.

We consider $o_t$ the network's forecast at time t. Because we take the entire sequence (for example, a complete phrase) as a single training sample, the total error is the sum of the errors at each point (each word).

The purpose is to compute the error gradient for the parameters U, V, and W and to learn good parameters through Stochastic Gradient Descent (SGD). In the same way that we totaled the mistakes, we summed the gradients at each step for a single training example.
RNNs struggle to learn long-term dependencies and interactions between distant words [158]. This is troublesome since the meaning of a statement is frequently dictated by distant words. If the activation function is a tanh or a sigmoid in this situation, we will have the vanishing gradient problem. This is why we employ the ReLu activation function, which does not have this issue, but there is an even more prevalent option, which is to use Long Short-Term Memory (LSTM) architectures. [159],[160] or Gated Recurrent Unit (GRU).
**– RNN application fields:**
RNNs are extremely effective in a wide range of natural language processing jobs. The LSTM architecture has done the greatest accomplishments of RNNs because it is considerably better at capturing long-term dependencies.

- Text creation and language modeling [161],[162],[163]: Given a string of words, we wish to forecast the likelihood of each word based on the preceding ones.

- [164],[165],[166]: Machine translation is similar to language modeling in that the input is a series of words in the source language (e.g. Arabic). We wish to produce a word sequence in a target language (eg English). One significant distinction is that our output begins only after we have viewed the whole input sequence because the first word of the translated phrase requires information acquired from the entire input sequence.

- Voice recognition [167] is a computer technology that analyzes human speech in order to convert it into machine-readable text.

- Image descriptions [168],[169],[170]: RNNs have been used in conjunction with convolutional neural networks to create descriptions for unlabeled photos. It's astonishing how well it appears to operate. Even created words are matched to visual attributes by the integrated model.

**– Long and Short Term Memory:**

The RNN is the foundation of long and short-term memory (LSTM) [129]. It was first developed to address issues that had been noted while using RNN, including disappearing or bursting gradients. It allows the network to handle lengthy time gaps between important time series of the processed data set. To do this, each unit receives a cell state (also known as a memory cell). It provides some statistical information (e.g., mean, variance) generated across a previously processed data time series. Depending on the importance of the stored information, this cell can be written on or deleted. A tiny neural network decides whether to write on the cell or clear it. This feature is particularly significant in the context of side-channel attacks, where the adversary must aggregate numerous delayed time samples to overcome masked implementations, for example.

#### 2.3.1.3 Deep generative models

While a discriminative model (for example, CNN, RNN, MLP) attempts to predict p(y|x) using y as the label and x as the input, a generative model describes how the data is formed, learning p(x;y) and predicting p(y|x) using Bayes' law [171]. A discriminative model must be employed if the aim is simply classification; however, generative models are capable of much more than simple classification, such as creating new data.
Here are some examples of a generative model:

- Boltzmann Machines[172].
- Restricted Boltzmann Machines [173],[174].
- Deep Belief Networks[175].
- Deep Boltzmann Machines [176],[177].
- Generative Adversarial Networks [178],[179],[180].
- Generative Stochastic Networks [181].
- Adversarial autoencoders [182].

### 2.3.2 The optimization approaches for Deep Learning

#### 2.3.2.1 Variants of gradient descent

Gradient descent is the most well-known approach for optimizing a neural network, in which the objective function $J(\theta)$ is minimized by updating the $\theta$ parameters in the opposite direction as the gradient of the objective function. The importance of the step we will take to attain the local minimum is determined by the learning rate $\theta$. This approach comes in three flavors. Depending on the amount of data, we will strike a balance between parameter update accuracy and update execution time.
**a- Batch gradient descent:**
We calculate the gradient of the cost function at the $\theta$ parameters for the full learning set using traditional gradient descent.

$$\theta = \theta - n.\nabla_{\theta}.J(\theta) \tag{2.9}$$

Because the gradient for the whole dataset must be calculated in order to conduct a single update, this approach can be exceedingly slow and unworkable if the data cannot be held in memory. If the cost function is convex, we will converge to the global optimum; otherwise, we will converge to the local optimum.
**b- Stochastic gradient descent:**
The parameters of the data set $x^{(i)}$ and label $y^{(i)}$ are updated through SGD (stochastic gradient descent).

$$\theta = \theta - n.\nabla_{\theta}.J(\theta; x^{(i)}; y^{(i)}) \tag{2.10}$$

This approach is quicker, but too frequent parameter updates generate oscillations in the objective function. These oscillations allow you to land in possibly better local minima, but they also make convergence more difficult. However, it has been demonstrated that by decreasing the learning rate n, SGD achieves the same convergence as batch gradient descent.
**c- Mini-batch gradient descent:**
This technique combines the best features of the previous two methods and changes the settings for each mini group of n examples:

$$\theta = \theta - n.\nabla_{\theta}.J(\theta; x^{(i:i+n)}; y^{(i:i+n)}) \tag{2.11}$$

This approach lowers the variance of parameter updates, resulting in more reliable convergence. Mini groups typically comprise 50 to 256 samples. It is a popular approach for training a neural network. Classical gradient descent may not always give satisfactory convergence and has various issues that must be addressed:

- The learning rate is difficult to choose; if it is too little, convergence will be too sluggish; if it is too large, the cost function will oscillate or will not converge at all.
- All parameters have the same learning rate. If the qualities of the observations differ in frequency, we may wish to apply a larger update to the parameters for the characteristics that occur less frequently.
- Another difficulty in minimizing the extremely prevalent non-convex cost functions in neural networks is avoiding becoming stuck in local optima. Dolphin et al. [183] contend that the issue stems from saddle points rather than local minima.

#### 2.3.2.2 Gradient descent optimization algorithms

In what follows, we will describe various methods often employed by the Deep Learning community to address the aforementioned difficulties.
**2.3.2.2.1 Momentum:**
SGD suffers in places where the surface is much more curved in one dimension than another [184], which are typically near local minima. SGD oscillates across the slopes of these areas in many circumstances, making only sluggish progress towards local optima. The momentum coefficient [185] is one way that can assist the network in escaping these traps.

$$v_t = y.v_{t-1} + n\nabla_{(\theta)} J_{(\theta)} \tag{2.12}$$

$$\theta = \theta - v_t \tag{2.13}$$

Alternatively, $y \notin [0; 1]$ denotes the momentum coefficient. A ball is moved down a slope using momentum. As it rolls down, the ball gains velocity and becomes quicker and faster. When you update your settings, the same thing happens. Momentum grows for dimensions with the same gradient direction and drops for updates with a different gradient direction. As a result, the convergence is faster and the oscillations are reduced.

**2.3.2.2.2 Nesterov accelerated gradient:**

Sutskever et al. (2013) presented the Nesterov momentum [186], which is a version of the classical momentum and is inspired by Nesterov's Accelerated Gradient (NAG) approach. The Nesterov momentum is to update the parameters as follows:

$$v_t = y.v_{t-1} + n\nabla_{(\theta)} J_{(\theta - v_{t-1}))} \tag{2.14}$$

$$\theta = \theta - v_t \tag{2.15}$$

While the classical momentum calculates the current gradient (small blue vector) before taking a large step in the direction of the cumulative gradient (big blue vector), NAG first takes a large step in the previously accumulated gradient direction (brown vector), measures the gradient, and then makes the correction (green vector). This technique has greatly improved recurrent neural network performance in a variety of applications.

**2.3.2.2.3 Adagrad:**

Adagrad [187] is a gradient descent-based optimization strategy that further basically keeps changing the learning rate following the parameters, going to produce a global footprint network given infrequent characteristics as well as fewer adjustments overall basic characteristics. frequent. Dean et al.[188] demonstrated discovered Adagrad reached the desired reproducibility of SGD and has been used at Google successfully training significant neural networks discovered for perceiving cats in YouTube videos, through these features [189]. Pennington et al. [190] have been using Adagrad to practice GloVe word representations as well because fewer important types necessitate highly updated information than many other commonly used ones.

Previously, the $\theta$ parameters were updated using the same learning rate N. Adagrad employs a unique learning rate for each parameter $\theta_i$ and each step t.

Let $g_{t,i}$ be the gradient of the objective function knowing the parameter $\theta_i$ at stage t:

$$g_{t,i} = \nabla_{\theta}.J_{\theta} \tag{2.16}$$

At step t, the update of each parameter $\theta_i$ is:

$$\theta_{t+1;i} = \theta_{t;i} - n.g_{t,i} \tag{2.17}$$

Adagrad updates the general learning rate n at each step t for each $\theta_i$ parameter in its update algorithm depending on the previous gradients computed for $\theta_i$.

Adagrad has the advantage of eliminating the need to manually modify the learning rate. The main drawback of Adagrad is the buildup of gradient squares in the denominator. As each term added is positive, the accumulation of sums increases throughout learning, lowering the learning rate until it becomes infinitesimally tiny, at which point the algorithm is no longer able to gain new information.

**2.3.2.2.4 RMSprop:**

Geoff Hinton suggested RMSprop, an unpublished adaptive learning rate algorithm, in Lecture 6e of his Coursera Class [191]. We limit the window of collected gradients to a set size w instead of accumulating all the squares of the preceding gradients.
An exponential moving average of the squares of the prior gradients is used instead of storing the w squares of the past gradients. At step t, the current average $E[g^2]_t$ is determined only by the previous average and the current gradient.

$$E[g^2]_t = y.E[g^2]_{t-1} + (1 - Y)g_t^2 \tag{2.18}$$

$$\theta_{t+1} = \theta_t - \sqrt{\frac{n}{E[g^2]_t + \epsilon}}.g_t \tag{2.19}$$

Hinton recommends that be set to 0.9, with 0.001 as a suitable default value for the learning rate n.

**2.3.2.2.5 Adam optimizer:** Another approach that produces an adaptive learning rate for each parameter is the Adaptive Moments (ADAM) optimizer [192]. ADAM retains an exponentially decreasing average of the past squared gradients $v_t$ as well as an exponentially decreasing average of the prior gradients $m_t$ as RMSprop.

$$m_t = \beta_1.m_{t-1} + (1 - \beta_1)g_t \tag{2.20}$$

$$v_t = \beta_2.v_{t-1} + (1 - \beta_2)g_t^2 \tag{2.21}$$

$m_t$ and $v_t$ are gradient estimates of first-order (mean) and second-order (variance), respectively. Because $m_t$ and $v_t$ are initialized as vectors of zero, the authors discovered that they are biased towards zero, particularly during the early steps and when the decay coefficient is minimal ($\beta_1$; $\beta_2$ near to 1), therefore they developed a bias adjustment for the first and second order estimations.

#### 2.3.2.3 Second-order methods

Another group of optimization methods in the context of deep Learning is based on Newton's method:

$$\theta = \theta - [HJ(\theta)]^{-1}\nabla.J(\theta) \tag{2.22}$$

$HJ(\theta)$ is the Hessian matrix, which is the square matrix of the function's second partial derivatives. The gradient is represented by the$\nabla$ $.J(\theta)$ word.
The Hessian matrix specifies the cost function's local curvature, allowing for more efficient updates. The multiplication by the inverse of the Hessian matrix, in particular, causes the optimization to take bigger steps in the direction of shallow curvatures and vice versa in the direction of steep curvatures. The lack of any learning rate in the update formula, according to proponents, is a significant benefit over first-order techniques.
However, because calculating (and inverting) the Hessian matrix into its explicit form is an extremely space and time-demanding procedure, the above update method is unfeasible for most deep learning models. time. A neural network with one million parameters, for example, would have a Hessian matrix of size [106; 106] that would take up around 3725 GB of RAM. As a result, a broad range of quasi-Newton techniques for approximating the inverse Hessian matrix has been devised. The most prominent of these approaches is Limited-Broyden-Fletcher-Goldfarb-Shanno (L-BFGS) [193], which leverages information in gradients across time to create the approximation implicitly. One downside of the naïve implementation of L-BFGS is that, unlike SGD mini-batch, it must be calculated across the whole training set, which might contain millions of samples. It is more difficult and takes more study to run L-BFGS on mini-batches.

### 2.3.3 The theoretical limits of optimization methods

Several theoretical studies suggest that each optimization technique devised for a neural network has a performance limit [194].
Some theoretical findings only apply when the network units provide discrete values. Theoretical studies show that there exist classes of intractable issues, but determining whether a given problem belongs to this class can be challenging. Other results suggest that finding a solution for a specific network is intractable, but in reality, we may easily discover a solution using a wider network for which many parameters equate to an acceptable solution.
Also, while training a neural network, we don't normally worry about finding the precise minimum of a function, but rather about reducing its value sufficiently to have a good error test. It is quite difficult to do a theoretical study to determine if an optimization method can attain this aim.

### 2.3.4 The Backpropagation method for loss-minimization

Backpropagation is a loss-minimization method, it is performed by altering the weights and biases progressively throughout the training phase. It influences the gradient descent procedure until the minimal tolerated loss is reached [129]. It allows a network to learn its perimeter by observing the effect of small changes in weights and bias values on the output. When a little change in the input leads to a minor change in the output, the network has only experienced a minor alteration. Furthermore, for the initial layers, the network output gradient becomes exceedingly low, giving birth to the name "vanishing gradient problem" ("What is the vanishing gradient problem? [186].
Hinton, Osindero, and Yee-Whye Teh published seminal work on the vanishing gradient problem in 2006 [175]. To put it another way, consider the slope to be a hill and the drive to be a wheel rolling down that hill to your intended destination. The wheel goes quickly when the inclination is severe, but slowly when it is level. It's the same with a deep network, which has a steep learning curve and slow development in the early stages. However, as the learning curve becomes steeper, the network accelerates [167]. The first layers of an image network give birth to a singularity. If the network's primary layers make mistakes, the future layers will as well. When a network wants to learn, it first studies the mistakes to see what weights and biases impact the output, then attempts to decrease the error by modifying the weights [175]. Backpropagation is a technique used to train networks. It resolves the gradient vanishing problem .
The other problem posed, in a deep neural network, exploding gradient is the very rapid increase in the values of the gradients during the backpropagation, which causes the capacity of the internal representation of the numbers to be exceeded and the learning to stop. Several techniques exist to counteract gradient explosion, beginning with improved initialization techniques, the selection of non-saturating activation functions such as the rectified linear function, batch normalization, and gradient clipping.

### 2.3.5 The Activation functions in deep learning models

The activation function is a mathematical function that is applied to a signal at an artificial neuron's output. The word "activation function" is derived from the biological counterpart "activation potential," a stimulation threshold that, when met, causes a neuron to respond. The activation function is frequently non-linear. Because increasing the weights of a hidden layer is essentially a linear modification, their purpose is to allow neural networks to learn more complicated functions than basic linear regression.

**– Features:**

- Nonlinearity [195]: When a function is nonlinear, a two-layer neural network can be used to approximate it.
- Differentiable everywhere: This characteristic enables the creation of gradient-based optimizations.
- Range [196]: When the activation range is finite, gradient-based learning approaches are more stable (impact on a limited number of weights). Learning is often more efficient when the range is limitless (impact on more weight).
- Monotone: The error surface associated with a mono-layer model is certified convex when the function is monotone.
- Monotone derivative: Functions having monotone derivatives have been demonstrated to generalize better in various circumstances. These functions enable the application of ideas such as Ockham's razor.
- Identity at 0: By randomly initializing the weights, these functions enable quick learning. If the function does not converge to the identity at 0, then the weights must be initialized with caution.

## 2.4 STAT OF THE ART OF THE NEURAL CRYPTANALYSIS

A simple internet search will reveal the presence of an extremely large collection of publications on the subject of cryptanalysis. In this section, we exerted a concerted effort to consolidate and organize various experiments depending on incredibly specific scientific criteria as well as the specific goals of the accomplishment of all these experiments.

### 2.4.1 Famous Cryptanalysis Attacks

In the quantum cryptanalysis technique, there are works of high scientific quality that have given good results and offer great scientific value. Exploring quantum-mechanical processes to solve computationally difficult problems has recently been a focus of academics, leading to the invention of Grover's search algorithm [76], Simon's algorithm [77], Shor's algorithm [78, 77], and others.
The adoption of such algorithms jeopardizes the security of cryptographic methods. The most noteworthy of them is Shor's algorithm, whose ability to solve the factorization issue and compute discrete logarithms in polynomial time has revealed the weakness of various public key cryptography methods, including RSA, ECDSA, and ECDH.
Because of the consequences of Grover's search algorithm on block ciphers, private key systems are subject to generic key recovery attacks [80]. Simon's algorithm's weaknesses on several specific symmetric key schemes have recently been explored [81, 82, 83, 84, 85, 86].
As a result of the inevitability of the advent of quantum computers, the security of cryptographic methods is at the point of being jeopardized. As a result of these circumstances, the National Institute of Standards and Technology (NIST) has issued a request for proposals for post-quantum cryptography standardization, intending to standardize new cryptographic algorithms that are secure against both classical and quantum attacks [87].
Thus, investigating the security of existing cryptographic methods in quantum computing models gives insights into their applicability in the post-quantum era.

Another very well-known and very famous category of attack known by the name of brute force technique, also widely recognized as an exhaustive key search attack, consists of the cryptanalysts constantly trying all possible combinations to interpret and respond to ciphertext. It is often used against any encryption algorithms [88], which would include stream ciphers, except for verifiable secured cryptosystems, such as one-time pad, which is not realistically practicable.

To break up a cipher with such key lengths of n bits, the adversary should indeed investigate $2^{n-1}$ combinations on averaging and $2^n$ values in the most extreme situation. An attack's computational cost is frequently understood as O $(2^n)$. [88] An approach is not regarded as a threat if its computation cost surpasses that required for an exhaustive key search.

Side Channel Analysis Attack is another extremely significant type of attack, in general, building any cryptographic primitive involves two phases. First, it is considered an abstract mathematical concept. Following that, this mathematical object must be realized in the form of a program, which in certain situations is then implemented in specialized hardware. Following implementation, these applications will be run in a computer environment on processing units. These executions will have certain distinct qualities. SCA refers to attacks that are based on physically observable aspects during execution. Some of the important physical characteristics used mostly for Side Channel Analysis usually involve energy and micro-controller speed required for the execution, electromagnetic waves, thermal management, and system disturbance.
Based on the properties listed above, there are many Side Channel attacks on ciphers in general and on stream ciphers in particular. Simple Power Analysis attack, Differential Power Analysis attack [89, 90], Timing Analysis attack [91, 92], Electromagnetic Analysis attacks [93, 94, 95], and Acoustic Cryptanalysis attack [96] are some of the strong techniques often employed for Side Channel Analysis attacks. Despite the absence of terms of appearance protective measure to such security breaches, several possible environmental security measures typically involve noise combination, output sequential buffering, physical trying to shield, signal reduction in size, and going to eliminate branch computation in constructed techniques to make the encrypted communication time equivalent [97, 98], and several more others.

A time-memory trade-off attack reaction is a form of code-breaking that attempts to break a strong encryption primitive from less interactive complexity than that of a try looking up the database and less offline complexity than from an exhaustive key search. By harmonizing processing time and memory specificity, the TMTO attack advances on the exhaustive key search approach[99].
This test is broken into two phases: offline (pre-computation) and online (attack). During the offline phase, a table is built using the lookup table approach, which involves picking distinct random keys and generating output for each one.
The output strings contain these pairs of output strings and keys in an indexed table. The attacker monitors the output created by unknown keys in the second phase, also known as the online phase. These results are then compared to the table outputs created during the offline phase. If a match is detected, the key from the matched output will be the appropriate key.
Amirazizi and Hellmen published the Time memory processor trade-off technique [99] on block ciphers early, whereas Babbage [100] and Golic [101] independently proposed TMTO on stream ciphers in 1995 as well as 1997, correspondingly. Biryukov and Samir later integrated the Babbage and Golic strategy with the Hellmen attack [102]. Birykov, Shamir, and Wagner developed this approach and used it on A5/1 [103]. To prevent TMTO on stream ciphers, Hong and Sarkar [104] proposed that the state size be equal to or higher than the sum of key size and IV size and that it be random. Babbage [100] and Golic [101] proposed that the state size be at least twice the size of the key.
In Distinguishing Attack, the most crucial condition for a successful stream cipher design is random key stream creation. A differentiating attack attempts to determine whether a given key stream is a random sequence or was generated by a cipher or generator. The distinguishing attack attempts to determine the relationships between internal state variables and the output keystream. To distinguish between attacks, the internal structure of a cipher must be thoroughly examined. A known keystream attack is a distinguishing attack.
Fluhrer and McGrew proposed this attack on an alleged RC4 key stream generator [105]. Ekdahl and Johansson [106, 107], Goli'c and Menicocci [108], Junod [109],

Watan-abe et al. [110], Englund and Johansson [111], Paul et al. [112], Rose and Hawkes [113], and many others have worked on this test. Paul and Preneel combined distinct tests into a single framework in [114]. Ciphers must employ sufficiently lengthy keystreams to prevent distinguishing attacks.

In the other hand, algebraic tests are relatively novel attacks for stream ciphers, and work is being made in this area at a rapid pace. Algebraic attacks on LFSR-based ciphers are quite effective [197]. The fundamental idea behind algebraic attacks is to describe a cryptographic system using algebraic equations. The first phase in this attack is to identify the set of algebraic equations that connect the starting state to the output keystream, after which keystream bits are monitored and their values are swapped into the equations. Attackers attempt to capture as many keystream bits as feasible. Finally, this set of equations is solved in order to obtain the beginning state and, from there, the secret key.
Courtois suggested an algebraic attack on stream ciphers in [197] against Toycrypt, and it was later utilized against LILI-128 [198]. This method was also successful against memory-based stream ciphers, which have been assumed to be more immune to this technique [199],[200]. Courtois later improved through this strategy and constructed the Accelerated Algebraic attack [201], which Armknecht refined [202]. A speedy algebraic attack performed by linearly aggregating equations before actually attempting to solve the systems of equations, including one that considerably positively impacted the attack's performance.

Correlation attacks are a type of well-known plaintext attack. These approaches are regulatory requirements to stream ciphers, including some dependent using feedback shift registers. A correlate exploit attempts to deduce knowledge about just the starting state from either the output keystream through carrying advantage of weaknesses in the development's combined mechanism.
Siegenthaler initially proposed the Correlation attack against a combination generator [203] in 1985, but Meier and Staffelbach developed it as the Fast Correlation Attack [204] in 1988. In [205], Zhang and Feng developed a faster correlation attack on stream ciphers. This attack was expanded upon and used in [206],[207],[208],[209]. To avoid such tests, correlation-immune functions must be built. In the case of LFSRs, one of the principles to prevent linearity that would aid in mitigating this attack is irregular clocking.

### 2.4.2 Machine Learning and Deep learning based cryptanalysis attacks

The concept of attempting to link cryptography with machine learning dates back to 1991 when Rivest first outlined how, at the time, each area was providing techniques and insights to the other in [229]. Following that, significant work has been done in this area, particularly in terms of using machine learning techniques to assist cryptographers in developing new good ciphers.
Any effort to breach a cryptosystem utilizing deep learning is referred to as neural cryptanalysis, Cryptanalysis is the study of cipher vulnerabilities and methods for exploiting them in order to compute the plaintext and/or the secret cipher key. Exploitation is not easy, and numerous problems have been demonstrated to be effective only on reduced versions of the ciphers. Machine learning algorithms are used in neural cryptanalysis to investigate cryptographic mechanisms. As computational power improves, using neural cryptanalysis to attack more complicated ciphers becomes a more viable alternative.
Deep learning (DL) models are capable of approximating any unknown function efficiently. DL applications have been expanded by researchers to every field, from consumer behavior prediction to video production. The most sophisticated forms of functions, particularly cryptographic primitives, have now reached the boundary. Neural cryptanalysis is the use of deep learning to attack cryptosystems. It has been a long time since the notion of using neural networks for cryptosystem design and analysis was first discussed [229]. Before the available processing power reaches a certain level, such systems and analytical methods must stay theoretically interesting. Now that the resource is abundant, theoretical improvements lead to more effective network construction and optimization strategies.

Among the numerous disciplines of ML-based cryptanalysis, many attempts have been made from the perspective of cryptanalysis of block ciphers. Differential cryptanalysis (DC) and linear cryptanalysis (LC) are two well-known cryptanalysis methodologies. The two attacks are noteworthy because they breached the security of DES. Other approaches, such as a strategy based on multivariate quadratic equations [230], have been presented.
Even though the target cryptosystem is the same, several attack scenarios exist based on different assumptions, resulting in different use cases for DL models in attacks. For example, key recovery attacks on block ciphers seek to recover keys used in the cipher, whereas cipher emulation attacks seek to emulate encryption or decryption oracles to recover plaintexts from ciphertexts without knowing the key. The encryption algorithm utilized for the ciphertext is determined through identification attacks. The ways in which deep learning models are employed in cryptanalysis differ greatly amongst tests. For example, when plaintext and ciphertext pairings are provided as training material, DL might be used for end-to-end key recovery, or DL could be included in traditional cryptanalysis approaches.

Attacks known as "key recovery" are attempts to determine the cipher's key with a non-zero advantage. Even if an attacker only learns a portion of the key, such as a round key, the knowledge nevertheless serves as a vital stepping stone toward full key recovery and, ultimately, the cipher's break.

- Attack on the Recovery Key (KR) The attacker attempts to nd k given a plaintext and ciphertext pair (p; c) fulfilling c = Enc(k; p). Training data consists of triples containing a random plaintext, a random key, and the associated ciphertext (ki; pi; ci) such that ci = Enc(ki; pi).
- Attack on Fixed Key Recovery: The attack has the same goal as regular key recovery tests, but only pairs of random plaintexts and ciphertexts (pi; ci) such that ci = Enc(k; pi) are used as training material. Because the key is not included in the training data, a single deep-learning model cannot offer an end-to-end solution. As a result, the attack is frequently combined with an analytical approach that necessitates knowledge of the algorithm. A popular strategy is to estimate the key and then test the model for each probable key.

One of the foundational research in the area of neurological cryptanalysis was produced by Albassal et al. [231]. In order to execute a key recovery attack on a fixed key for an n-round cipher, an attacker may guess the round key r and train the neural network for (n - 1)-round cipher. The faulty key randomization hypothesis will result in an equally distributed result for the test data if r is wrong [232]. The error rate will be much lower than the total number of inaccurate guesses if r is accurate.
Wabbersen [233] utilized analogous but somewhat different strategies and network designs against another toy cipher with an SPN structure for his master's thesis.

Key recovery attacks and other kinds of emulation attacks were divided by Alallayah et al.[234] and both were applied to SDES. An analogous task was done by Danziger et al. [235] using thinner layers and more training data. The results of the neural cryptanalysis, according to the authors, effectively corrected a potential differential vulnerability on the S-box for SDES.

A real-world lightweight encryption called Speck32/64 was included to the key recovery attack thanks in large part to Gohr [236]. He adopted a systematic strategy, initially applying the model for key retrieval after training a real-or-random distinguisher on a selected plaintext assumption based on certain known attributes from differential cryptanalysis. The erroneous key randomization hypothesis is not appropriate for lightweight ciphers, according to the author, who also provided a number of methods for removing some bias from the hypothesis (called "wrong-key response problem") and utilizing the bias for key recovery.
Only when we successfully completed a simple comparison, did toy ciphers have been the target of the vast bulk of key recovery efforts?
The designers of [231] utilized their own toy cipher called HypCipher as the target. A Feistel-structured toy cipher called Hyp-Cipher can accommodate an 8-bit key, a 16-bit plaintext, and a 16-bit ciphertext and borrows one of AES's S-boxes. Attacks against full-round ciphers were effective because toy ciphers only had a small number of rounds.

Gohr, on the other hand, used Speck32/64, the only non-toy target block cipher recovery attack that could be used with the key. The lightweight nature of Speck32/64 should be highlighted. block cipher with 32-bit blocks and 22 rounds. He began the attack from lower rounds and increased it up to 11 rounds, which is half of Speck32/64's total rounds, using a bottom-up strategy.

Several of the most evident and measurable strict limitations, On 2-round SDES, which still maintains a linear connection between plaintexts and ciphertexts, it has been found that explicitly training a neural network to predict key from plaintext and ciphertext pairs are successful.
We argue that the method is only useful for breaking toy ciphers due to the difficulties of developing lightweight and realistic block ciphers. The model must be created for each round key in the round key space according to the round key guessing approach. While DES and AES use 48-bit and 32-bit round keys, respectively, Speck32/64 uses a 16-bit round key. The authors would therefore be unable to extend their approach to practical ciphers with enormous key spaces.
In emulation attacks, emulating a target cipher's encryption or decryption oracle is the goal of cipher emulation attacks.

- Plaintext Restoration Attack: Given a ciphertext c, the attacker tries to infer bits of the corresponding plaintext p with a strong advantage such that c = Enck(p). Random plaintext and ciphertext pairs (pi; ci) are used as training data, with ci = Enc (k; pi). The oracle is delivered to the attacker if they want to employ the chosen plaintext attack.

- Bitwise Plaintext Restoration Attack: A plaintext restoration attack may be launched using specific plaintext bits. In place of (pi; ci), pairs of (pi[k]; ci) are given as training data, where pi[k] denotes the kth bit of the plaintext pi.

- Attack using encryption emulation: Given a plaintext p and a matching ciphertext c, the attacker tries to gain a significant advantage by guessing bits of the matching ciphertext so that c = Enc(k; p). Random plaintext and ciphertext pairs (pi; ci) are used as training data, with ci = Enc (k; pi).

In the related works, findings from a 12-bit SDES cipher emulation attack were given by Alallayah et al. [234]. The exploit emulated encryption while also restoring plaintext. A lightweight block cipher with 31 rounds and 64-bit blocks called PRESENT was used by Mishra et al [237] to encrypt the ciphertext. The authors broke the plaintext restoration problem down into more manageable pieces. The researchers tried to build 64 binary classifiers that predicted each bit on the recovered plaintext rather than building a single large network with 64 output neurons. As a result of the models' failure to provide any relevant predictions of plaintext, the authors concluded that testing reduced-round PRESENT or other successful attacks would be a future study. A similar outcome was obtained when Jain et al. [238] attacked FeW, another lightweight encryption. DES was targeted by Xiao et al. up to two rounds [239]. For the purpose of restoring plaintext, they tested several network types (deep and thin, shallow and fat, and cascade) and activation functions. The researchers found that activation functions had no effect on the results and that the shallow and fat networks outperform the other networks. Three rounds of DES were not subjected to the attack by the authors.
Key recovery attacks and cipher emulation attacks work on toy ciphers but not on practical or full-round lightweight ciphers. For Xiao et al. [239], the attack on 2-round DES was successful, but not on 3-round DES. However, it is obvious that 2-round DES is insufficient to produce the avalanche effect. The avalanche effect, which explains how the one-bit plaintext difference enlarges over the whole bits of ciphertext as rounds go, is a fundamental requirement of block ciphers.

It is frequently believed that deeper neural networks, as opposed to shallow and fat networks, are better at expressing complex tasks. While Alallayah et al. [234] relied on deep and thin networks to effectively attack the toy cipher, Xiao et al. [239] studied whether shallow and fat networks are preferable for cracking two-round DES.
There are two claims [240],[241] regarding plaintext restoration attacks on practical ciphers in addition to the studies mentioned above. According to Alani [240], DES and 3DES recovery only required a simple, fully interconnected network with 4 to 5 entirely connected layers.
Hu and Zhao [241] attacked AES with a similar tactic. Alani's work, however, was shown to be irreplaceable by Xiao et al. [239], and Lagerhjelm came to the same result in his Master's thesis [239]. We suggest that these results are the consequence of over-fitting rather than success in neural cryptanalysis after making an effort to duplicate the findings of Alani [240] and Hu and Zhao [241].
Identification Attacks ara is another aspect of the research work carried out, the work of Chandra et al.[243] to determine whether the plaintext was encrypted with a stream cipher or a block cipher (Enhanced RC6) (SEAL). Although the authors had excellent training accuracy, they did not assess test data. The authors then conducted a larger experiment by combining two stream ciphers, and two block ciphers, and enhancing RC6 and Serpent [244]. (RABBIT and LILI-128). This time, after analyzing test data for each pair of ciphers, the authors were able to differentiate RABBIT ciphertexts from other ciphertexts with an accuracy that was substantially higher than 50%. The two different block cipher types couldn't be distinguished, though. De Souza et al. [245] looked at attacks to identify cipher systems as well as plaintext types. As practice data, the writers used the Bible's words in eight distinct languages. Since clustering may be accomplished using a self-organizing map of neurons, the researchers used unsupervised learning as opposed to earlier experiments. The training data was divided into 6- to 8-kilobyte 'collections,' and the encryption was changed to a vector instead of identifying each ciphertext block.
The study of de Souza et al.[245] stands out since it uses an approach to job training that is unsupervised. Neural network topologies like self-organizing maps are required for unsupervised learning problems. Due to the attack's use of block collections rather than individual blocks, it was possible to identify the same block cipher and plaintext by taking advantage of duplication in ciphertext blocks.
The neural network clustering method in [245] uses the ECB mode of the ciphers, encrypting the same plaintext blocks with the same key, but it is not strictly a deep learning model because there are no hidden layers. The neural network clustering approach in [245] uses the ECB mode of the ciphers, encrypting the same plaintext blocks into the same ciphertext blocks, despite not strictly being a deep learning model as there are no hidden layers. As a consequence, a model may plausibly infer that two ciphertext collections are from the same cipher algorithm and plaintext class with a low probability if they share at least one ciphertext block. The cosine angle between vectors in two collections grows lower than 90 degrees if and only if this is true. The outcomes are attacks on the ECB mode of operation rather than cryptanalysis on block ciphers themselves. This is in line with earlier research using decision trees [246] and support vector machines [247],[248], two machine learning approaches that are not deep learning.

### 2.4.3 Permutation-based cryptanalysis related works

Researchers have long been interested in the security of permutation encryption schemes. The cipher text-only attack (COA) affects several permutation-only broadcast-TV systems because the specific design of analog video signals restricts the flexibility of generating sophisticated permutations. Bertilsson et al [249], for example, presented another approach to Matias and Shamir's famous architecture, in which every structure of a video is examined over to the other multiple pseudo-random space-filling curves to selectively retrieve the contents of the video by using the connection between subsequent frames. The issue appears to improve considerably whenever secret permutations are applied to encrypt multimedia content information.

Furthermore, if certain correlations exist between the elements to be permuted, a cipher text-only attack (COA) can still be effective. Li et al. demonstrated a COA attack on row-column shuffled images in [250] by exploiting any link between separate rows and columns. The above technique was developed in [251] to break permutation-only image encryption of pixel bits.

However, if the atoms of each element $L$ are not low and the entropy contained in each element is large, it is clear that discovering all ($L$!) possibilities is exceedingly difficult, and COA is therefore virtually impossible. As a result, some research strategies advocate developing more difficult methods to create secret permutations in order to provide increased security while also meeting a variety of additional application-dependent demands. [252, 253, 254, 255].

Despite efforts to make permutation-only ciphers more resistant to cipher text-only attacks, most cryptographic algorithms of this sort are vulnerable to plaintext attacks. Making every part of the plaintext unique from one another (input difference) increases the impact of a chosen-plaintext attack (CPA), in which the adversary obtains the cipher text of a selected plaintext.

[256] proved that the lowest limit on the number of selected plaintext to completely extract the underlying permutation pattern is ($Logrl$), where r appears to be the number of possible intensities.

It was challenging to determine how frequently these well-known plaintexts would be required in the event of a targeted attack (KPA), which varies from a CPA only in that the adversary would be unable to choose the plaintext arbitrarily.

$L$ is often substantially higher than r in multimedia data. According to the pigeonhole principle, certain values in 0, 1, ..., $r$1 must appear more than once.

The identical pixel quantity of 0 should appear around 512 times within the permutation encrypted ciphered image when a known plain image of size (512x512) has a uniform distribution. There must be ([512]!) possibilities for one item in the permutation pattern whose pixel value corresponds to zero [257] because of this clear image and the related encryption.

[257] The uncertainty in such interactions should eventually disappear as more pairs

of known plain images and encrypted images are seen.

To offer a quantitative analysis of the known plaintext attack on the permutation-only multimedia techniques in [258], Li et al. built on the work of [266]. Their approach is separated into two steps: identifying the intersection of the sets between various pairings and grouping a permutation sequence according to the atom values in each pairing of plaintext and ciphertext. By using a tree structure, [259] was able to reduce the storage and computing complexity of this method. Both of these approaches came to the same conclusion: there are about ($LogrL$) [257] known plaintexts. These two articles are well recognized for the development and study of lightweight multimedia encryption systems [260],[261],[262],[263],[264],[265] due to its universality.

Reexamining the KPA attack on permutation-only ciphers from the perspective of composite representation Leo Yu Zhang. Bianchi et al.[267],[268] confirmed it in a series of studies, suggesting that it is used to shrink the size of secret text and quicken linear operations on ciphered data produced using additive homomorphic cryptograms. On permutation-only systems with composite representation, he gives a complete theoretical study of the KPA cryptanalysis. The composite representation suggests many KPA algorithms, one of which performs faster processing with the same storage as the well-known "optimal" method.

## 2.5 Conclusion

In this section, an overview of machine learning and its architectural differences, as well as deep learning approaches and these different approaches, has been presented. We have witnessed several milestones in its progress and accomplishments. We have discussed different deep learning community approaches and described the principles behind each along with the properties of each. We also discussed the three main families of models, namely convolutional networks, recurrent networks, and generative models, the last of which is currently in full development. In the field of the application of these techniques of artificial intelligence in cryptanalysis, a very detailed study of the state of the art has been presented, where we have tried to recover all the aspects targeted by experts in this field. and their research purpose. We discussed the perspectives of this research work, the major challenges, and the different points of view of each approach.

# Part II

# Contributions and Validation

# MULTIMEDIA P-BOX PRIMITIVES SECURITY ASSESSMENT

3

CONTENTS

This chapter provides research that contributed to the publication of the results of the study in a published article titled "Automated Deep Learning BLACK-BOX Attack for Multimedia P-Box Security Assessment," throughout the IEEE Access journal, volume 10, pages 94019–94039, 2022, with the following authors: Zakaria TOLBA, Makhlouf Derdour, Mohamed Amine Ferrag, S M Muyeen, and Mohamed Benbouzid.[301]

## 3.1 Introduction and Motivation

Deep learning has lately attracted the attention of cryptography professionals, interestingly inside the discipline of cryptosystem cryptanalysis, in which the vast majority of research concentrates on differential distinguisher-based black-box approaches. This study addresses a deep learning-based decryptor to better understand the permutation data structures used throughout multimedia block cipher encryption techniques. We would like to leverage ciphertext couple characteristics in conjunction with palette convolution neural network architectures that establish significant relations within and between permutable elements to achieve maximum information retrieval within moderate restrictions. These could make it possible for us to acquire all the plaintext from the ciphered text without being required to understand anything whatsoever about P-boxes. This skill set and theoretical background can be used in deep learning to better assess P-box recombination techniques and methodologies, which are becoming more common in multimedia encrypted communications.

This same significant number of published papers in the literary works which satisfy this situation either attempt to recover the whole straightforward form of a specific cryptosystem that would use conventional research methodology as well as numerous different optimization techniques to distinguish the key had been using, or those who depend exclusively on black-box attacks for the vast number of the recovery.

The fundamental objective of various attacks, on the other extreme [301]., is constrained to obtaining just the cipher understanding and also its basic meaning, which could also be accomplished by employing black-box-based deep learning attacks that do not necessitate additional sophisticated cryptanalysis as well as complex proposed method detailed information. On the other hand, because they can go into additional considerations and requirements of the optimization technique and specifications of techniques employed, such strategies require it impracticable throughout many circumstances. Besides this, standardizing those kinds of approaches is exceptionally hard.

In way of comparison, our method provides a comparatively simple procedure that can then be just recommended in testing requirements. Image files, except for text files, include characteristic features often including large amounts of data volume, redundancies, as well as remarkable adjacent pixel causal connection, which requires significant the use of specific techniques throughout most of the encryption process to break down the significant relation between pixels in the image. Non-linear systems and permutation techniques founded on chaos theory provide different types of the above. This methodology appears to be advantageous for having to deliver multimedia content, and significantly increased pictures, across the insecure channel.

We take into account the potential advantages of deep learning cryptanalysis methodologies on such assessment process by using convolution neural network parameters together in a black box attack to determine the relationships among both permutable entities to more effectively synthesize the plaintext from either the ciphered text without demanding any P-box understanding [301]. We illustrate an automatically generated decryptor made up of deep convolutional neural networks which outperform previous attempts. A decryptor is indeed a neural network system that really can decrypt plaintext or plain image without such precondition for specific information such as the number of iterations, the negligible difference distribution, or P-box production patterns.

Throughout this context, we designed to obtain information with either a non-uniform distribution. There exists an achievement massive imbalance among both theoretical methodologies and demonstrated experimental successful implementation because even though the data-driven specializes in working in the existing literature to highlight potential claimed potential benefits are primarily focused on that same unified distribution of elements, that either completely resembles the hypothetical assessment. We demonstrate that our neural model satisfactorily leverages ciphertext combination characteristic features however with poor distribution of the data to enable greater retrieval under moderate restrictions and limitations whereas previous differential approach approaches were unable to address [258],[259],[260],[261].
The principal objective of this research is either to implement machine learning methodologies to enhance the much more often deployed differential cryptanalysis methodology. We contributed to this field of investigation by trying to investigate whether deep learning can indeed enhance differential cryptanalysis, which is employed to assess the validity of cryptographic algorithms. The security of block ciphers was judged by employing artificial neural networks as well as a wide variety of criteria including input and output distinctions, repetition frequency, and P-boxes spending patterns [301].

## 3.2 BACKGROUNDS

Before going deeper into our topic, it is crucial to clarify some concepts and principles that are fundamental to properly understanding the methodology and contribution of this study, as well as judging the advantages provided to science in this subject.

### 3.2.1 Principle permutation primitive (P-Box)

The substitution (S-box) and permutation (P-box) phases are the two principal choices for achieving Shannon's confusion and diffusion metrics in a simple encryption system. While the permutation step changes atom positions, the substitution stage changes atom values. These two options appear to be capable of dealing with digital multimedia encryption effectively while minimizing distraction, accidental deletion, and obfuscation caused by the encryption operation, and it should be noted that many block-cipher multimedia cryptographic structures involve independent processes for all permutation and substitution procedures. Many lightweight block ciphers use public permutation, including AES, substitution-permutation networks (SPN), and generalized Feistel structures (GFS), and it is also a symmetric cipher for secret permutation, with the key being the secret for creating the permutation sequence.

### 3.2.2 Notation and description

Definition:
The function $P : B^n \rightarrow B^m$
P is called P-BOX if there exists a sequence $(i_k)_K^m = 1$,
Where $i_k \in \{1, ..., n\}$
such that for all $b \in B^n$ and $k = 1, ..., m$
we have: $P(b)_k = b_{ik}$.

- It simply states that the $K^{-th}$ item of $P(b)$ reflects the $i_K^{-th}$ item of b.
- S-BOX has a unique sub-type called P-BOX.
- The elements of the input are permuted, repeated, or discarded by P-BOXES but not altered.
- We signify that $n = m$" because P-BOXES is a specific kind of S-BOXES.".

The picture $M$ is represented as a two-dimensional array of positive integers with a size of $(MN)$ for the multimedia permutation.

- Each pixel in a picture is represented by the entry variable $M_s$, which is defined as $M_{ls}(i, j)$.
- Where $(i, j)$ and $l$, respectively, represent the pixel coordinates in $M$ and the intensity value
- In order to recreate the ciphered picture without distortion, each element's position must be moved to a different location during the permutation stage.
- A two-dimensional table with each table member carrying the new element location might be imagined as the secret permutation phase.

A permutation table (P-box) of size $MN$ with the definition $Prm$ is as follows:

$$Prm = \begin{bmatrix} p_{11} \ldots\ldots\ldots\ldots\ldots\ldots p_{1n} \\ p_{21} \ldots\ldots\ldots\ldots\ldots\ldots p_{2n} \\ \ldots\ldots\ldots\ldots\ldots\ldots\ldots\ldots \\ \ldots \;\; \ldots \;\; \ldots. \;\; \ldots \;\; \ldots \\ p_{m1} \ldots\ldots\ldots\ldots\ldots\ldots p_{mn} \end{bmatrix}$$

- The permutation $Prm$, represented by the function $T_k$ in the following : $C = MT_k = T_k(M)$.

- $T_k$ is a bijection function that converts each element $e_{ij}$ of M with location $(i, j)$ to $e'_{i'j'}$ a new location $(i', j')$ referring to a key k over a set number of rounds with $i \neq i'$ and $j \neq j'$ and $i, i' \in \{1, \cdots m\}$ and $j, j' \in \{1, \cdots n\}$.

$$C = \begin{bmatrix} e_{11} \ldots\ldots\ldots\ldots\ldots\ldots e_{1n} \\ e_{21} \ldots\ldots\ldots\ldots\ldots\ldots e_{2n} \\ \ldots \;\; \ldots \;\; \ldots. \;\; \ldots \;\; \ldots. \\ \ldots \;\; \ldots \;\; \ldots. \;\; \ldots \;\; \ldots \\ e_{m1} \ldots\ldots\ldots\ldots\ldots\ldots e_{mn} \end{bmatrix} T_k$$

$$C = \begin{bmatrix} e'_{11} \ldots\ldots\ldots\ldots\ldots\ldots e'_{1n} \\ e'_{21} \ldots\ldots\ldots\ldots\ldots\ldots e'_{2n} \\ \ldots \;\; \ldots \;\; \ldots. \;\; \ldots \;\; \ldots. \\ \ldots\ldots\ldots\ldots\ldots\ldots\ldots\ldots \\ e'_{m1} \ldots\ldots\ldots\ldots\ldots\ldots e'_{mn} \end{bmatrix}$$

- The matrix $Prm[MN]$, which represents the permutation relationship, contains the locations of each clear image's cipher image pixels.

- This idea assumes that the same method may be used to crack permutation-based encryption schemes.

- As a result, the inverse function $T_k^{-1}$ must be used to find the beginning point of pixel $p$.

- The encryption private key $k$ forms the $Tk$ function and its inversion $T_k^{-1}$, which has the same dimension as the considered plaintext and is the block cipher size (P-box size).

- A symmetric block cipher with an input size of $(MN)$ and a key size of $(MN)$ is what this framework shows picture encryption to be.

- We conclude that all permutation approaches will be included in the $[MN]!$ possible scenarios, which also represent the greatest number of chosen plaintexts that results in a conclusion. Permutation cryptanalysis should therefore focus on those dimensions as a problem space, according to [283].

Insecure channels where the requested security measures and the associated impacts of a security threat are typically low are where permutation methods are frequently used as multimedia content encryption techniques. These methods are also strongly advised as an alluring choice in the formulation of marketing and security engineering disciplines.
Because of its simplicity, adaptability, ability to save space, and effectiveness in concealing evident visual information, permutation is a common strategy in many secure multimedia systems. Public permutations and secret permutations are the two categories depending on whether or not they are connected to a secret key.
Efforts in this field are based on a direct alteration of image intensities. The encryption technique is applied directly to the picture frame in the spatial dimension. The pixel correlation is eliminated by the encryption in this field. Using the opposite method, the image's pixel intensities may be completely and flawlessly recreated.
The goal of schemes in the field of cryptographic algorithm frequencies is to alter the image's frequency through modifications. As a result, information loss and distortion frequently happen while attempting to retrieve the original picture's pixel intensities during the decryption process.

The permutation transitional period for discrete chaotic maps is widely implemented using various chaotic and non-chaotic techniques, such as the logistic and Arnold's maps. The hyper-chaotic system of Chen and the Lorenz attractor is used to create chaotic permutations with continuous attractors. Different tactics, such as the chess-based horse movement and the trajectory of a water wave movement, may be handled to provide non-chaotic options.

## 3.3 DEEP LEARNING MODEL IMPLEMENTATION

The basic aim is to discover a suitable model that can imitate the permutation method in its encrypted image decryption phase, regardless of the type of algorithm, pattern, or number of rounds utilized. The main goal is to find a DL model that can emulate the permutation algorithm in its encrypted image decryption phase, whatever the type of algorithm and the pattern used, and with any number of rounds.

**– Neural distinguisher specification:**

The model calculates the difference between a variety of inputs and outputs depending on the dataset, with the correlation between close pixels in two-dimensional space now serving as a critical characteristic. According to the official statement, the goal of the decryptors is to identify the visual distinction between the inputs and outputs:

- $\Delta d = Input_1 \otimes Input_2$
  where $Input_1$ and $Input_2$ are two distinctive plain images and $\otimes$ denotes to the dissimilar function.
  The dissimilar function reflects the distance between two photos because we operate in two dimensions. For additional information, if we lay two photos of the same size on top of each other, the function represents the number of pixels in the same position in the two images of different hues, as seen below:
  $Input_1 = Img_1$ and $Input_2 = Img_2$.
  The dissimilar function represents the number of pixels $P_{ij}$ with the condition $P1_{ij} /= P2_{ij}$ and $P1_{ij} \in Img_1$ , $P2_{ij} \in Img_2$ .

- $\Delta d' = Output_1 \otimes Output_2$, For an outputting distinction, $\Delta d'$ can be alternatively constructed by exploiting a pair of relative cipher images $Output_1$ and $Output_2$.

- The underlined I-iteration differential pathway like the propagation of $\Delta d$ to $\Delta d'$ after i iteration of permutation is represented by : $\Delta d \xrightarrow{i} \Delta d'$ .

- Every differential direction must have a specific probability of holding: $Pr(\Delta d \xrightarrow{i} \Delta d') = a^{-p}$ in the case of unified distribution of the intensity I with $I \in \{0, \ldots, a\}$.

### 3.3.1 The experimental selection of model, architecture, and parameters

Test assessment is viewed as a regression issue for a supervised model, with layers of the model being trained by a variety of factors including changes in input and output, number of iterations, and P-box generating patterns. Deep learning techniques are utilized to locate a decryptor because, in addition to the necessity for explicit purposeful feature extraction engineering, they can uncover latent structures in digital information. We tested different types of neural networks, including the basic Multi-Layer Perceptron (MLP), deep neural network (DNN), convolutional neural network (CNN), and long-short-term memory network (LSTM). We explored the width (number of neurons for each layer) and depth (number of hidden layers) of the latter to get the maximum accuracy and learning speed.
We also ran trials with various types of activation functions and the weights initialization approach.[301] After several testing, we discovered that the CNNs is suitable for detecting a decryptor. The key rationale for this selection is that CNNs are meant to identify patterns in input data, which facilitates the differential process, and it works for any input in which two-dimensional data is related in any way.
Convolution is built on three main strategies that can assist improve a machine-learning approach, which is as follows:

- Sparse interactions : Matrix multiplication by a table of characteristics with different parameters indicating the connection between each incoming and outgoing unit is used in classical neural networks. This means that each output element connects with each entering element, which is not the case with convoluted neural networks. This is performed by lowering the core below the entrance. Matrix multiplication by a table of characteristics with separate parameters indicating the connection between each incoming and outgoing unit is used in traditional neural networks. This means that, unlike complex neural networks, each element of output connects with each element of the entrance. This is achieved by lowering the core below the entrance.

- Parameter sharing: This refers to using the same feature for several functions in a model. Each core element in a convolution network of neurons is employed at each entry point. The parameter sharing done by the kernel size guarantees that rather than learning a set of various parameters for each location, we just find one set, significantly reducing the model's data needs.

- Equivariant representations: Because of the specific trait of parameter sharing, the layer in the case of convolution has equivariant interpretations. When a function is considered to be equivariant, it appears to suggest that the output changes regardless of the input. A function $H(x)$, in particular, is equivalent to a function $K$ if $H(K(x)) = K(H(x))$. This attribute allows our suggested model to build a strong connection between the plain and cipher pairings.

### 3.3.2 Hyperparameters configuration

The hyper-parameters that make the most difference for a certain job must be picked when deploying machine learning algorithms. These characteristics are frequently obtained experimentally by studying numerous network topologies while following best practices. There are automated methods for tweaking hyperparameters, but they need large resources that can be difficult to duplicate. The findings of the manual architectural search are then shown.
The following are the remaining hyper-parameters that were appropriately applied in our fascinating experiments:

- Initial Learning rate: 0.1.
- Batch Size: 2000.
- Epochs: fixed in 1500.
- Trainable parameters: 679 338.
- Weights initialization : Xavier Initialization [61] , also known as Glorot Initialization is a neural network initialization strategy. Biases are set to zero, and for each level, the weights $W_{ij}$ are established as:

$$W_{ij} = Ds \quad -\sqrt{\frac{1}{Prv}}, \sqrt{\frac{1}{Prv}} \tag{3.1}$$

Where $Ds$ is a uniform distribution and $Prv$ is the dimension of the preceding layer (the number of columns in $W$).

- Optimizer: As an optimizer, we used the Adam algorithm [296]. Since it slightly differs from the classical gradient descent we presented before, we give a brief explanation here. We denote two sequences:

$$x_t = \gamma_1 x_{(t-1)} + (1 - \gamma_1) f_t \quad (3.2)$$

$$y_t = \gamma_2 y_{(t-1)} + (1 - \gamma_2) f_t^2 \quad (3.3)$$

  $x_t$ and $y_t$ are respectively $1^{st}$ order (mean) and $2^{nd}$ order (variance) gradient estimates.
  and $f_t = \nabla E\theta^{(t-1)}$
  where $\theta(t)$ represents as before our trainable parameters, E is our loss function, and $\gamma_1, \gamma_2$ are constants.

- Error function :
  - The mean squared error of a regression line (MSE) By squaring the distances between the values and the regression line, determines how close it is to a set of values (the "errors"). To remove any negative parameters, squaring is necessary. Larger inequalities have also been shown to be more significant. The word "mean squared error" comes from the fact that we are computing the average of a sequence of errors. The smaller the $MSE$, the more accurate the prognosis.
    Below is the description of the mean squared error:

$$MSE = \frac{1}{a} \sum_{i=1}^{n} (y_i - \hat{y}_i)^2 \quad (3.4)$$

    With :
    * $n$ is the number of items.
    * $\Sigma$ is summation notation.
    * $y_i$ represent original ground of truth or observed y-value.
    * $\hat{y}_i$ is the predicted y-value from the model.
  - The quantile loss function is used to anticipate intervals or ranges of forecasts rather than single-point predictions. As the title and subtitle imply, the quantile regression loss function is used to predict quantiles. A quantile is a value from which a certain number of observations in a group are determined.

- The coefficient of determination ($R-squared$) is a metric that indicates the strength of fit of a model. This is a quantitative metric in the regression architecture that reflects how well linear regression duplicates the correct data. It is absolutely critical when using a quantitative model to better estimate results or validate assumptions. There are various varieties (see below), but this is the most common:

$$R_squared = 1 - \frac{SSR}{SST}, \quad (3.5)$$

$$= 1 - \frac{\sum(y_i - \hat{y}_i)^2}{\sum(y_i - \bar{y})^2}. \quad (3.6)$$

with $SSR$ as the sum squared regression and $SST$ representing the total sum of squares. The coefficient of determination ($R-squared$) is a metric that indicates the strength of fit of a model. This is a quantitative metric in the regression architecture that reflects how well linear regression duplicates the correct data. It is absolutely critical when using a quantitative model to better estimate results or validate assumptions. There are various varieties (see below), but this is the most common:

$$\text{Residual} = \text{actual } y \text{ value} - \text{predicted } y \text{ value}, \quad (3.7)$$

$$r_i = y_i - \hat{y}_i. \quad (3.8)$$

A negative residual indicates that the desired value was too high, whereas a positive residual suggests that the received value was too low. The purpose of a regression line is to minimize the sum of residuals.

For calculating residuals , recognizing that :$r_i = y_i - \hat{y}_i$ and understanding that the regression contains the equation : $\hat{y}_i = a + bx_i$

The residual of observation is calculated as follows: $r_i = y_i - \hat{y}_i = y_i - (a + bx_i)$

- Activation function: The linear activation function was chosen for any situation in which activation is roughly proportional to the input. It is also known as "no activation" or "identity function".

  The function makes no changes to the weighted combination of the parameters; it just returns the value that was supplied. In our case, utilizing this function correctly retains the parameters provided by the Adam optimizer and improves the efficacy of the convolution features.

### 3.3.3 The architecture details

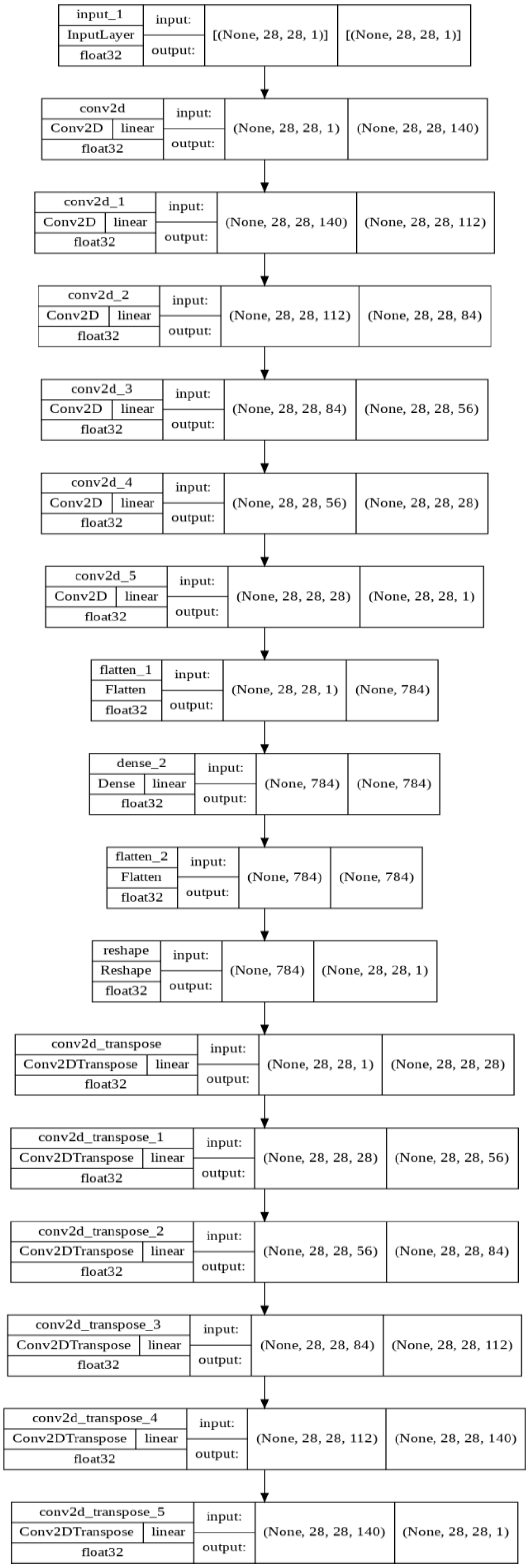

Figure 3.1 – *The model structure*

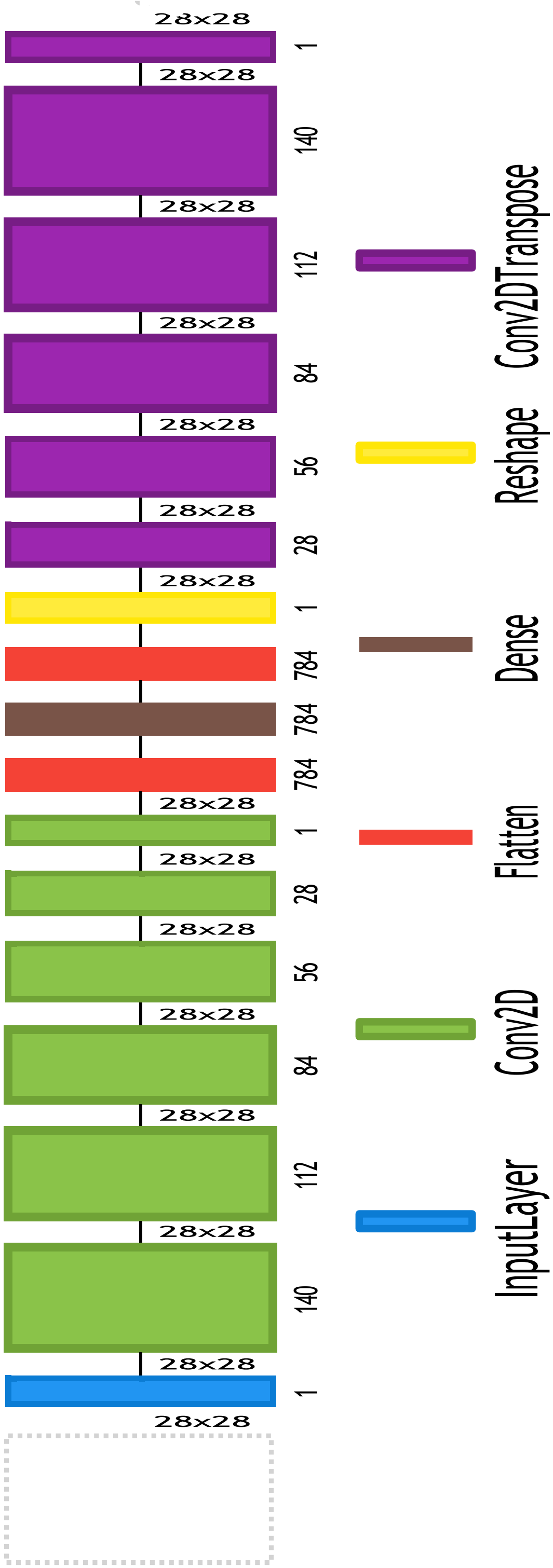

Figure 3.2 – *The model architecture*

Table 3.1 – *Convolutional groups parameters.*

| Conv | Filters | Kernels | Strides | Padding | Parameters |
|---|---|---|---|---|---|
| 1 | 140 | (1,1) | (1,1) | valid | 280 |
| 2 | 112 | (1,1) | (1,1) | valid | 15792 |
| 3 | 84 | (1,1) | (1,1) | valid | 9492 |
| 4 | 56 | (1,1) | (1,1) | valid | 4760 |
| 5 | 28 | (1,1) | (1,1) | valid | 1596 |
| 6 | 1 | (1,1) | (1,1) | valid | 29 |

Table 3.2 – *Dense layer parameters.*

| Dense Layer | Neurons | Trainable parameters |
|---|---|---|
| 1 | 784 | 615440 |

**Parameters for padding, strides, kernels, and filters**

- The first important Conv-2D measurement is the total of filters that the convolutional layer should receive.

- The depth of the kernel, which is a 2-tuple indicating the size of the 2D convolution frame, is the next essential factor that must be supplied to the Conv-2D class. The kernel size must be an integer value as well.

- The strides configuration is a pair of integers that describes the movement of the convolution along the input volume's x and y dimensions.
- The padding argument of the Conv-2D class could have one of the two possible parameters: valid or the same. By using the valid measurement, the entry dimension is not zero-padded, so the spatial perception has been restricted naturally through the use of convolution.

Figure 5 depicts the model architecture, while tables 1, 2, and 3 provide the parameters of the convolutional groups, dense layer, and de-convolutional groups, respectively.

Table 3.3 – *De-convolutional groups parameters.*

| Deconv | Filters | Kernels | Strides | Padding | Parameters |
|---|---|---|---|---|---|
| 1 | 28 | (1,1) | (1,1) | valid | 56 |
| 2 | 56 | (1,1) | (1,1) | valid | 1624 |
| 3 | 84 | (1,1) | (1,1) | valid | 4788 |
| 4 | 112 | (1,1) | (1,1) | valid | 9520 |
| 5 | 140 | (1,1) | (1,1) | valid | 15820 |
| 6 | 1 | (1,1) | (1,1) | valid | 141 |

## 3.4 DATASETS AND TRAINING OBJECTIVES

Choices are fundamental in the discipline of deep learning. If the model's architecture, number of layers, hyperparameters, and training base are adequate, it tends to result in a well-trained model with outstanding prediction; otherwise, it converges to underfitting and overfitting difficulties and sometimes vanishing problems. In this section, we will go through the reasons for selecting the dataset in depth.

### 3.4.1 Datasets selection arguments

The MNIST [297] and FASHION MNIST [298] data sets are used for training and cracking tests.
We use the remaining 10,000 ciphered photographs from the same dataset as the test set, and the remaining 60000 plain photos from the MNIST and FASHION MNIST data sets as the training set, to create our encrypted images. The main argument for selecting a dataset in which the intensity of the image is stored as a number between 0 and 255, providing each pixel with 256 different possibilities, is the non-uniform distribution of colors in each image. The intermediate values are grayscale levels spanning from black to white, whereas the integers 0 and 255, respectively, represent black and white. A histogram is a helpful tool in image processing because it shows how an image's intensity (or color) distribution is distributed. The figures below show the distribution of color intensity produced by the MNIST and Fashion MNIST datasets. We can see the table values and the grayscale tones that, in the absence of a uniform distribution, constitute the image in this picture. Tables 3 and 4 display the intensity distribution of the relevant samples from the two data sets, whereas Figures 1 and 2 display 16 examples from the fashion MNIST and MNIST data sets, respectively.

which the present work seeks to reduce since differential patterns can be used as a statistical or quantitative distinguisher for permutation cryptanalysis attacks. There is a performance gap in recent works [257],[258],[259],[205] because every differential direction has to have a specified probability of holding in the case of a non-unified distribution. In this study, we show that our neural decryptor successfully utilizes properties of ciphertext pairings not addressed by earlier differential attempts, even in the absence of data distribution.

### 3.4.2 The deep learning model's training objectives

The reduction of correlation between close pixels is one of the main objectives of picture encryption systems,because images include valuable visual information that may be observed by superimposing correlations between nearby pixels. One of the key characteristics of our model [301] is its ability to rediscover and reconstruct this correlation by training the model with a variety of input and output (plain/cipher) pairs for permutation feature extraction. This correlation is used to predict clear images from encrypted images through implicit exploration of this correlation.

Training a model to predict a clear picture from its ciphered counterpart is the main objective. However, to differentiate between the visible picture and its encrypted counterpart, a model that can be trained with varied attributes is necessary, regardless of the permutation approach or the number of rounds.

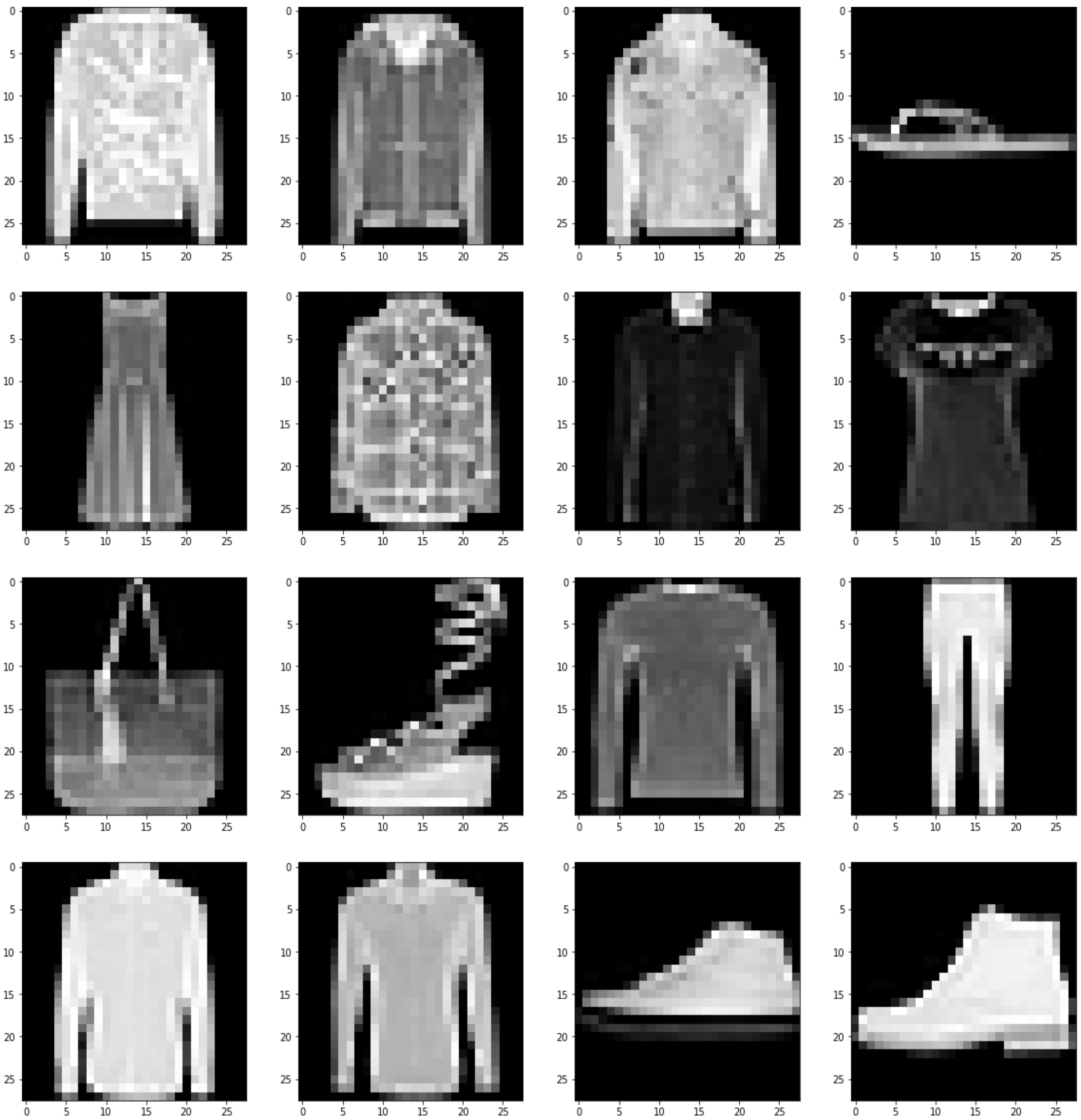

Figure 3.3 – *Samples of FASHION MNIST*

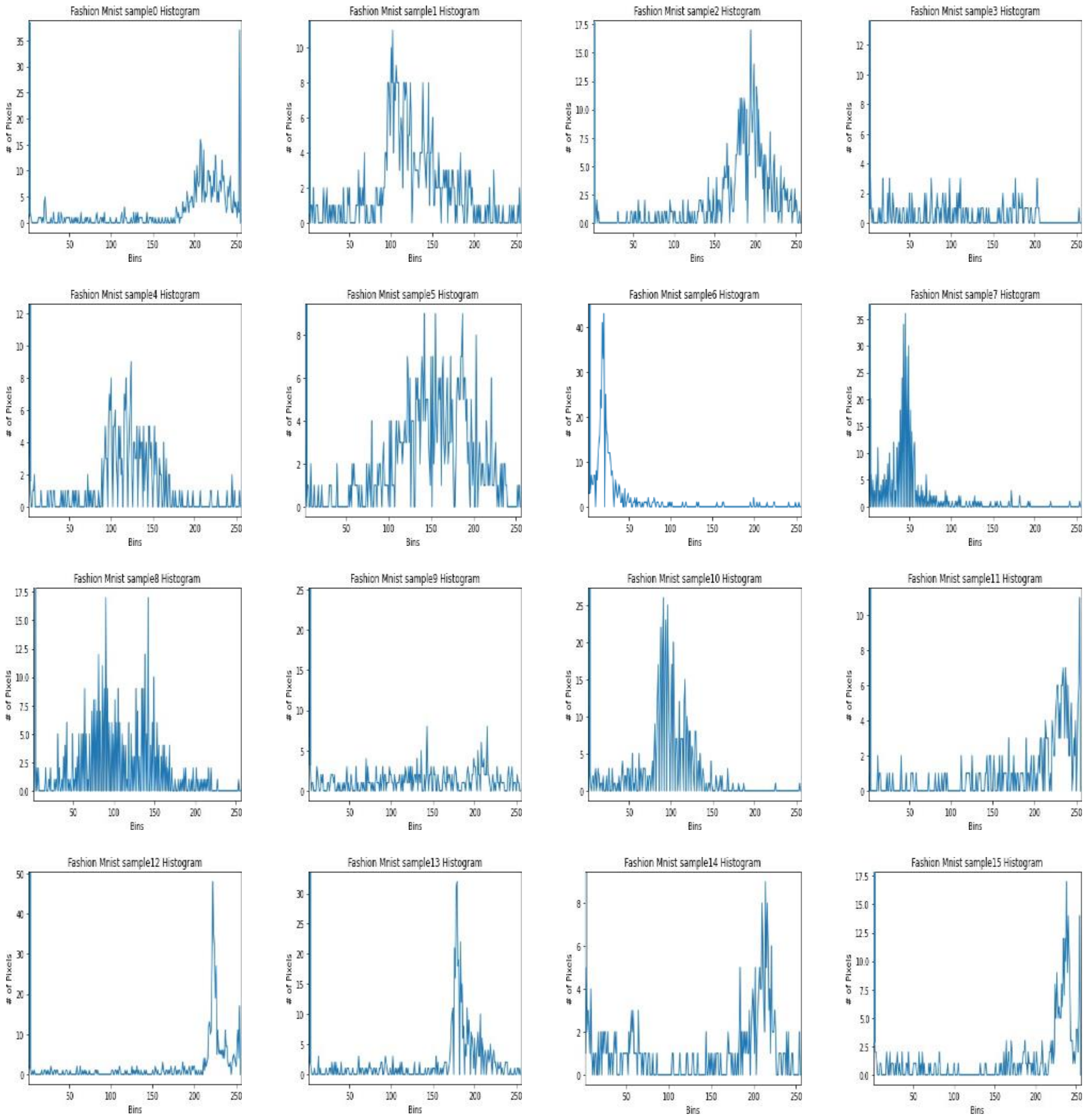

Figure 3.4 – *Color intensity distribution in Fashion Mnist samples*

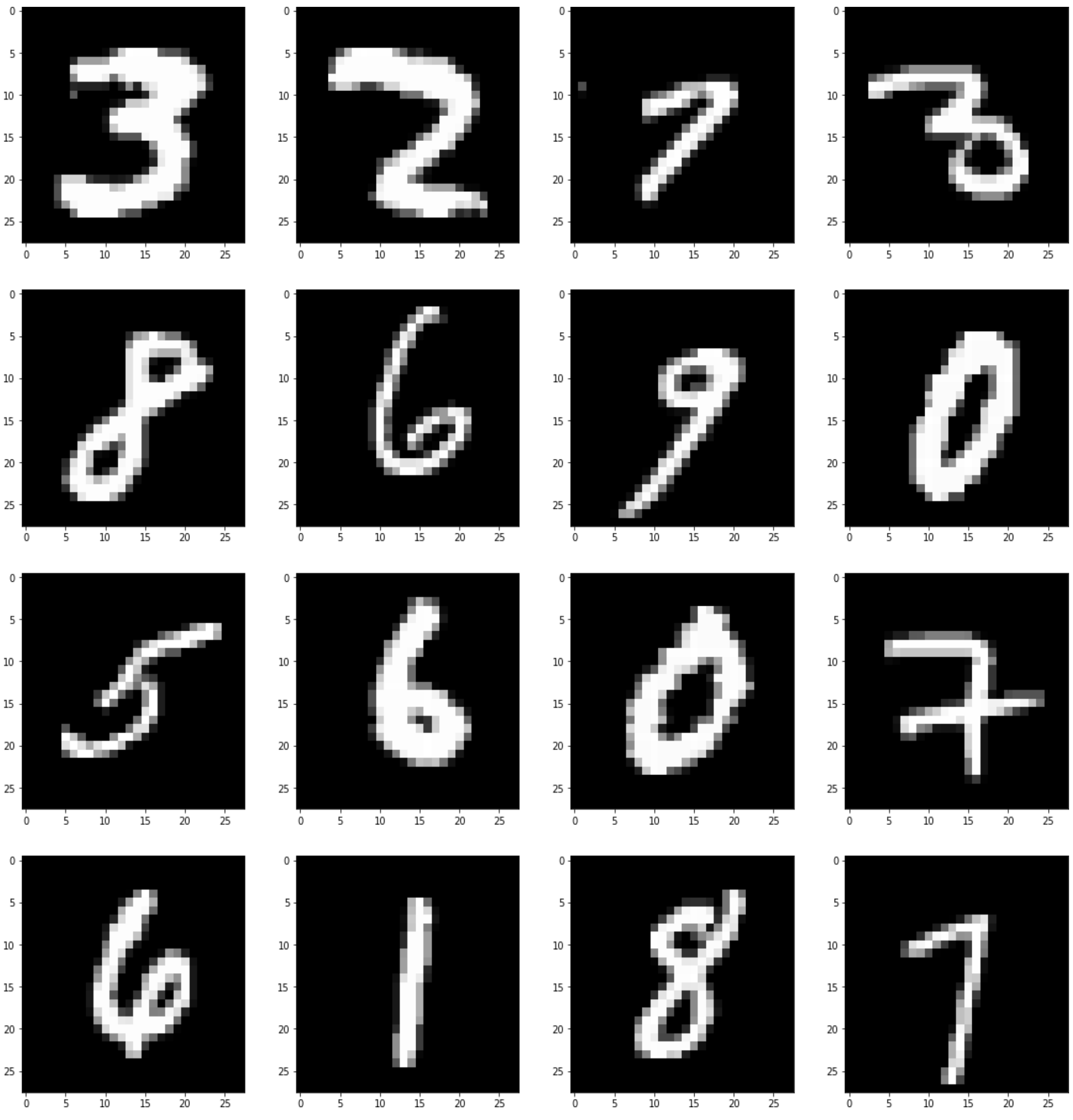

Figure 3.5 – *Samples of MNIST*

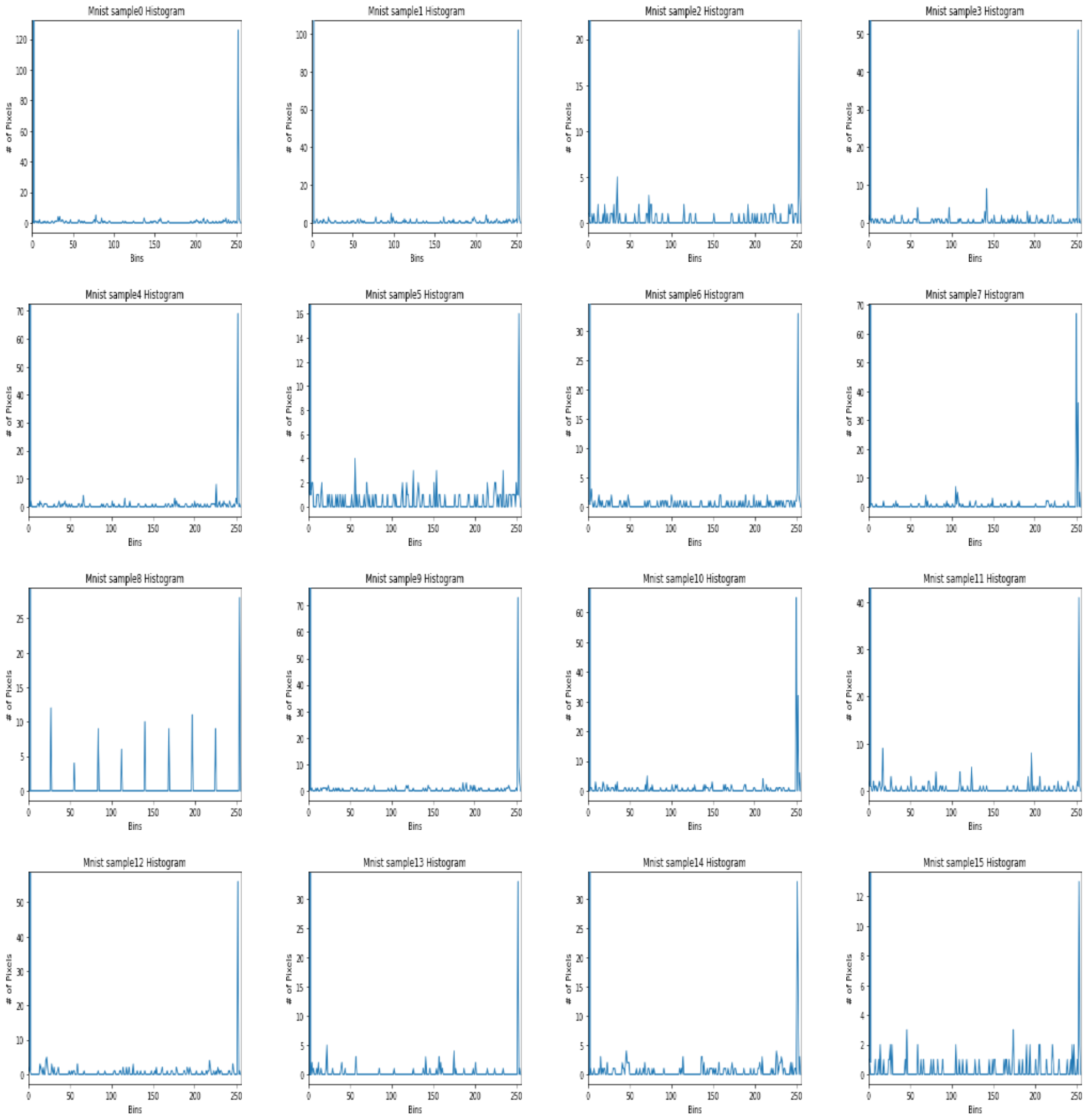

Figure 3.6 – *Color intensity distribution of corresponding MNIST samples*

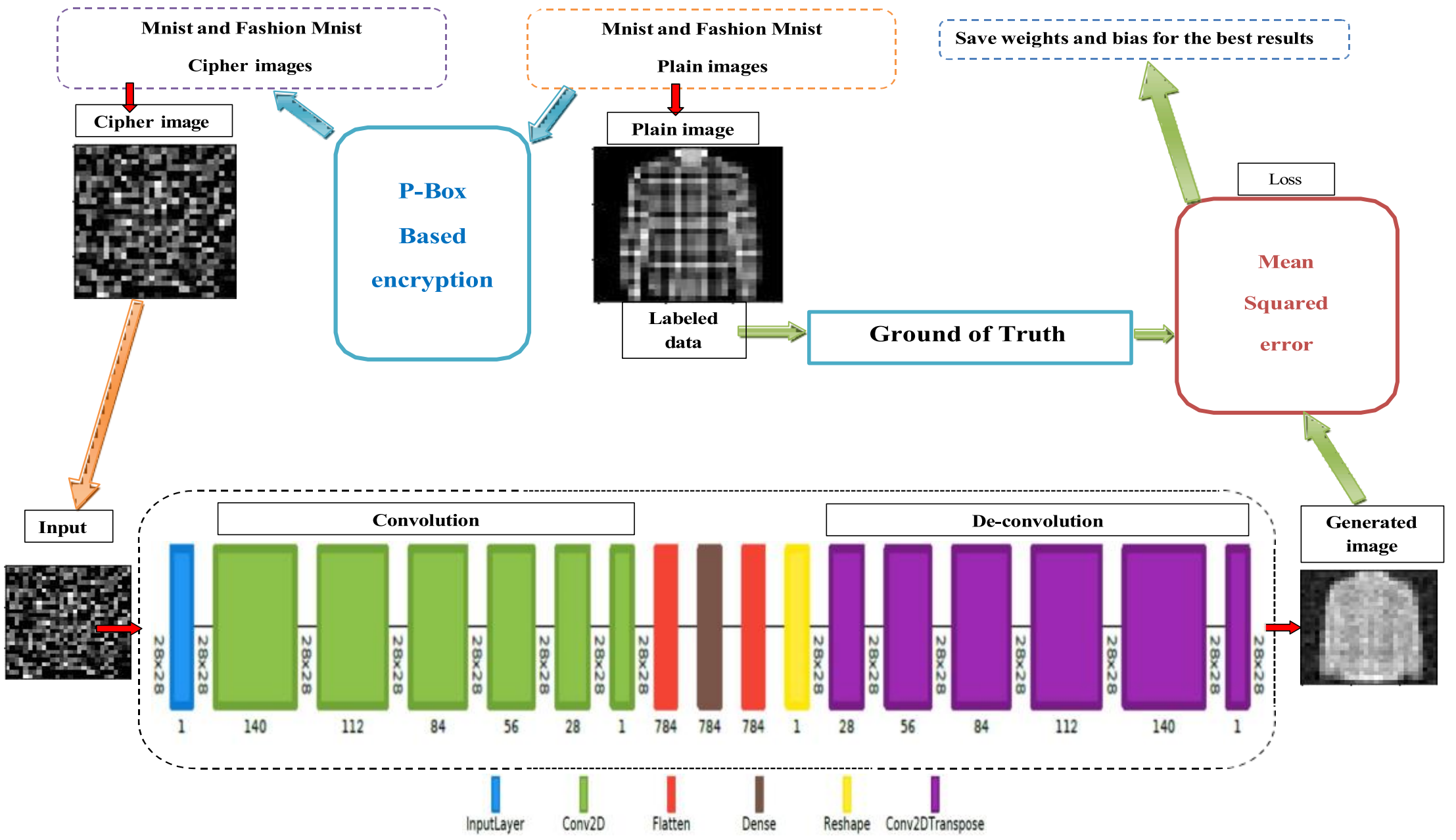

Figure 3.7 – *The model training procedure.*

The task was supposed to be a regression challenge for a supervised model. It means that the model will be trained to predict the clear picture from its matching encrypted image; in other words, the model's inputs are encrypted images and the model's outputs are images created to seem like the original plain images. In the output, the acquired pictures will be compared to the labeled data, which are the original image of Mnist and fashion Mnist, using an error function that will measure the model's outcomes. Then, after each iteration, the model's parameters will be updated following the optimizer of choice. The system will preserve the weights and biases associated with the best outcomes produced after each iteration, and the training will loop until the number of iterations (epochs) is reached. We used four chaotic and non-chaotic system permutation algorithms on images from the Mnist and Fashion mnist data sets to examine permutation pattern measurements in the context of visual cryptography, and we used these patterns as permutation key generators to strengthen our study and better validate the established model.
Following the generation of these permutation keys, these patterns encrypt the datasets Mnist and Fashion mnist to generate 60000 encrypted photos of each, which are used as inputs to the trained model, and the remaining 10000 images are encrypted for use in the model validation step. To make training easier, the encrypted pictures in the Mnist and Fashion mnist datasets are ordered in the same order as the clear original images. The model also uses real photos from the Mnist and Fashion mnist databases as label data.[301] The training approach is depicted in Figure 6, and table 4 has a pseudo-code of the training algorithm's strategy.

Table 3.4 – *The model training algorithm's pseudo code.*

| **Algorithm** |
|---|
| **Input: C** stands for Ciphered Image Data Set, **P** for Plain Image Data Set, **L** stands for Loss Functions, **r** stands for Initial Learning Rate, **E** stands for Number of Epochs, **O** stands for Optimizer, **I** stands for Initializer, and **B** is for Batch Size.<br>**Output:** Produced picture **G** .<br>Using the **I** approach, initialize the algorithm's weights and bias.<br>**For e = 1 to E do**<br>    **For i = 1 to \|C\| do**<br>      -Extract the $i^{ith}$ sample $c_i$ cipher image from the dataset **C**.<br>      -Extract the $i^{ith}$ sample $p_i$ plain image from the dataset **P** corespondent to the ciphered image $c_i$ .<br>      -Forward propagate the sample $c_i$ through the model M to obtain the output generated image **G**.<br>      -Compute the loss L using the output generated image**G** and the labeled $p_i$ plain image from the dataset **P** .<br>      -Back-propagate the loss L through the M model with **O**.<br>      -Update the weights and bias of the M model using the **O** and **B**.<br>      -Update the learning rate **r**.<br>      -Save the best weights and bias of the M model .<br>    **end for i**<br>**end for e**<br>return the produced picture **G** |

## 3.5 EXPERIMENT FINDINGS

### 3.5.1 Permutation patterns and primitives for P-boxes generation

To examine permutation pattern measurements in the context of visual cryptography, we used four chaotic and non-chaotic system permutation algorithms on images from the Mnist and fashion mnist data sets. The following methodological technique was utilized to construct P-box permutations of overall images with dimensions of 28x28:

**– Discrete chaos:**

In this example instance, the logistic map is used to generate a sequence of integers; however, any discrete chaotic map may be used in the same way.
After ascendingly sorting these values, the scoring system for each integer in the sorted series is utilized to populate the permutation P-BOX. The classic logistic map with *lambda* parameter looks like this:

$$r_{n+1} = \lambda r_n (1 - r_n) \tag{3.9}$$

The discrete chaotic system was iterated $(fracMNspc)$ rounds for a P-Box of size $MN$, where $spc$ indicates the lowest integer greater than or equal to $(fracMNspc)$.

**– Continuous chaos:**

In this example instance, a sequence of numbers is generated using the Lorenz system, however same technique may be used in any continuous chaotic system.

$$\begin{pmatrix} \dot{a} \\ \dot{b} \\ \dot{c} \end{pmatrix} = \begin{pmatrix} -10 & 10 & 0 \\ 8 & 4 & 0 \\ 0 & 0 & -8/3 \end{pmatrix} \begin{pmatrix} a \\ b \\ c \end{pmatrix} + \begin{pmatrix} 0 \\ -ac \\ ab \end{pmatrix} \tag{3.10}$$

To begin, short-term reliability is removed from the three output frames. The variables from each sequence are then merged to generate a single sequence. This series is sorted in ascending order to complete the permutation matrix, and a scoring system is provided for each value in the sorted series. The chaotic system is iterated $(fracMNspc)$ times for a P-Box of size $MN$.

**– Method for permuting using gray coding(*GCBPM*):**

The bijective non-linear map denoted by the following equation serves as the foundation of the [299] approach, which employs a straightforward Gray-code-based permutation strategy.

$$\gamma = \theta \oplus (\theta \gg (\beta + 1)) \tag{3.11}$$

Where $\theta$ is a number with k bits, $\gamma$ is a number with k bits of Gray code $\oplus$ is a binary XOR operation, $\beta$ is an integer, and $\gg$ is a right shift in binary, these terms are used interchangeably. The Gray-code of a k-bit number is also a k-bit number. This code performs the permutation operation by converting the picture into a one-dimensional array of pixels.

This method takes four digits $\beta_1$, $\beta_2$, $\delta_1$ and $\delta_2$ as input. It is worth noting that $\delta_1$ and $\delta_2$ are k-bit integers. Two Gray-code values, $I_1$ and $I_2$, are calculated for each pixel location. Where : $I_1 = GRAY(\theta, \beta 1) \oplus \delta_1$ and $I_2 = GRAY(\theta, \beta 2) \oplus \delta_2$

Next, move the pixel at location $X1$ to position $X2$ in the permuted picture.

**– Coupled Map Lattice(CML):**

The coupled map lattice [300], a dynamical system with discrete time and discrete space, is utilized in the manner described by (12). This system's period is sufficient for usage in cryptosystems, and its output is transformed into integer numbers using (13). The generated numbers are then used to carry out right cyclic shifts to the picture's rows and up cyclic shifts to its columns.

$$f_{d+1}(k) = (1 - \epsilon)\,\tau\,(f_d\,(k)) + \epsilon\tau\,(f_d\,(k - 1)) \tag{3.12}$$

$$s_d = mod \quad f_d \quad k \quad 10^{16}, X \tag{3.13}$$

Where $k$ = 1, 2, ..., $L$ denotes the lattice site indices, $L$ the lattice width, and $fd(k)$ the constant variable for the K th site at time $d$. For row shifts, X = N, and for column shifts, X = M, the coupling parameter *epsilon* is either one or zero. When the map *tau* is chaotic, the entire system is chaotic. Then, given a P-box of size $MN$, the linked map lattice system is repeated $Max(MN)$ times.

### 3.5.2 Experimentation in training, transfer learning, and prediction

The following tests were carried out using Python 3.7.13, TensorFlow 2.8, and Keras API on Google Collaboratory with the Google Compute Engine backend (free GPU NVIDIA Tesla K80) and 12 GB RAM. On each iteration, we were maintaining the outcomes that were much more wanted as well as the weight and bias of the CNN model by using the Keras checkpoint known as Call Backs. To demonstrate the breadth of training a machine learning-based decryptor using significant distinctions between (Plain images) and (cipher images), we set up an experiment in which DL decryptors are trained in a single round, eight rounds, and sixteen rounds using the following settings:

The results of our initial model training utilizing information from the Mnist data set are displayed below.

#### 3.5.2.1 Single round of P-box-based encryption tests

Before moving on to larger 8 and 16-round equivalents, we conduct exploratory tests on one-permutation ciphers of a lower size to gauge the effectiveness of the experimental assessment. The following pattern requirements are intended to automatically produce one round permutation keys, and they are used to generate samples using four different permutation approaches using images from the Mnist data set:

- Chaotic system 1 : The logistic map with $r_0$ = 0.448 and $\lambda$ =3.988.
- Chaotic system 2 : The Lorenz system with $a_0$ =6.293 ,$b_0$=-6.749 and $c_0$=2.886.
- Non chaotic system 1 : Coupled map lattice with $x_1$ =0.31457 ,$y_2$ =0.6532 and $\epsilon$=0.94.
- Non chaotic system 2 : Gray code-based permutation with $d_1 = 1, d_2 = 28, \delta_1 = 29493, \delta_2 = 23749$ .

As a result, it is possible to produce a large number of encrypted images quickly: 60,000 encrypted images for training and 10,000 encrypted images for testing. The two-dimensional array of elements in the ciphered pictures' structure has elements in the ranges 0 to 255. The intermediate values indicate gray scale levels from black to white, whereas the numbers 0 and 255 represent black and white, respectively.

Each dataset sample used to train a deep learning model has block cipher-related characteristics. In this initial experiment, we use four models. The labeled data for the model comprises distinct pictures that link to the data sets.

The encrypted photos are arranged in the same sequence as the clear original images from the Mnist data set to simplify training. As label data, the model also uses the original images from the Mnist dataset. Our study is divided into four main sections:

Create the permutation keys for the four permutation patterns built for the block cipher's encryption procedure in a single encryption cycle.

-Using the generated keys, create a Cipher Mnist training dataset with 60,000 samples for training and 10,000 samples for model validation.
-The same produced keys were used to create ciphered Fashion Mnist encryption training datasets for the four permutation patterns, with 10,000 samples for each model's prediction and 60,000 samples for a transfer learning test.

-Using the encrypted MIST data sets to train the four models and save the best outcomes for each.
-Use the four models developed using photos from the encrypted fashion mnist data set as deployment models for images anticipated from those images.

The method for producing data sets is shown in Figure 7, and the training results for the Mnist data set are displayed in Table 6.

#### 3.5.2.2 Experiments with 8 rounds of P-box-based encryption

To make the operation of permutation more challenging, effectively break the connection between the pixels, and force the established models to anticipate the text from the more challenging ciphertext, we increase the number of rounds of permutation

Table 3.5 – *Pseudo code for the P-box-based permutation encryption technique.*

**Algorithm**

**Input:** Plain image data set **P**, Number of rounds **R** , P-box Generation pattern **F** .
**Output:** Generated Ciphered image data set **C** .
**For r = 1 to R do**
  -Generating the permutation key $K_r$ from the **F** permutation patterns.
  **For i = 1 to |P| do**
    -Extract the $i^{ith}$ sample $p_i$ plain image from the data set P.
    -Encrypt the $i^{ith}$ sample $p_i$ image from the data set P by the key $K_r$ to get $c_i$.
    -Save the encrypted $c_i$ as the $i^{ith}$ sample $C_i$ image of the generated data set **C**.
  **end for i**
  -Use the encrypted data set **C** as the plain data set **P** for the following round **P <– C**.
**end for r**
return the generated ciphered images data set **C**.

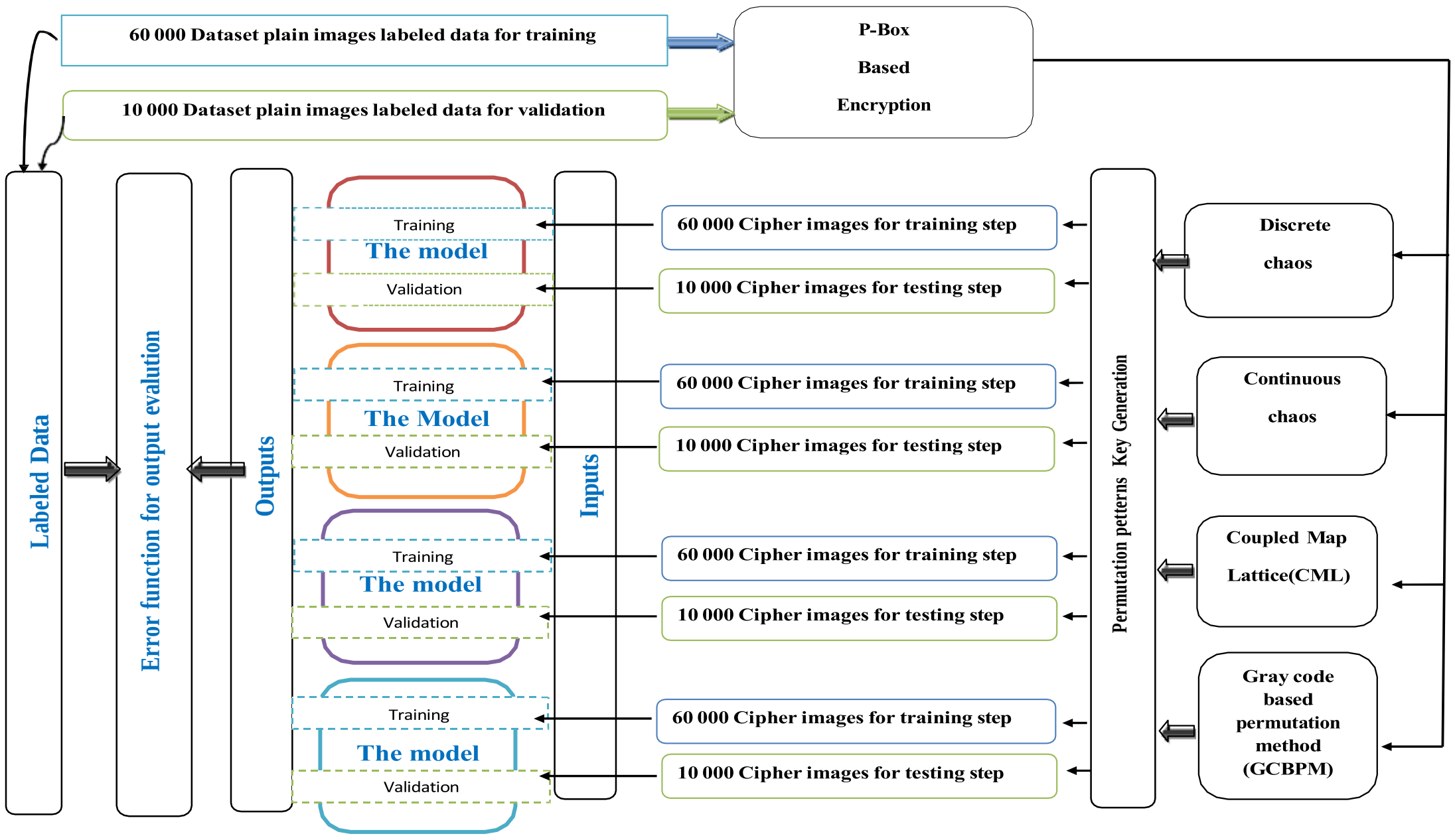

Figure 3.8 – *Creation of data sets and labeled information for single-round studies*

Table 3.6 – *One training cycle was conducted using the ciphered Mnist data.*

| Patterns | Loss | $R^2$ | MsE |
|---|---|---|---|
| Discrete chaos | 0.0842 | 1 | 0.1365 |
| Continuous chaos | 0.0708 | 1 | 0.3144 |
| Gray code based permutation | 0.0239 | 1 | 0.0052 |
| Coupled map lattice | $8.7997e^{-04}$ | 1 | $1.8217e^{-04}$ |

Table 3.7 – *Mnist ciphered data set training results for 8 rounds.*

| Patterns | Loss | $R^2$ | MsE |
|---|---|---|---|
| Discrete chaos | 0.0471 | 0.9997 | 1.5738 |
| Continuous chaos | 0.0278 | 0.9994 | 3.8839 |
| Gray code based permutation | 0.0294 | 0.9998 | 1.3939 |
| Coupled map lattice | 0.0655 | 1 | 0.0721 |

from one to eight. Although it should be noted that if many rounds are employed, the key generation step will be carried out following the number of rounds, the settings for automatically producing the first round of permutation keys are the same as for the initial trials. If there are eight rounds, the baseline characteristics listed above are used to construct the first key, the second key from the first, the third key from the second, and so on until the last round. If there are eight rounds, the baseline values described above are used to produce the first key, the second key from the first round, the third key from the second, and so on until the eighth and final rounds. The three main stages of this inquiry are as follows:

-The encryption process detailed in Table 5 allows us to acquire:

-Using the generated keys, create a Cipher Mnist training data set with 60,000 samples for training and 10,000 samples for model validation.

-The same produced keys were used to create ciphered Fashion Mnist encryption training data sets for the four permutation patterns, with 10,000 samples for each model's prediction and 60,000 samples for a transfer learning test.

-Using the encrypted MIsT data sets to train the four models and save the best outcomes for each.

-Using the four models developed using photos from the encrypted fashion mnist data set as deployment models for images anticipated from those images.

Table 7 displays the training results for the Mnist data set, and Figure 8 illustrates the method for producing data sets.

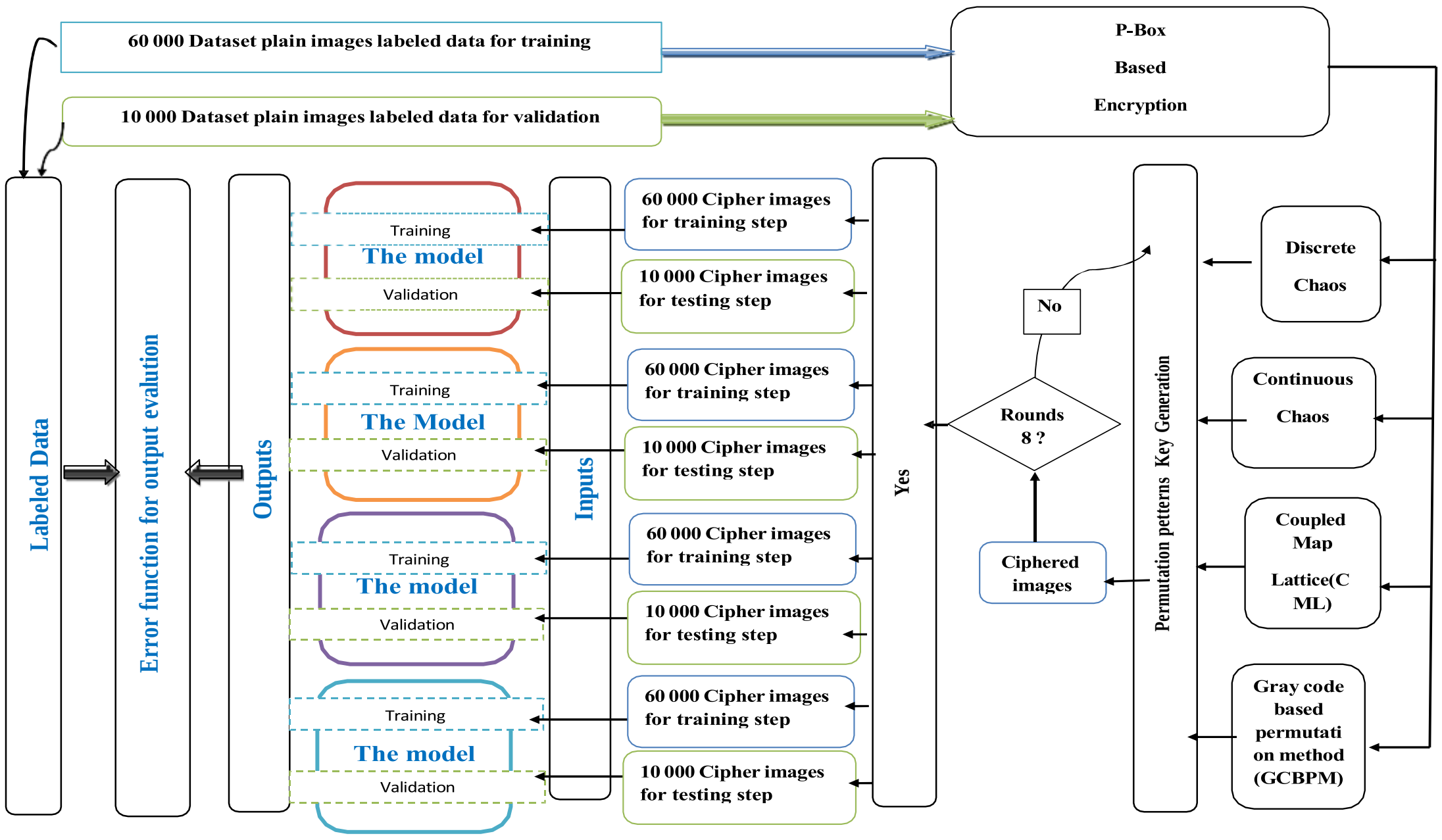

Figure 3.9 – *Generation of data sets and labeled samples for 8 rounds of tests*

### 3.5.2.3 16-round P-box-based encryption experiments

To make the permutation process more challenging and break the link between the pixels, we increase the number of rounds of permutation from "8" to "16." By doing this, we force the existing models to predict the text from the more challenging ciphertext. Although it should be noted that if many rounds are employed, the key generation step will be carried out in accordance with the number of rounds, the settings for automatically producing the first round of permutation keys are the same as for the initial trials. With 16 rounds, the baseline values described above are used to produce the first key, the second key from the first, the third key from the second, and so on until the last round.

Following are the stages of this experiment:

-Following the use of the encryption technique outlined in Table 6, we receive the following:

-Using the generated keys, create a Cipher Mnist training data set with 60,000 samples for training and 10,000 samples for model validation.

-The same produced keys were used to create ciphered Fashion Mnist encryption training data sets for the four permutation patterns, with 10,000 samples for each model's prediction and 60,000 samples for a transfer learning test.

-Using the encrypted MIST data sets to train the four models and save the best outcomes for each.

-Use the four models developed using photos from the encrypted fashion mnist data set as deployment models for images anticipated from those images.

The method for producing data sets is shown in Figure 9, and the training results for the Mnist data set are displayed in Table 8.

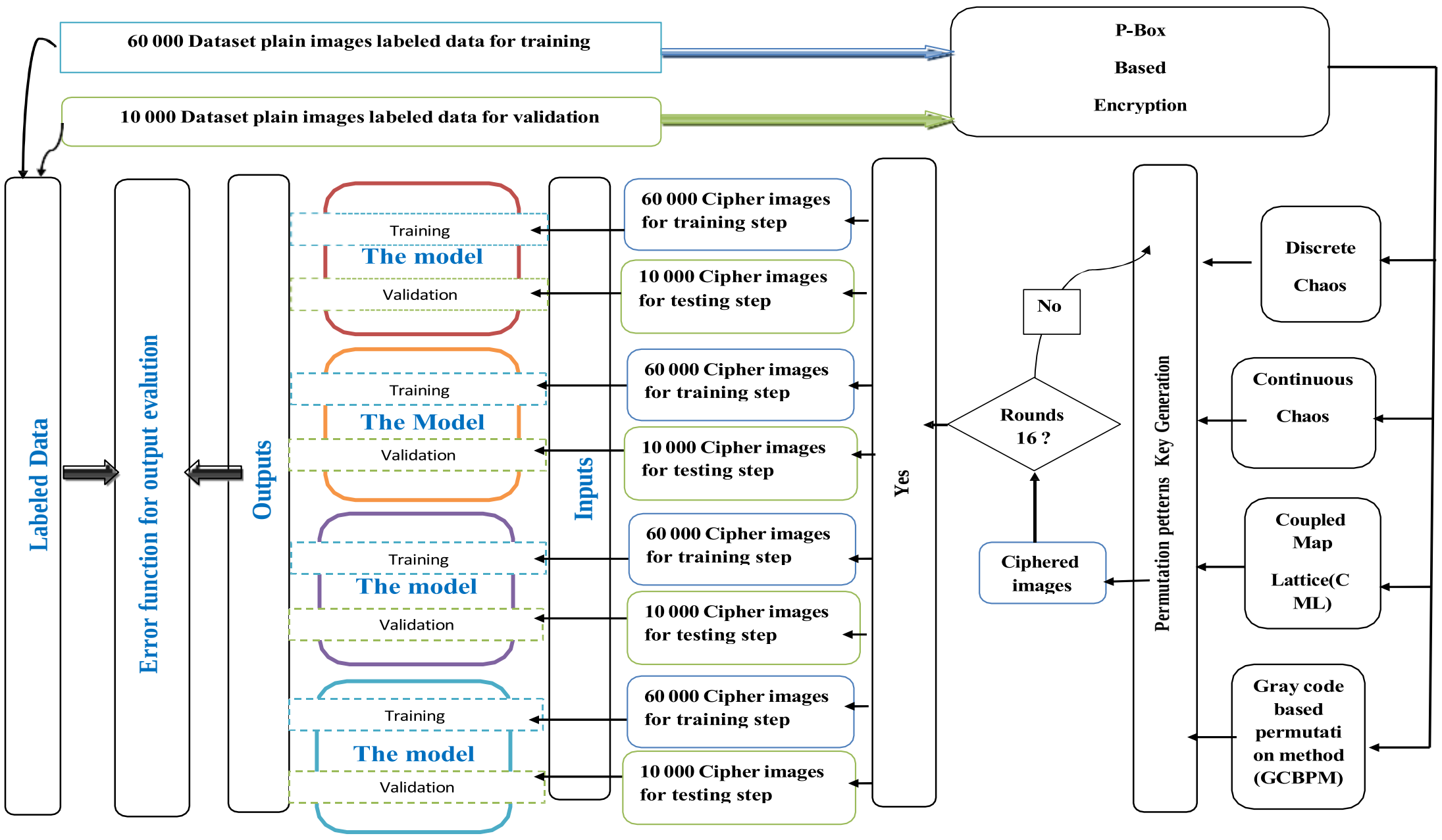

Figure 3.10 – *Data set creation and labeling for trials with 16 rounds*

Table 3.8 – *Mnist ciphered data set for 16 rounds of training outcomes.*

| Patterns | Loss | $R^2$ | MsE |
|---|---|---|---|
| Discrete chaos | 0.1061 | 0.8507 | 922.6311 |
| Continuous chaos | 15.1446 | 0.6793 | 1957.4821 |
| Gray code based permutation | 0.2193 | 0.9998 | 1.3684 |
| Coupled map lattice | 0.0724 | 0.9996 | 2.3448 |

Table 3.9 – *Experiment with one cycle of P-box-based encryption for transfer learning.*

| Patterns | Loss | $R^2$ | MSE |
|---|---|---|---|
| Discrete chaos | 0.1867 | 1 | 0.3265 |
| Continuous chaos | 2.4533 | 0.9782 | 175.4175 |
| Gray code based permutation | 0.0480 | 1 | 0.0725 |
| Coupled map lattice | 0.0014 | 1 | 0.0015 |

Table 3.10 – *Experiment with transfer learning using 8 rounds of P-box-based encryption.*

| Patterns | Loss | $R^2$ | MSE |
|---|---|---|---|
| Discrete chaos | 0.0270 | 0.9993 | 5.8473 |
| Continuous chaos | 0.1995 | 0.9994 | 5.0487 |
| Gray code based permutation | 0.0769 | 0.9994 | 4.6592 |
| Coupled map lattice | 0.0530 | 1 | 0.1907 |

#### 3.5.2.4 Experiments on transfer learning

The p-box production patterns are intended to produce permutation keys that appear random (pseudo-random generators), but they also permit the inverse operation, which is decryption without data loss. The model itself is not distinctive, but it helps identify the decryptor. In other words, a CML one-round encryption method based on P-boxes and a model trained on one-round CML encrypted data may be distinguished from one another. Although all of the models have the same architecture, layers, and hyperparameters, the parameters gathered throughout the training phase are what distinguishes them (weights and bias).

Using the weights and bias of the first model trained with the Mnist data set for one, eight, and sixteen rounds as deployment models for the Fashion Mnist models for one, eight, and sixteen rounds with the same permutation patterns and algorithm parameters, respectively, we attempted to implement learning transfer. The evaluation process converges toward desired results and the error function shrinks without training, which is the most unexpected finding.

Table 3.11 – *Experiment with transfer learning using 16 rounds of P-box-based encryption.*

| Patterns | Loss | $R^2$ | MSE |
|---|---|---|---|
| Discrete chaos | 0.0992 | 0.7599 | 1940.4117 |
| Continuous chaos | 8.3224 | 0.9223 | 623.6430 |
| Gray code based permutation | 0.5570 | 0.9994 | 4.9375 |
| Coupled map lattice | 0.0714 | 0.9991 | 7.6776 |

It should be noted that using the transfer learning techniques presented in this research, it is also feasible to build acceptable decryptors from scratch by combining transfer learning with the best prior cryptographic knowledge. Tables 9, 10, and 11 show the results that were achieved, accordingly. The most potent example of the idea of distinguishing ability stressed in this study is the experience of learning transfer re-usability to enhance model performance.

#### 3.5.2.5 Model prediction experiments

Specifically, models trained on the Mnist data set were estimated using the Fashion Mnist data set, with both models projected using identical weight and bias settings. The prediction was then made by merging the search results. The images below serve as an example of our findings and a sample of the best outcomes from the model of the CML-encrypted images from the two data sets and the prediction of the associated images. The discrepancy in color distribution that the projected model of simple images predicts is exactly what we can see in the encrypted photos (Figs. 10 and 12). (Fig11, Fig13).

Additionally, the photos (Figs. 14 and 15) demonstrate the prediction of the encrypted images of the worst-case scenarios of the chaotic discrete models after 16 iterations, with the visible degradation in image quality in the forecast being rather evident.

## 3.6 DISCUSSIONS

### 3.6.1 Results evaluation

We need a tool that compares the visual original plaintext from the Mnist and Fashion mnist data sets to the prediction results in order to fully analyze and interpret the research findings and present the test in a more reliable manner. We employed a pre-trained deep learning model with sufficient accuracy to differentiate between the outputs in order to evaluate the effectiveness of our efforts. The model was trained using the Adam optimizer and the Sparse Categorical Cross entropy error function. The Mnist and Fashion Mnist data sets are accurately recognized by this model. Its design is rather simple, and it has an accuracy of (98.05%) for Fashion Mnist and (99.00%) for Mnist.
It is suitable for use in experimental studies as well. Figure 16 shows the architecture of this pre-trained model. Before utilizing the model to monitor the effectiveness of our predicted encrypted photographs, we first assessed the model's prediction performance on the original MNIST and Fashion MNIST test sets. The visual results of the quantitative prediction analysis are displayed in Figures 17 and 18.

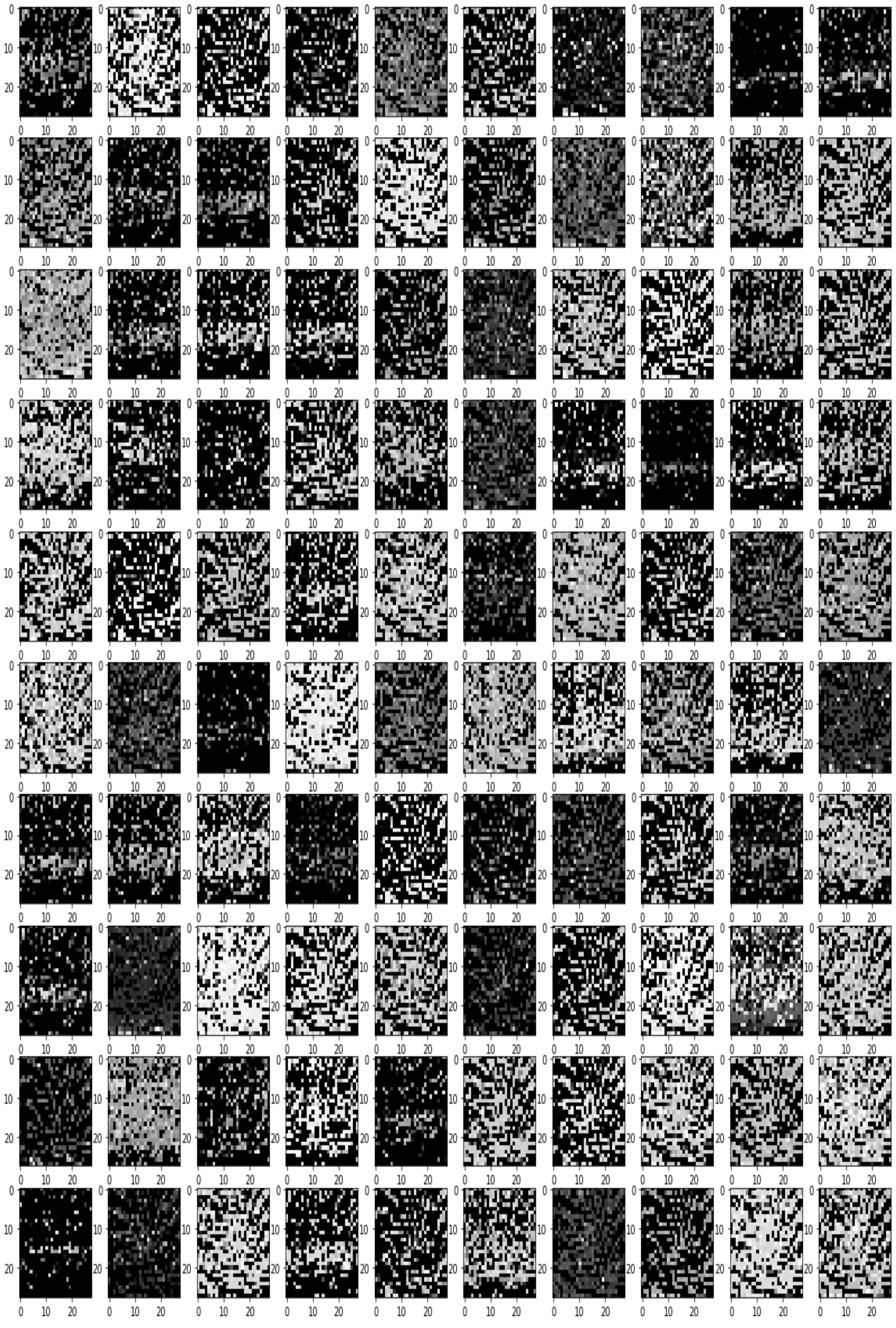

Figure 3.11 – *Fashion Mnist ciphered pictures with a single round CML.*

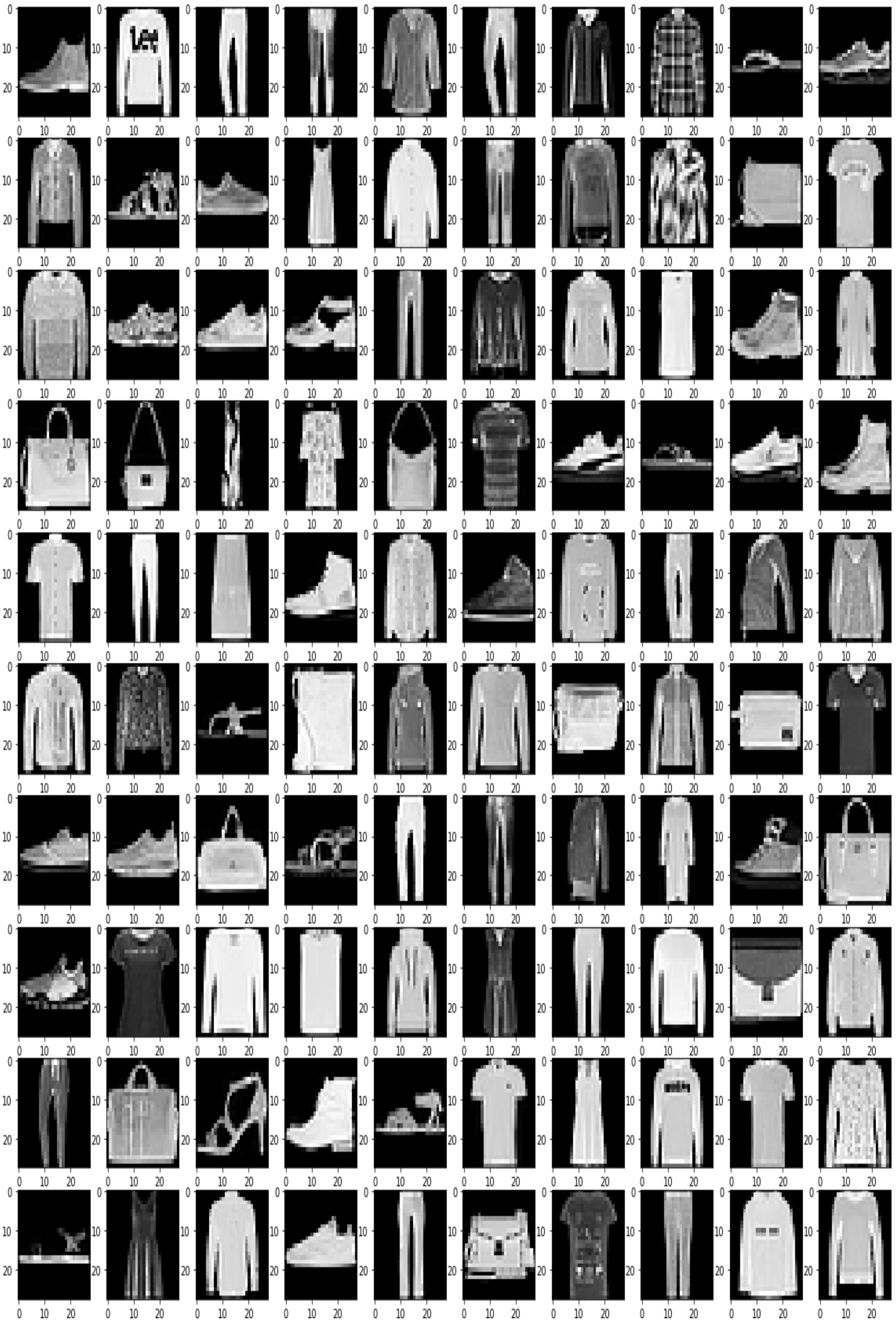

Figure 3.12 – *Fashion Mnist anticipated visuals for one round of CML.*

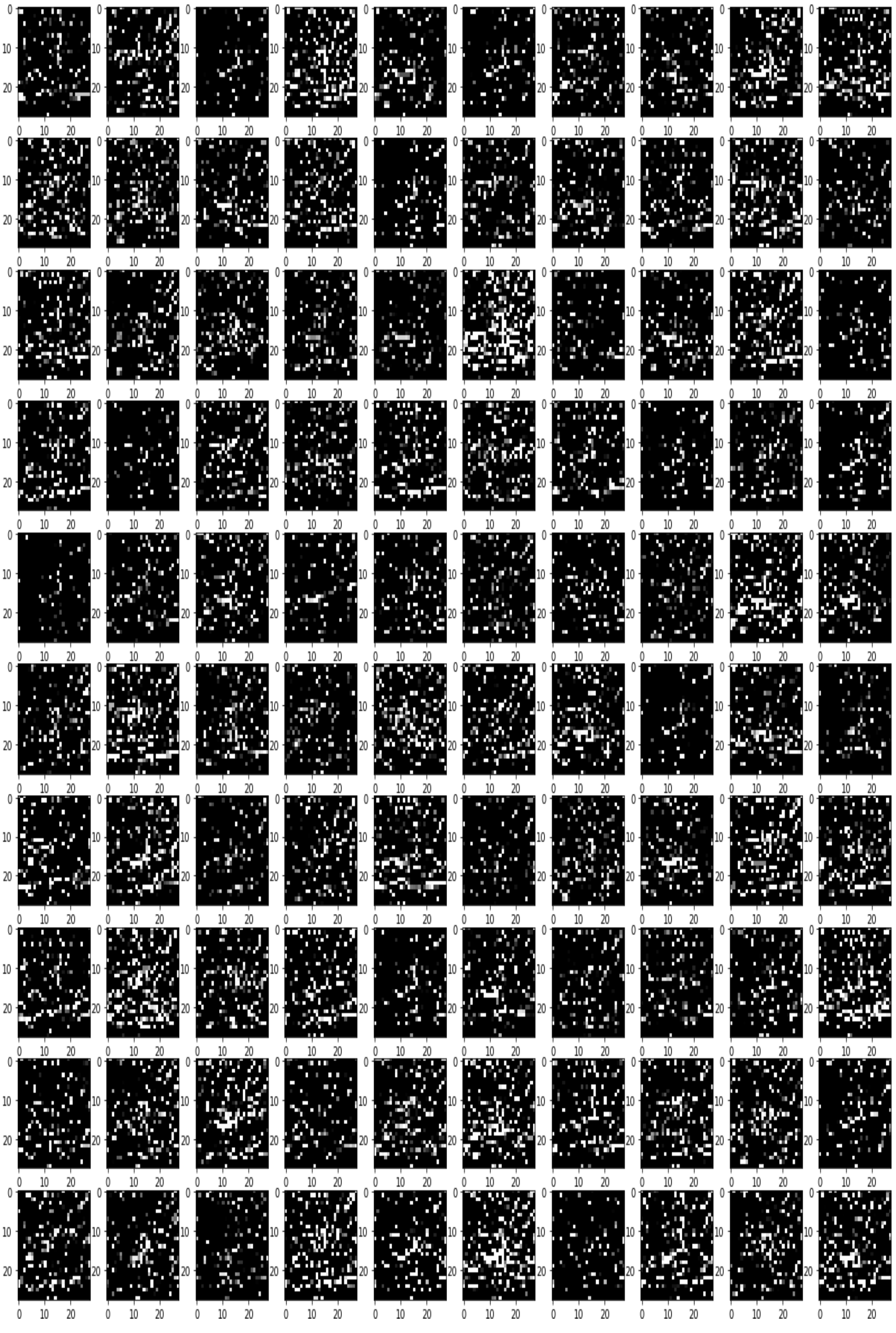

Figure 3.13 – *Mnist ciphered pictures using a single round CML.*

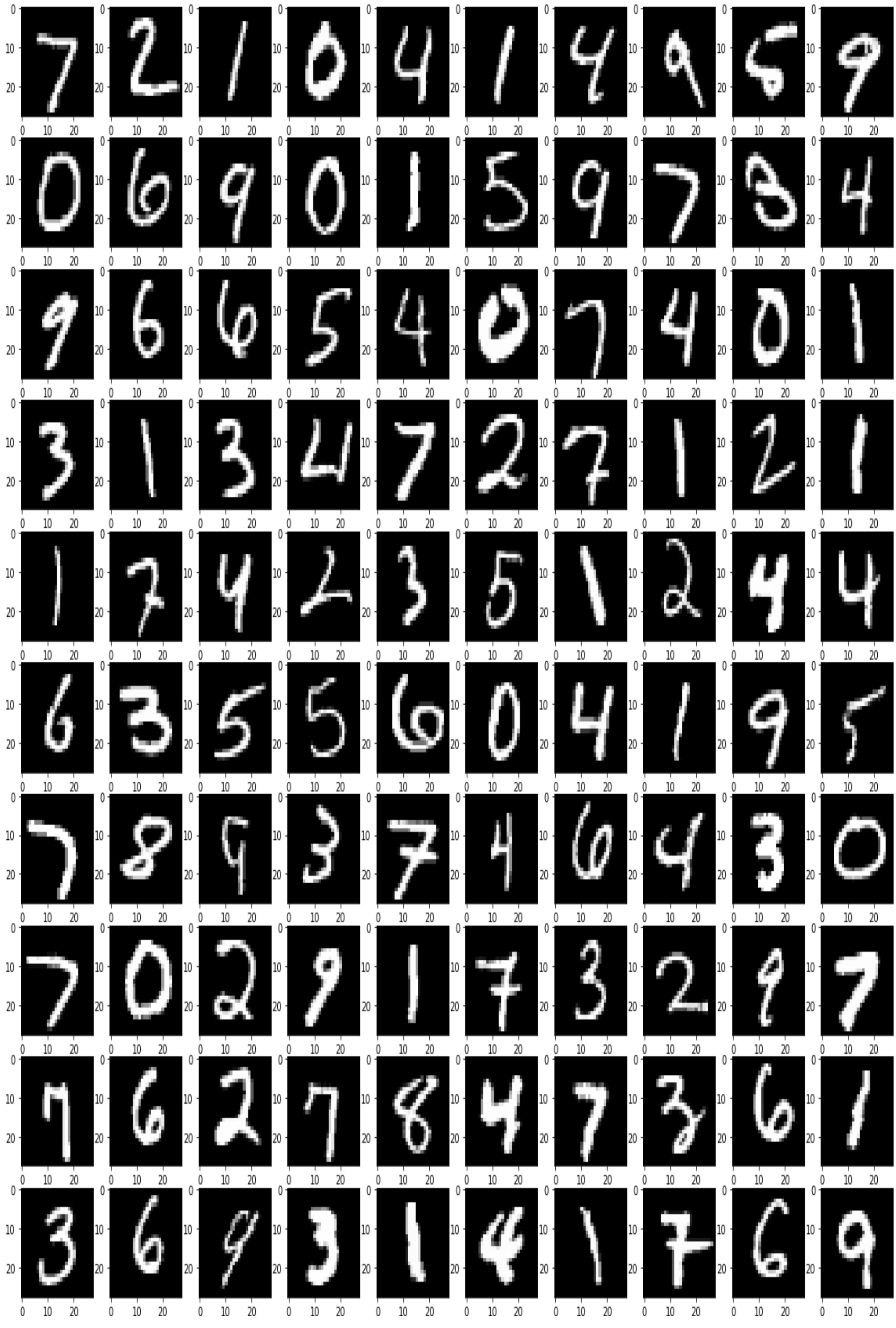

Figure 3.14 – *Mnist expected photos of a single cycle of CML.*

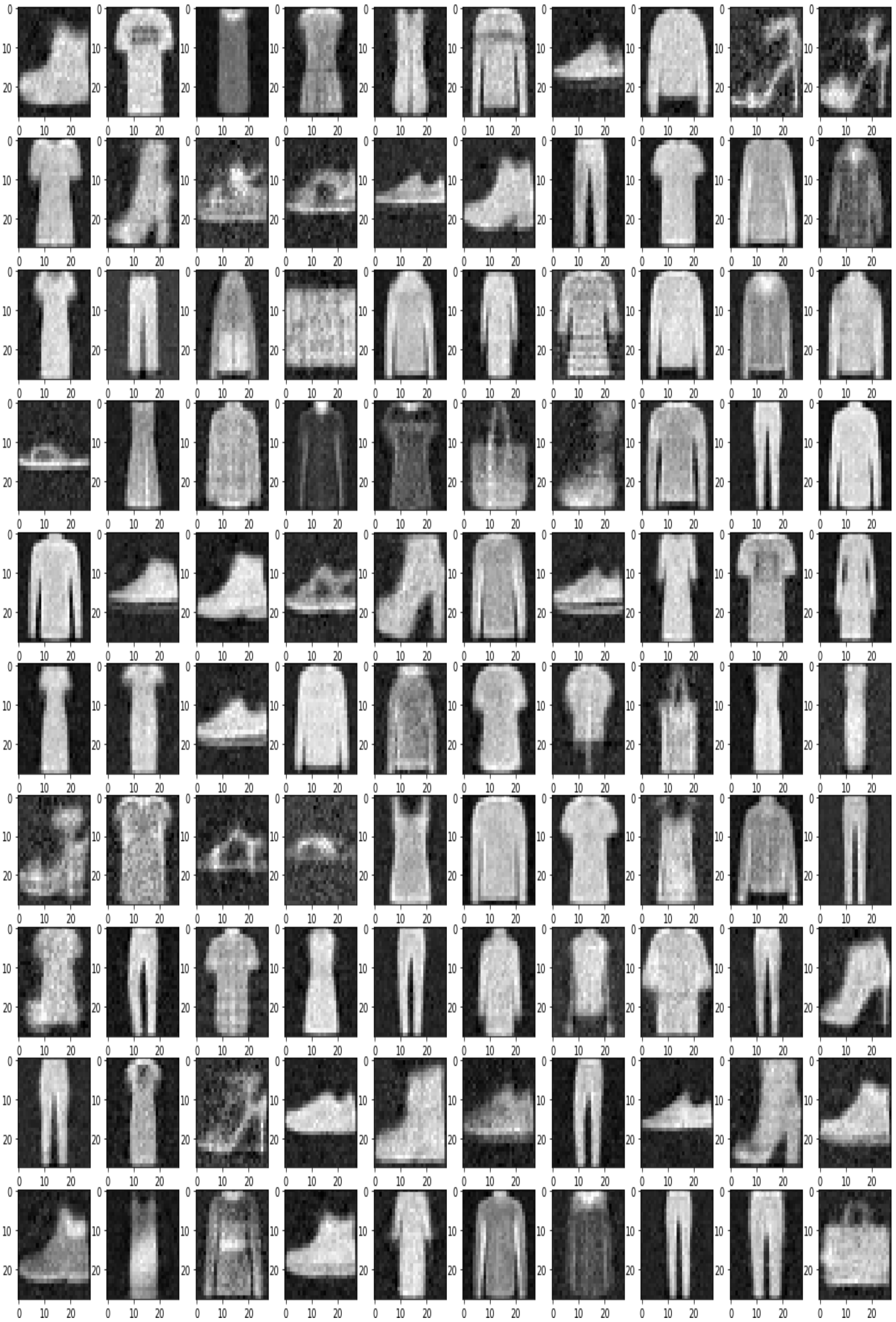

Figure 3.15 – *Mnist predicted pictures for the first 16 rounds Distinct chaos.*

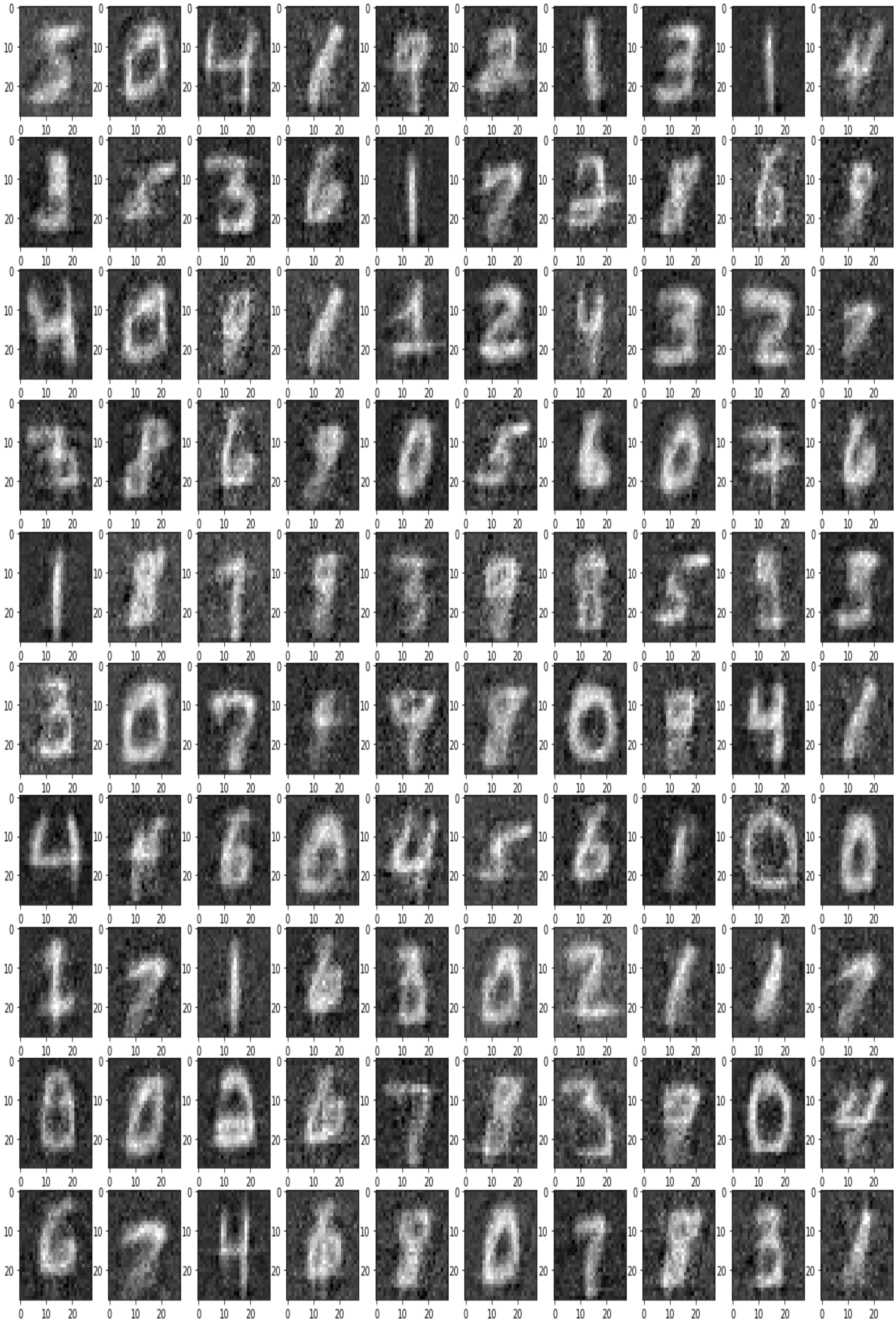

Figure 3.16 – *Mnist matching generated pictures of 16 rounds Discrete chaos.*

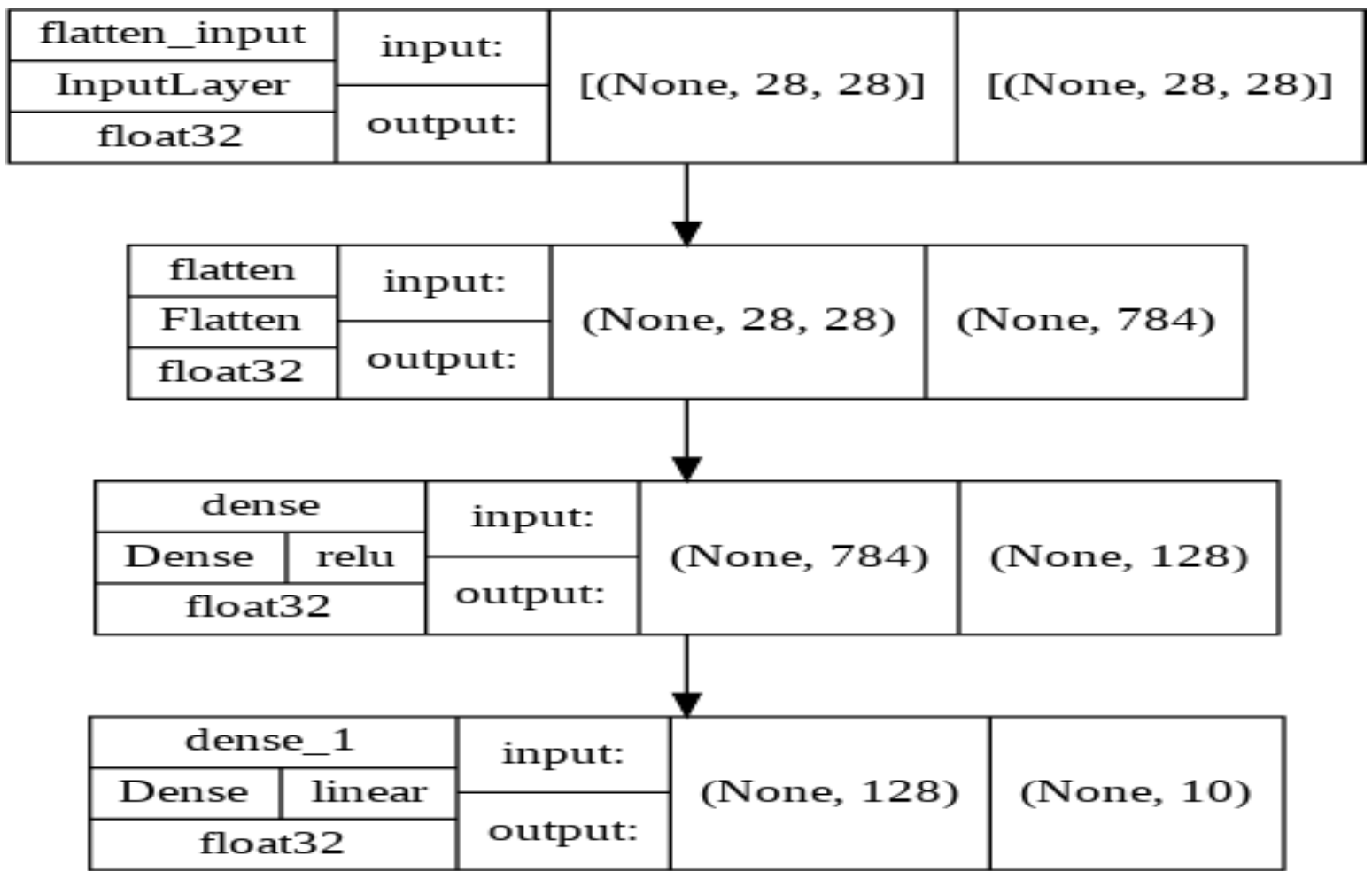

Figure 3.17 – *the architecture of the measurement model.*

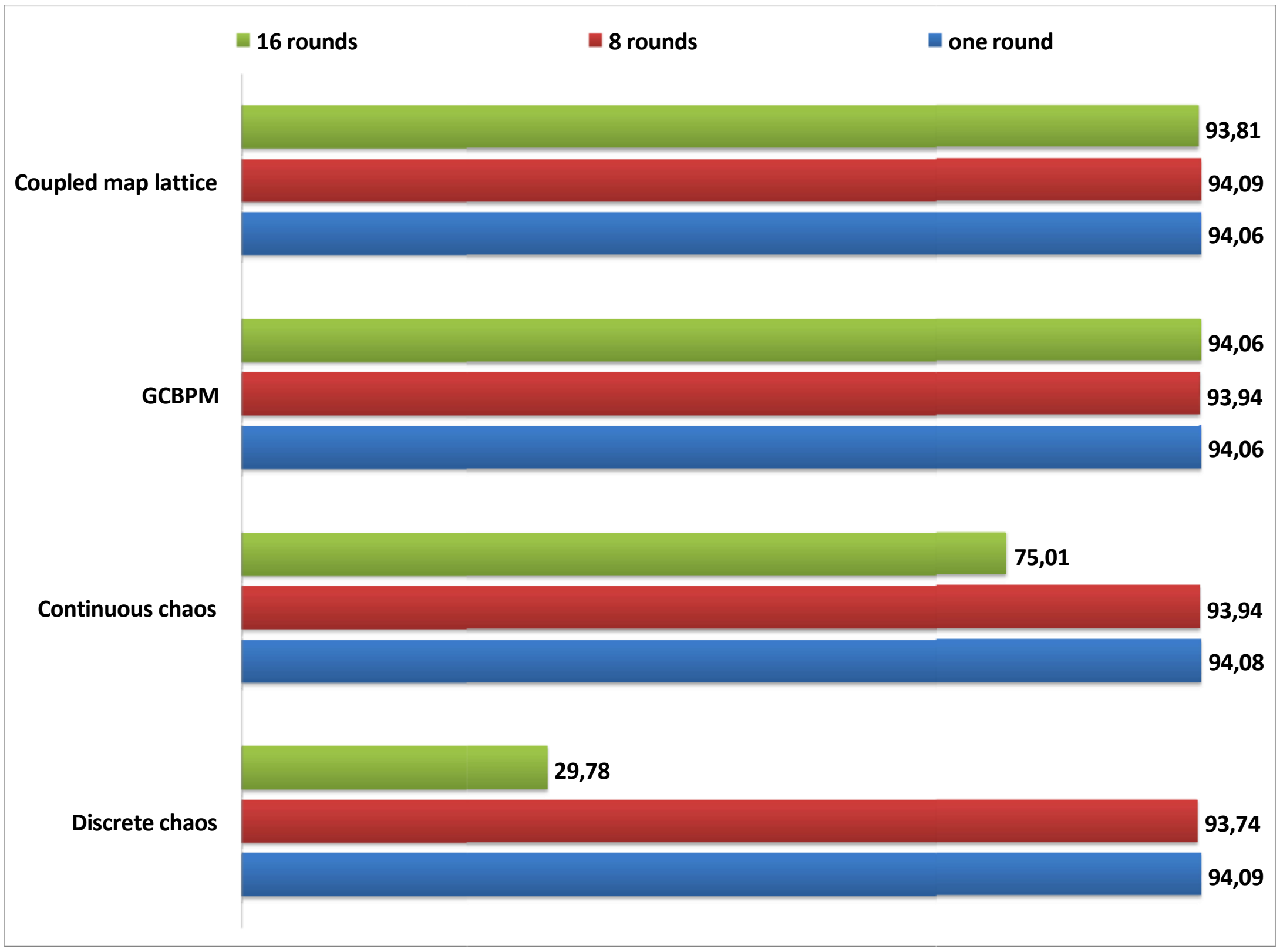

Figure 3.18 – *Results in numbers for the Mnist dataset.*

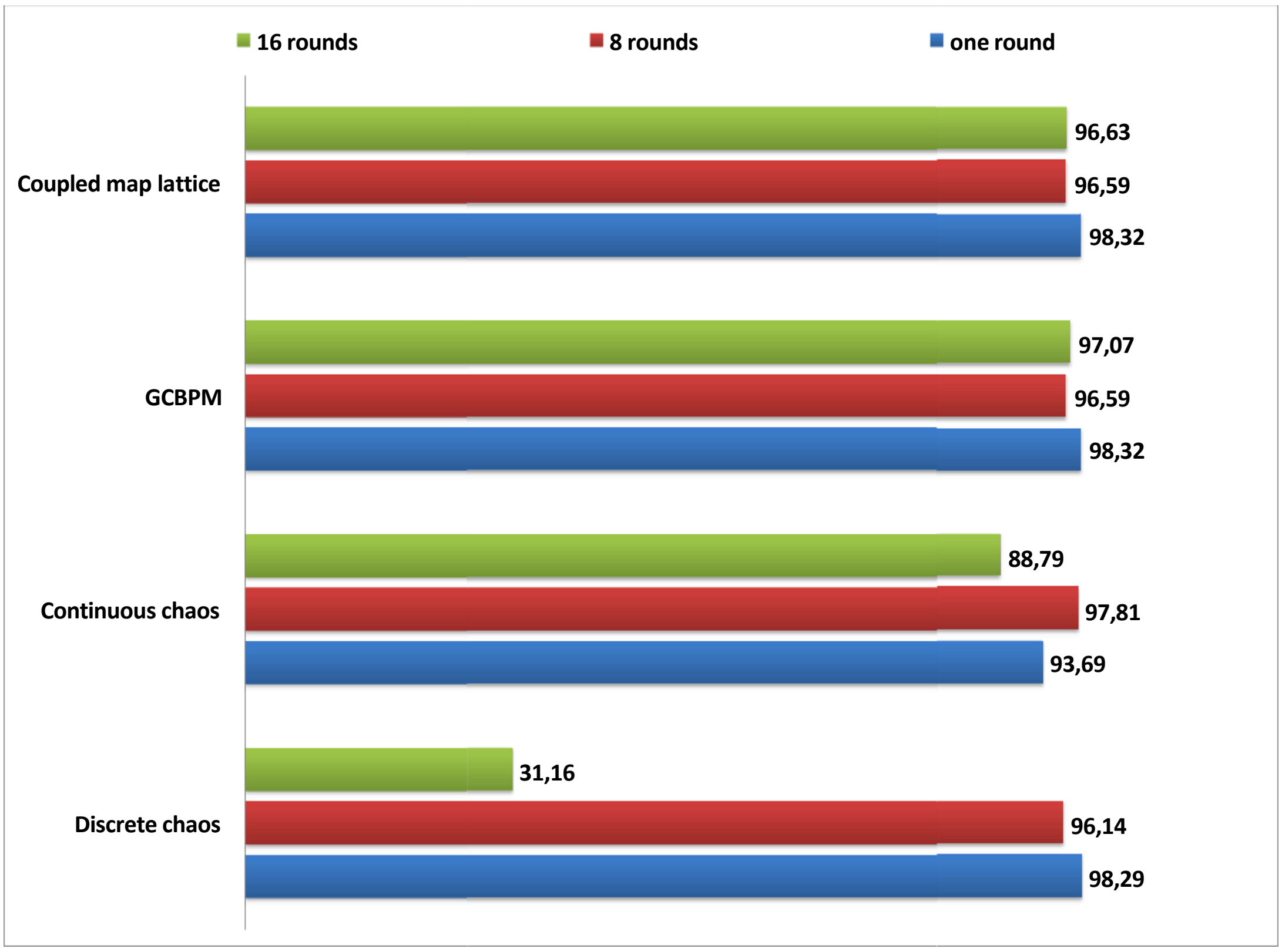

Figure 3.19 – *Results in numbers for the Fashion Mnist dataset*

Table 3.12 – *Related literature works Comparison*

| Comparable linked works | Black Box Attack | Distribution | KPA (Cipher/Plain) Pairs | Attack complexity | Rounds consideration | retrieved data | Contribution |
|---|---|---|---|---|---|---|---|
| Shujun Li at al[43] | Yes | Uniform | $O(n(MN)^2)$ | $O(Log_L(MN))$ | It has not been addressed | 50 % of the key | General quantitative concept |
| Chengqing Li at al [44] | Yes | Uniform | $O(32(MN))$ and $O(16.n_0.(MN))$ | $O(Log_L(MN).MN)$ and $O(MN)$ | It has not been addressed | More than 50 % of the Key | Optimal quantitative concept |
| Alireza Jolfaei at al [46] | Yes | Uniform | $Log_L(MN)$ | $O(n(MN))$ | It has not been addressed | 100 % of the key | Recovering completely the key |
| Leo Yu Zhang at al [54] | Yes | Uniform | $Log_L(MN)$ | $O(n(MN))$ | It has not been addressed | 100 % of the key | Concept of composite representatio |
| Our work | Yes | NoN Uniform | More pairs for more best results | Depends on the trained model and parameters | Taken into consideration | Plain image | Concept of non uniform distribution |

### 3.6.2 Quantitative findings and comparisons

The first thing we saw was that the effectiveness of the deep learning technique [301] rose as the number of rounds did. On the other hand, this degeneration is brought on by a range of factors and varies depending on different permutations. For instance, linked map lattice is more secure and long-lasting than Gray code-based permutation, and discrete chaos permutation patterns are more resistant to tests than continuous chaotic patterns.

As a result, we discover that as the number of rounds increases, discrete chaos becomes more resistant to our attack, followed by some minor resistance from continuous chaos; the scientific interpretation of the resistance is the discrete generation of permutation patterns, which makes the attack more difficult by more efficiently destroying the correlation between the swappable atoms. Despite the enormous number of rounds, CML and GCBPM do poorly in this test.

Another important aspect to emphasize is the non-uniform distribution of atoms (pixels) in plain text. Despite the non-uniform distribution of colors in the Mnit and fashion mnist datasets, the attack findings differ dramatically; the quantity of duplicated atoms is self-evident.

It should be noted that the type of data we utilize has a significant impact on our work. For instance, if we use a picture with all intensities set to zero and a total number of pixels between zero and 255, the model can only learn the loss function, which is symbolized by the zero difference constraint in differential cryptanalysis.

### 3.6.3 Comparison of related works in the literature

To extract the permutation key for use in the encryption of pictures encrypted only by permutation, Shujun Li and associates [257] developed a quantitative cryptanalysis method. Based on the color box color black test, which makes use of KPA pairings and is mainly concerned with the uniform distribution of colors in the pairs involved, this way of expression is based on the color black. The tested permutation key could be recovered using this approach to a 50% success rate. However, the success of their work is entirely dependent on powerful computing and storage facilities. By lowering the number of KPA pairs need to operate, as well as the computational and geographic complexity, Chengqing Li et al. [258] improved the findings of [259]. They enhanced the attack and successfully recovered more than half of the permutation key used in the test. Their method may be adapted to execute any attack as a permutation cipher using merely KPA pairs of uniform distribution. By demonstrating that the correct permutation mapping is fully recovered and selected in all permutation-only picture ciphers, irrespective of cipher design, Alireza Jolfaei et al. [256] improved on earlier studies. The quantity of necessary KPA pairs, as well as the computational and complexity requirements, have all been decreased by him.

By discussing how to balance storage cost and computational complexity while using the KPA approach, Leo Yu Zhang et al. [205] improve on earlier studies. To close the KPA gap between artificial noise-like pictures that fully fit the theoretical model and corresponding actual images, he also used a unique idea of the composite representation. Regardless of their applicability, all of these algorithms have the same significant flaw: there is no consistency in the distribution of colors in KPA pairings. One of the drawbacks is that none of these works considered the potential of employing several rounds of permutation. All of these studies are based on the black box test using conventional research methodology and various optimization methods. Since uniform distribution cannot be found in real data, its recurrence during the evaluation of the permutation approach continues to be problematic in actual, practical settings. Table 12 compares and contrasts each of these works. In contrast to earlier studies, we took a fresh approach to the issue in our contribution. We chose the absence of homogeneous color dispersion as our primary target. The round count and key generation techniques for this topic were also looked at. We believe that our method's greatest benefit is how easily it can be applied again and again. This approach might be used in the deployment phase to assess the efficacy of picture encryption techniques or in the development, phase to identify the most effective permutation scheme.

The design team can use the same dataset as in our study to generate images encrypted by the various algorithms under test, and then train our model on these generated data with different cycles, algorithm specifications, and combinations to select the best permutation methodology for using, as well as its best option of the best possible algorithms' characteristics and the adequate number of cycles.

Encrypting photos several times and then evaluating the findings from the models to change the number of times and comparing other approaches, may also help with decision-making. There are two facets to the idea of reuse. Using the same datasets, model architecture, and parameters as outlined in this article, the chosen permutation strategy must first be put to the test. The second step is using the suggested encryption method to create encrypted 28x28-pixel pictures from various datasets while preserving the same model architecture.

We advise concentrating on our convolution and de-convolution algorithms, which are employed to create deep learning algorithms with more appropriate layers if designers wish to evaluate cryptosystems in a bigger area with larger pictures. if the hyper-parameters are kept constant. Considering this, our approach exposes the flaws of the approach by requiring pre-treatment dataset processing, a big amount of computer resources, computing time, and a huge number of tests.

## 3.7 CONCLUSION

In this chapter, we present a unique approach for identifying decryptors on symmetric permutation primitives employing deep convolutional neural networks. The methodologies are applied to any number of (non-zero) input adjustments. We use numerous divergences to solve the challenge of discriminating in a two-dimensional domain. The offered research is intended to be utilized independently of how encryption systems operate. It ought to be utilized beforehand to investigate the most powerful permutation approach to be employed during the cipher architecture construction process.

Alternatively, it could be employed to assess and contrast various permutation pattern solutions based on established scientific principles. However, the duration it takes to start generating those assessment methods seems to have a substantial impact on their convenience. We provide no potential that deep-learning technology will replace conventional cryptanalysis one day. Moreover, our results demonstrate that deep learning models are sometimes capable of performing cryptanalysis at a level that cryptographers deem advantageous, and also that deep learning methods could be a necessary addition to the armory of cryptographic auditors.

# LIGHTWEIGHT CRYPTANALYSIS AND REGIONS OF INTEREST DETECTION

# 4

CONTENTS

## 4.1 Introduction

TODAY'S world, the globe is in the midst of the IoT age, in which all data travels from one personal device to another, along with personal and secret information. Normally, this type of information necessitates private security. Cryptography is the current science of systematically exploring advanced strategies for effectively safeguarding sensitive information in communication networks or in proper data storage [269]. Famously used cryptograms such as AES and DES crucially demand a large number of resources for effective implementation; nevertheless, these cryptograms are not practical in particular IoT devices [270],[271],[272],[273] due to the probable restrictions of different performance metrics. To overcome these inherent limits, lightweight cryptography played a key role in dealing with devices that often have limited memory space. For IoT, lightweight block ciphers work properly on a specific block of sophisticated data for fixed-length specific bits and a symmetric key with a profound change.

In most situations, these profound alterations share simple bit operations, such as hypothetical substitution and permutation networks (SPN) or private Feistel networks. Lightweight ciphers are mostly symmetric ciphers that are made lightweight in practical terms of modest size, small storage, local memory, potential limited energy, and processing time. They are widely used in smart healthcare sensors, intelligent wireless multimedia surveillance networks (SWMSN), radio frequency identification tags (RFID), self-driving vehicles (SDV), drones surveillance systems (DSS) modern cars, bio-chip remote farm animals surveillance, cyber-physical surveillance, and cyber-physical surveillance.

## 4.2 Techniques for Implementing IoT Lightweight Ciphers

Data encryption and decryption are totally handled by terminal devices in the majority of IoT systems. However, they have intrinsic restrictions, such as:

- Obviously, their limiting size, storage memory, and power consumption limit their particular utility in the secure transmission of critical data when reasonable development of lightweight ciphers meets these feasible restrictions [270].
- IoT junctions are vulnerable to both active and passive cyber-attacks, as well as external threats, because of the absence of effective authentication on IoT devices, weak encryption, and inadequate key management.
- Specific protocols established in IoT devices enable network data transfers and control information, and each of these protocols has its own set of error detection/correction systems, key, and identity management systems, and compression and encryption algorithms.

For these considerations, the framework among these cryptographic techniques in the existing situation is founded on two significant configuration generations that achieve light encryption with benefits for IoT. The following approaches are used to create IoT lightweight ciphers:

### 4.2.1 Hardware technique

In this obvious situation, the algorithms are easily implemented in specified hardware [274], and the primitive's practical efficiency is appropriately judged by the following metrics:
–The economic power consumed by the specific hardware circuit is known as energy consumption efficiency. The economic performance is beneficial for low power consumption.
– Latency is typically defined as the amount of time taken by the hardware circuit to create a certain output. It is properly priced in the time required. If latency is reduced, the outstanding performance will be even more satisfying.
–The Gate Equivalents (GEs) are the physical memory areas that are normally required to efficiently implement an algorithm primitive. If the specific region is smaller, the outstanding performance will be superb.

### 4.2.2 Software technique

The lightweight technique may also be implemented in specialized software [275], primarily for practical usage on microcontrollers. The performance metrics that will be adequately examined in this specific sort of successful implementation are:
– Random Access Memory (RAM) utilization, program length, and throughput.
The throughput efficiency is the measured number of valid messages processed correctly per time unit. It is measured in basis points (bps). If the private message is normally handled quickly, the ideal performance will be more outstanding. The Random Access Memory (RAM) usage is related to the reasonable amount of necessary information carefully scripted to the appropriate storage IoT specialized device and Program length represents the fixed quantity of necessary information ordinarily required to validate the economic performance without typically relying on its essential input.

## 4.3 THE EXPERIMENTS WITH KATAN BLOCK CIPHER

This section provides research that contributed to the publication of the results of the study in a published conference paper titled : "Deep Neural Network Based TensorFlow Model for IoT Lightweight Cipher Attack", International Conference on Artificial Intelligence and its Applications (AIAP) in El Oued, Algeria in 2022, with the following authors: Zakaria TOLBA, Makhlouf Derdour. [303]
Our work [303] is correctly established as a regression challenge in this illustrated study, where we tentatively offered a deep learning technique to KATAN 32-bit [276] lightweight block cipher security study. Typically, we train fully connected deep neural network models to successfully predict the plain text from the chosen ciphertext. TensorFlow Framework is used to improve fully connected deep neural networks in a Google Cloud environment. Using cloud technologies, we thoroughly examine the economic feasibility of the proposed test.

### 4.3.1 Problem framing for experiments tests

The proposed work's [303] overall purpose is to train neural network models to predict plaintext from the ciphertext. We structured the problem as a regression challenge using supervised learning since the aim is to forecast the Block cipher consisting of non-negative integers. Professional experiments were carried out for both the KATAN 32-bit cipher. A cryptanalyst's ultimate objective is to reduce the distinguisher model from diverse block cipher data inputs.

### 4.3.2 Model architecture details

We used fully connected neural networks in all of the following tests. We first used hyper-parameter tuning to determine the best number of layers, neurons per layer, loss function, optimizer, number of epochs, and batch size for the regression job. We carefully picked a neural network with seven hidden layers based on scientific evidence and practical trials. The number of neurons per unique layer often varies depending on the layer. There are 32 neurons in the input layer (corresponding to the specified number of input features) for Katan 32-bit ciphers, seven hidden layers with 24, 20, 16, 12, 16, 20, and 24 neurons each, and an output layer with 32 neurons to represent the projected plaintext, as shown in Fig.5.1.

```
Model: "sequential"
_________________________________________________________________
Layer (type)                 Output Shape              Param #
=================================================================
dense (Dense)                (None, 32, 32)            64
_________________________________________________________________
activation (Activation)      (None, 32, 32)            0
_________________________________________________________________
dense_1 (Dense)              (None, 32, 24)            792
_________________________________________________________________
activation_1 (Activation)    (None, 32, 24)            0
_________________________________________________________________
dense_2 (Dense)              (None, 32, 20)            500
_________________________________________________________________
activation_2 (Activation)    (None, 32, 20)            0
_________________________________________________________________
dense_3 (Dense)              (None, 32, 16)            336
_________________________________________________________________
activation_3 (Activation)    (None, 32, 16)            0
_________________________________________________________________
dense_4 (Dense)              (None, 32, 12)            204
_________________________________________________________________
activation_4 (Activation)    (None, 32, 12)            0
_________________________________________________________________
dense_5 (Dense)              (None, 32, 16)            208
_________________________________________________________________
activation_5 (Activation)    (None, 32, 16)            0
_________________________________________________________________
dense_6 (Dense)              (None, 32, 20)            340
_________________________________________________________________
activation_6 (Activation)    (None, 32, 20)            0
_________________________________________________________________
dense_7 (Dense)              (None, 32, 24)            504
_________________________________________________________________
activation_7 (Activation)    (None, 32, 24)            0
_________________________________________________________________
dense_8 (Dense)              (None, 32, 32)            800
_________________________________________________________________
activation_8 (Activation)    (None, 32, 32)            0
=================================================================
Total params: 3,748
Trainable params: 3,748
Non-trainable params: 0
_________________________________________________________________
```

Figure 4.1 – *Architecture parameters details*

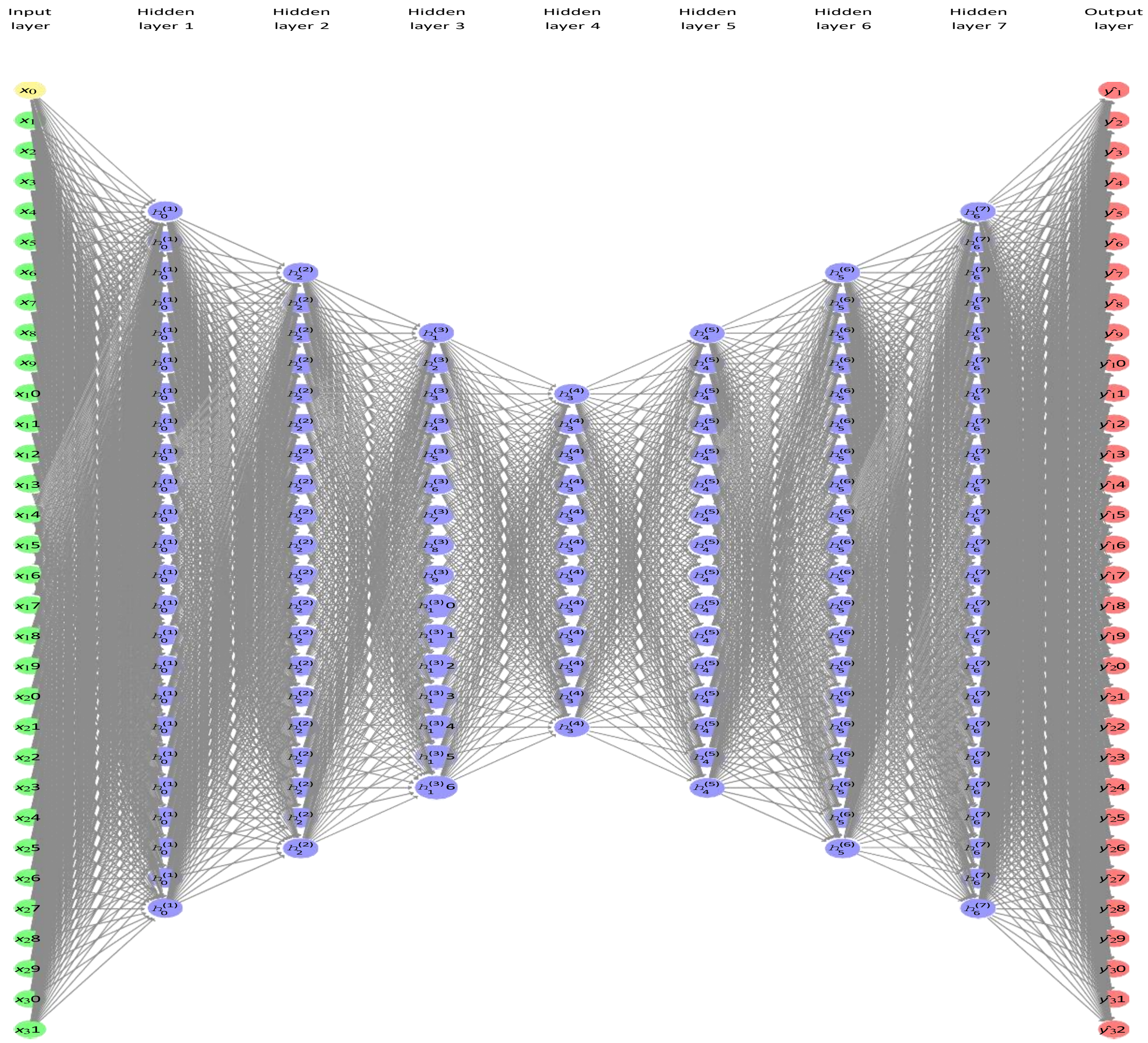

Figure 4.2 – *Model layers details*

### Hyperparameters used in the experiment

The following are the remaining hyperparameters that were successfully employed in our tests:

- Initialization of weights: Glorot Uniform.
- "SGD" stands for Stochastic gradient descent Optimizer.
- learning rate = 0.001.
- Rectified Exponential linear unit is the activation function (RELU).
- Three thousand epochs.
- 64 batch sizes.
- 3 748 trainable parameters.
- There are zero non-trainable parameters.
- MSE (mean squared error) is the error function.
- R-squared accuracy function.

## 4.3.3 Datasets

We were able to construct enormous datasets in a reasonable time by appropriately leveraging the Katan 32-bit ciphers for proof-of-concept studies. When compared to the PRESENT cipher, the KATAN block cipher is more compact and achieves a smaller visual footprint of 802 GE. It is a hardware-oriented block cipher that employs a basic key scheduling method and a specific type of fundamental Feistel structure.[303] We economically created datasets of 1 470 000 samples for the training step and 245 000 samples for the model validation utilizing Python lightweight cryptography tools in the Google collaboratory settings.

## 4.3.4 Experiment outcomes for training and prediction

R2 is always between 0 and 1, inclusive. An R2 of 1 implies that the regression predictions fully fit the data. Our work is a regression issue, and the model's accuracy is characterized by the following extra metrics: cosine closeness, root mean squared error, and absolute squared error. Figure 2 depicts the trained model's Graph dependencies.
We used the Keras checkpoint called Call Backs to store the most satisfying results after each iteration as well as the weight and bias of our trained model. After 3000 epoch iterations in 11 hours and 43 minutes, the most spectacular results were produced in the 2759 epoch. Mean squared error = 0.0087, according to the error function. The accuracy function is equal to 0.89. The high R-squared value, which normally ranges from 0.89, also shows that the anticipated plain text and ciphertext have an effective association. The findings are shown in Fig. 4 below.

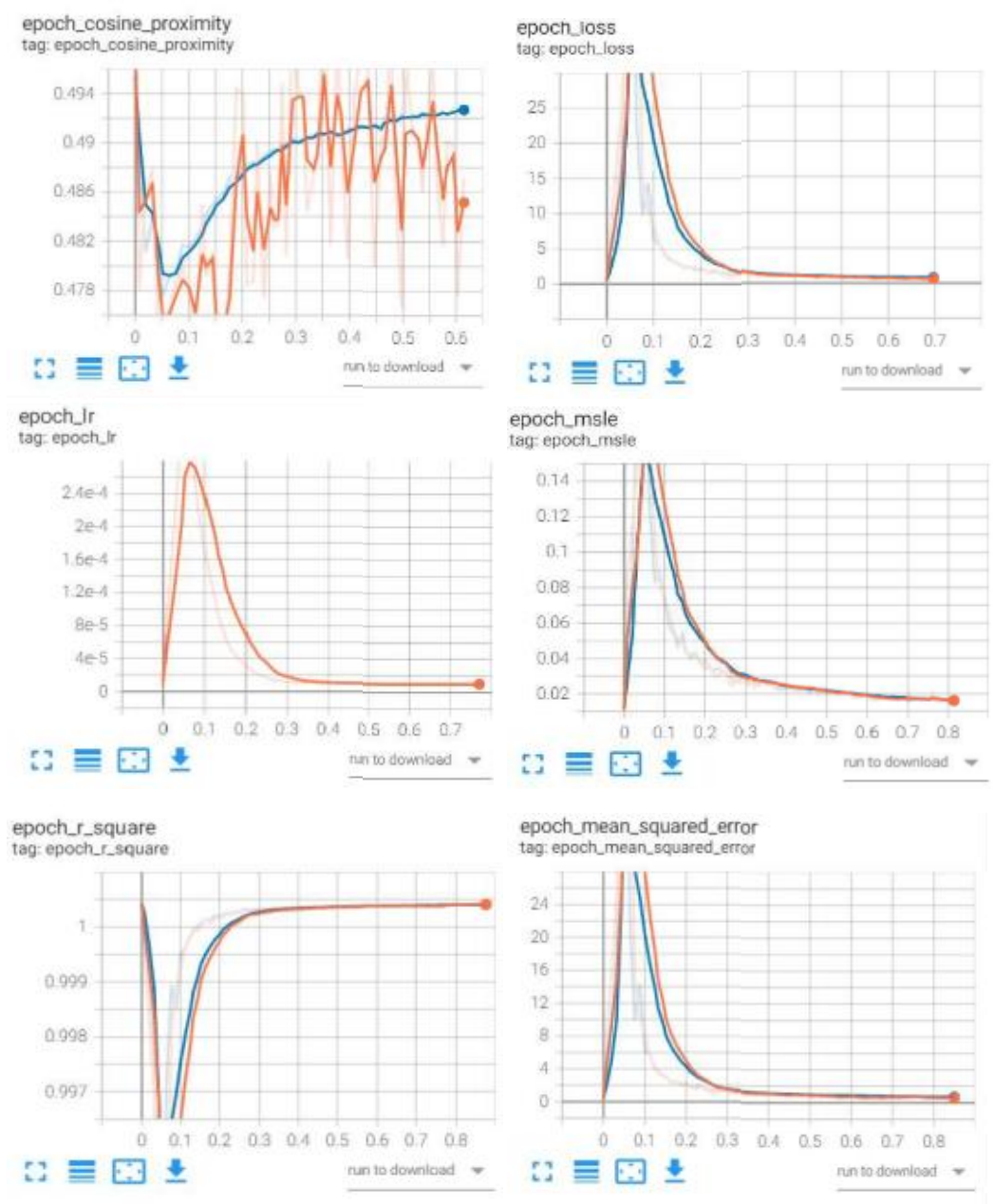

Figure 4.3 – *The experiments results for accuracy and loss function metrics details*

## 4.4 THE EXPERIMENTS WITH SIMON BLOCK CIPHER

This section provides research that contributed to the publication of the results of the study in a published conference paper titled: "Deep learning for cryptanalysis attack on IoMT wireless communications via smart eavesdropping", 5th International Conference on Networking and Advanced Systems (ICNAS) in Annaba, Algeria in 2021, with the following authors: Zakaria TOLBA, Makhlouf Derdour. [304]

The Internet of Medical Things (IoMT) technology is present in all practical aspects of our modern life, and its extraordinary utilization is rising dramatically. It has efficiently absorbed extensive interest in recent periods as a productive system typically consisting of associated hospitals with many patients and doctors by medical dispositive, cloud systems, and clinical networks as a productive system typically consisting of associated hospitals with many patients and doctors by medical dispositive, cloud systems, and clinical networks. IoMT can not only provide practical benefits to appropriate patients such as medicine anywhere and real-time monitoring, but it can also assist medical teams in achieving remote medical treatments by using artificial intelligence tools to accurately predict any potential diseases for potential patients .

However, as the number of patient health information and medical information system applications grows, the healthcare information system is frequently confronted with fundamental challenges such as the operational efficiency of storing, retrieving, and dealing with big health data patient information. As demonstrated in figure 1, cloud computing is a good solution to this issue, with considerable storage and computing resources that can generally handle big data applications .

In a sophisticated IoMT system, the medical wearable dispositive controls the physiological data from the selected cases and forwards it to a cloud database storage via the web, during which official health cares exponents such as a civil emergency, assurance, doctor, nursemaids can enter and cultivate medical knowledge conformable with the collected physiological information from wearable IoMT devices of correlated patients [278],[279] .
Because of flaws connected to the leaking of sensitive data controlled by detectors and uploaded to the cloud via open wireless connections, the Internet of Medical Things (IoMT) poses major security problems.
Patients' information is encrypted by terminal devices in most IoMTs, and published cryptanalysis works show obvious limitations because they are performed for the theoretical model without any real-world implementation scenario on wireless encryption protocols, resulting in a performance gap between the hypothetical investigation and practical application of these attacks.
To address these shortcomings, we propose a method based on a deep learning model for an attack simulation on specific IoMT protocols that discovers the private key of IoMT wireless communications using a smart eavesdropping conceptual scenario on multilevel layers of the IoMT protocol stack for its implementation.[304] We describe the attack technique for creating datasets from wireless packet protocols to train the suggested cryptanalysis model.
Data encryption and decryption are handled by terminal devices in the majority of IoMT systems. However, they have intrinsic restrictions such as:

- The obvious limitations in size, storage memory, and power consumption purposely confine their specific utility in the secure transfer of critical data, where the reasonable development of lightweight ciphers reacts tolerably to these feasible bounds.
- The lack of strong authentication on IoT dispositive, lightweight encryption, and poor key management leave IoT wireless junctions vulnerable to foreign active and passive attacks and threats.
- The network transfers of data and control information are provided by particular protocols defined in IoT devices, and each of those protocols has its own unique set of error checking/correction, key, and identity management systems, compression, and encryption algorithms.

The published cryptanalysis works on lightweight encryption algorithms used in IoT protocols may be impractical or convincingly demonstrate obvious limitations because they are performed perfectly to the theoretical model understudy, without any conceptual implementation scenario, posing a performance gap between theoretical research and the practical application of these common attacks on real wireless communication protocols. This work presents a deep learning (DL) model-based methodology for an attack simulation that discovers precisely the private key of IoMT wireless communications using a comprehensive multilevel smart eavesdropping scenario on specific IoMT protocols such as Bluetooth, 3GPP, Zigbee, and Wi-Fi for its implementation, and we properly present a trained model to scientifically verify the operational security of IoMT.

### 4.4.1 IoT encryption protocols background

The Internet of Things protocol stack is not the same as the TCP/IP or OSI protocol stack. The most advanced IoT security mechanisms are permitted to operate in mixed pile levels to provide secrecy. Figure 2 depicts the IoT protocol stack and the technologies employed at each tier. The wireless security protocol operates in collaborative layers, utilizing numerous keys and confidants of the traffic by ciphering messages and managing communication identity administration.

Encryption can be used at the device-device, device-gateway, device-cloud, or gateway-cloud levels. The encryption scheme used is determined by the architecture and communication model of the IoT system (figure 3 depicts such communication modes [280]). The device-gateway model is commonly employed in most IoMTs.

#### Intelligent eavesdropping using a friendly jamming technique

In his published paper, Jongyeop Kim [281] correctly introduces a sensible eavesdropper, which can mitigate the friendly jamming impact by typically using some antenna techniques tools and signal processing approaches, where his smart eavesdropper improved model and thus the combined solution method between transceivers arbitrate the sensible possibility to scarcely hold the jamming signal power to realize the optimal performance.[304]

- In practice, eavesdroppers are typically passive and silent to conceal their espionage activities. [281] .

- To naturally lower the perceived channel signal quality of potential eavesdroppers, the friendly jamming strategy is used against eavesdropping attacks. In this technique, artificial noise is typically generated and transmitted at the same time as the transmission mechanism of the confidential data signals of the IoMT devices.

- Additionally, passive listeners were unable to stop or effectively counter the friendly jamming[281].

### 4.4.2 Problem framing for experiments tests

The overall goal of our proposed effort [304] is to train a model to read the secret key from plaintext and hence ciphertext.
We defined the problem as a regression assignment since the goal is to utilize supervised learning to forecast the secret key corresponding to non-negative values. A cryptanalyst's ultimate objective is to weaken the distinguisher model from particular block cipher data inputs.

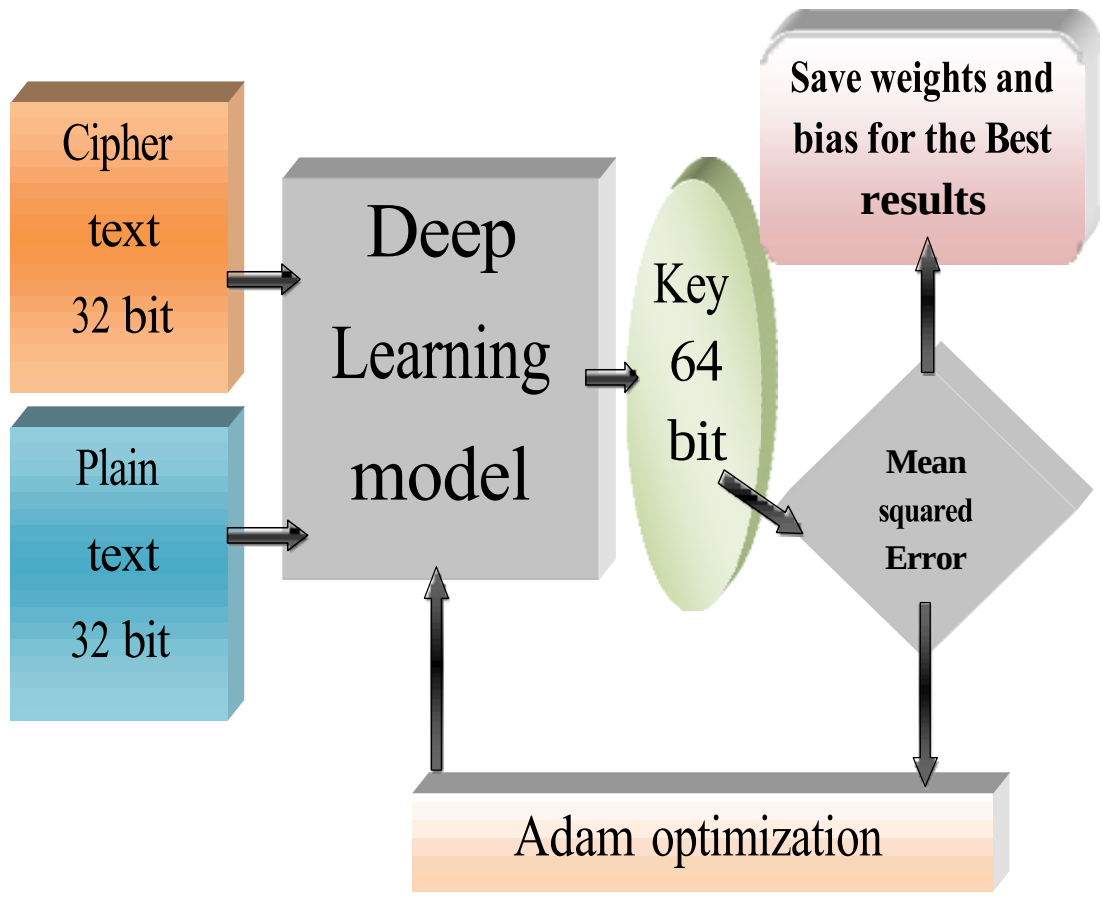

Figure 4.4 – *The training experiments goals*

Table 4.1 – *Model architecture*

| Layers | Neurons | Drop Out |
|---|---|---|
| Input | 64 | - |
| 1 | 16 | 0.9 |
| 2 | 32 | 0.8 |
| 3 | 64 | 0.7 |
| 4 | 128 | 0.9 |
| 5 | 16 | 0.8 |
| 6 | 32 | 0.7 |
| 7 | 64 | 0.9 |
| Output | 64 | - |

### 4.4.3 Model architecture details

The table 4.1 provides information on the architecture. :

#### Hyperparameters used in experiment

Below is a list of the remaining hyper-parameters correctly employed in our experiments:

- Weights initialization: Normalized Xavier.
- Optimizer: Adam.
- Initial Learning rate: 0.1.
- Activation Function: segmoid.
- Epochs : 5000 .

- Batch Size:128.
- Trainable parameters: 21 024.
- None Trainable parameters : 0.
- Error function: Mean squared error
  $$MSE = \frac{1}{n}\sum_{i=1}^{n}\left(y_i - \hat{y}_i\right)^2$$
- Accuracy function: R-squared
  $$R^2 = 1 - \frac{Unexplained-Variation}{Total-Variation}$$

### 4.4.4 Datasets

If we employ Simon 32-bit [277] effectively for proof-of-concept testing, we should be able to create substantial volumes of data in a fair period. Instead of complex software, SIMON is specifically built for optimal performance in hardware executions. It has been offered as a widely used universal numbering scheme that is not domain-specific. This Feistel method, which has been published in several application classes [282], has effectively promoted the use of incredibly inexpensive materials. Using Python cryptography tools in Google collaborative settings, we were able to create data sets of 2,000,000 simple for the training stage, 245 000 simple for the model validation, and 50 000 simple for the model prediction testing.

### 4.4.5 Experiment outcomes for training and prediction

– **Scenario of multi-level smart eavesdropping:**

- To begin, the active test will be effectively created by wireless through the use of antenna methods or signal processing, as well as the practical usage of the Kali Linux workstation. This latter is a robust software application designed specifically for network penetration testing and information system security audits.

- The IoMT network topology that we use for our attack is a device-gateway connection.

- The first level attack is a spoofing active attack performed on a single device to gather as much (plain / cipher) text and the matching private key as feasible to create the database for deep learning model training.
  This attack may be used on the Data Link layer for any wireless dispatch protocol homologous as 3GPP, Bluetooth, Zigbee, and Wi-Fi. Because IEEE802.15.4 is used to provide banger-hierarchical protection for MAC structures protected by cryptographic techniques like Zigbee, 6LoWPAN, and Wireless-HART.

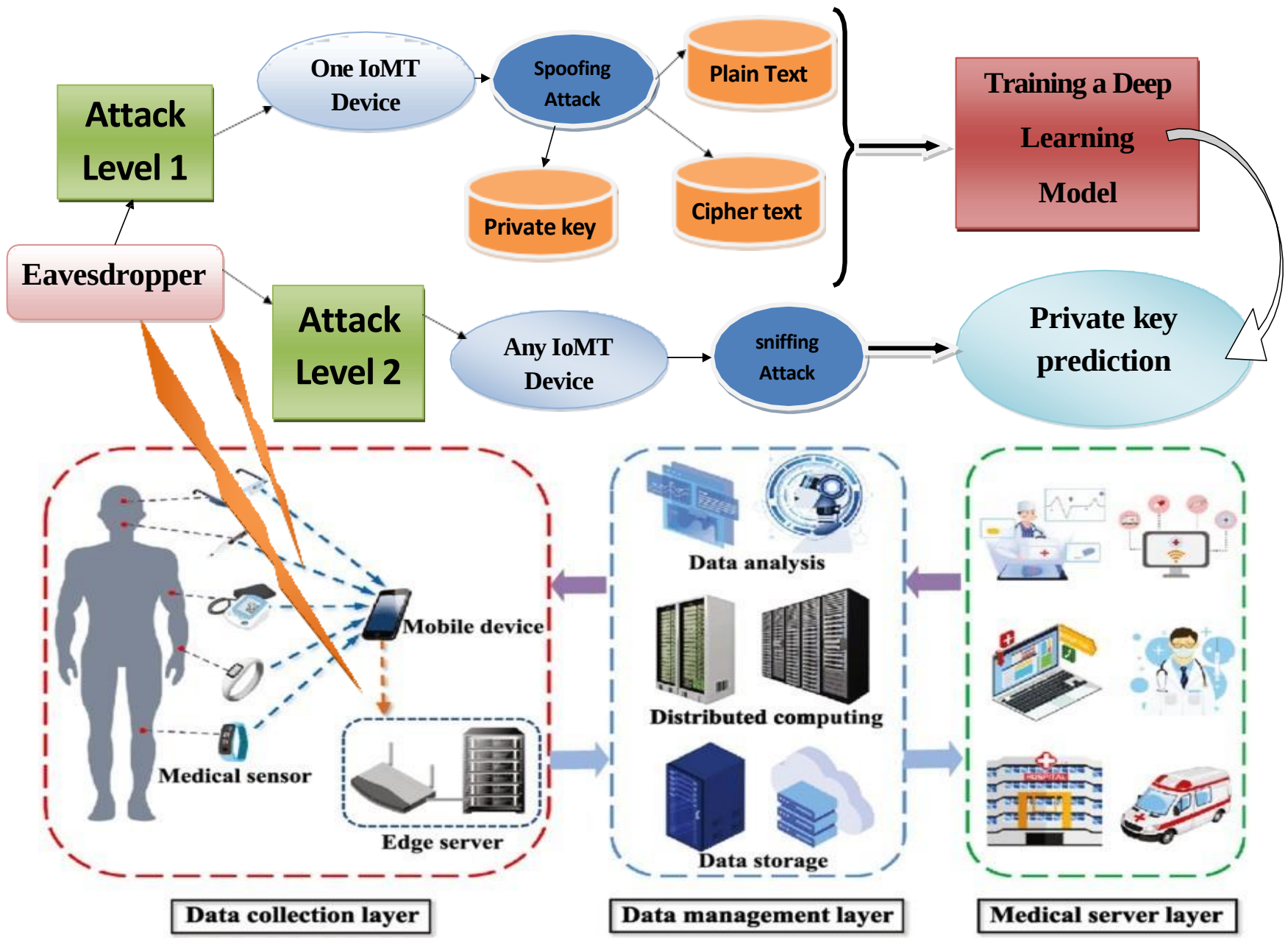

Figure 4.5 – *Attack levels activity diagram*

The Extensible Authentication Protocol allows many authentication methods and operates independently of IP, making spoofing attacks possible with Kali Linux tools.

- The second level attack is a sniffer passive attack on any personal IoMT device; a well-trained deep learning model requires just one pair of (cipher/plain) text to properly guess the utilized private wireless key.
  This attack is appropriate on a Physical level since it is data-driven and is responsible for regulating data from IoT disposition.
  The devices attached to this level are sensitive to security issues such as device tampering, eavesdropping, and data changing. Low Power Wide Area Network (LPWAN) is the protocol used in IoMT for transmitting little amounts of data over vast distances while remaining battery productive. It employs modulation schemes comparable to the ultra-narrow band, narrow-band, and wide-band attacks, and the attack will be sufficiently created by the operation of antenna systems or signal processing, as well as the practical usage of Kali Linux machine instruments.

## 4.5 Deep Learning technique for detection regions of interest in cryptanalysis platforms

This section provides research that contributed to the publication of the results of the study in a published conference paper titled: "Towards a Novel Cryptanalysis

Table 4.2 – *Training results*

| | |
|---|---|
| Mean squared error | 0.0074 |
| R-squared | 0.87 |
| Mean absolute error | 0.00175 |
| Mean absolute poucentage error | 0.000503 |
| Root mean squared error | 0.0198 |

Platform based Regions Of Interest Detection via Deep Learning models", International Conference on Recent Advances in Mathematics and Informatics (ICRAMI), in Tebessa, Algeria 2021., with the following authors: Zakaria TOLBA; Makhlouf Derdour; Rafik Menassel. [305]

The kind of permutable units differs from one cryptogram to the next depending on the structure of the multimedia encryption technique. These units might represent bits, bit planes, pixels, transformation coefficients, pixels blocks, tree nodes, variable-length code words, motion vectors, prediction errors, and their numerous combinations in the spatial or frequency domains.

- The encryption strategy is applied to the image plane directly in the spatial domain and works in this category are based on the direct modification of the image pixels. The encryption in this domain eliminates the correlation between the pixels. The pixels in the image may be restored entirely without any loss of information using the reverse technique.
- Schemes in the frequency domain encryption are based on modifying the frequency of the picture using transformations; hence, reconstruction of the original image pixels in the decryption process frequently results in information loss.

There are three advanced approaches to this unique mechanism for picture permutation in the spatial domain based specifically on the deep change of specific site of an essential portion of the image:

- Binary permutation (bit permutation): Consider the picture to be an array of successive bits, with each pixel containing eight bits for a total of 256 unique grayscale colors. The bits in a set of keys are swapped with the selected key using a pseudo-random index generator in this approach. All of these permuted bits combine to generate the encrypted picture. [283].
- Pixel permutation: Visible pixels meticulously collected from the visible image are purposely swapped using a key chosen from a collection of keys. The key encryption and decryption procedures are the same as a bit exchange mechanism. The visible size of the pixel group is the same as the length of the private keys, and all keys have the same length. Perceptual information is reduced when the length of the private keys exceeds the correct size of the pixels group [283].

- Block permutation: This advanced approach divides the picture into little chunks. A predetermined group of key blocks is normally chosen from the visible image and swapped in the same manner as bit and pixel permutations. The block size should be reduced to improve the concealment of obvious perceptual information. If the key blocks are exceedingly small, the visible items and their distinct boundaries will not be readily seen, which positively supports the information's concealment. In this approach, particular blocks in the image [283] are swapped horizontally and vertically.

Although the associated studies have offered essential answers to different cryptanalysis issues, they have several limitations, including:

- In these systems, the key assessment process is not automated; the naked eye is used to analyze the quality of the attack on the picture, and the search for the key process must then be halted manually if required; this necessitates an interactive attack to review the findings.
  Each time, precise human intervention is required to recognize the relevant sections that emerge immediately after the created keys are applied to the encrypted picture. This judgment may be delayed if the human abandons the observation of the emergence of crucial portions. Furthermore, it is ineffective in the situation of frequent emergence of tiny or fragmentary components that are difficult to recognize precisely by humans. The issue is that the phase of evaluating cryptanalysis findings is critical in this process, and it permits convergence towards good results if it is based on improved criteria and tools since it does not allow for any solution, despite thousands of iterations.

- The images used by these approaches to scientifically prove their potential effectiveness represent artificial images with a unified distribution of colors, which correspond perfectly to the theoretical model under study, but the real world lacks this distribution, creating a performance gap between theoretical models and practical application of these possible attacks.

- These techniques need the use of pairs of (plain/cipher) pictures, which are not always accessible.

### 4.5.1 The models re-used by the platform :

Our platform [305] uses (Figure 4.7):

### 4.5.2 The the genetic algorithms parameters details:

- Individuals $K_i$ correctly represent potential decryption keys. Each key is accurately represented in the form of a matrix, with each box $K_i$[x, y] defined as follows:$K_i$[x, y] = ($P_{xy}$, E) in which:
  –$P_{xy}$: indicate the new location of the box C [x, y] in the box $C_i^{'}$ [x, y]proposed by the key $K_i$.
  – E: indicate the state of the point $P_{xy}$. It is true to claim that the location $P_{xy}$

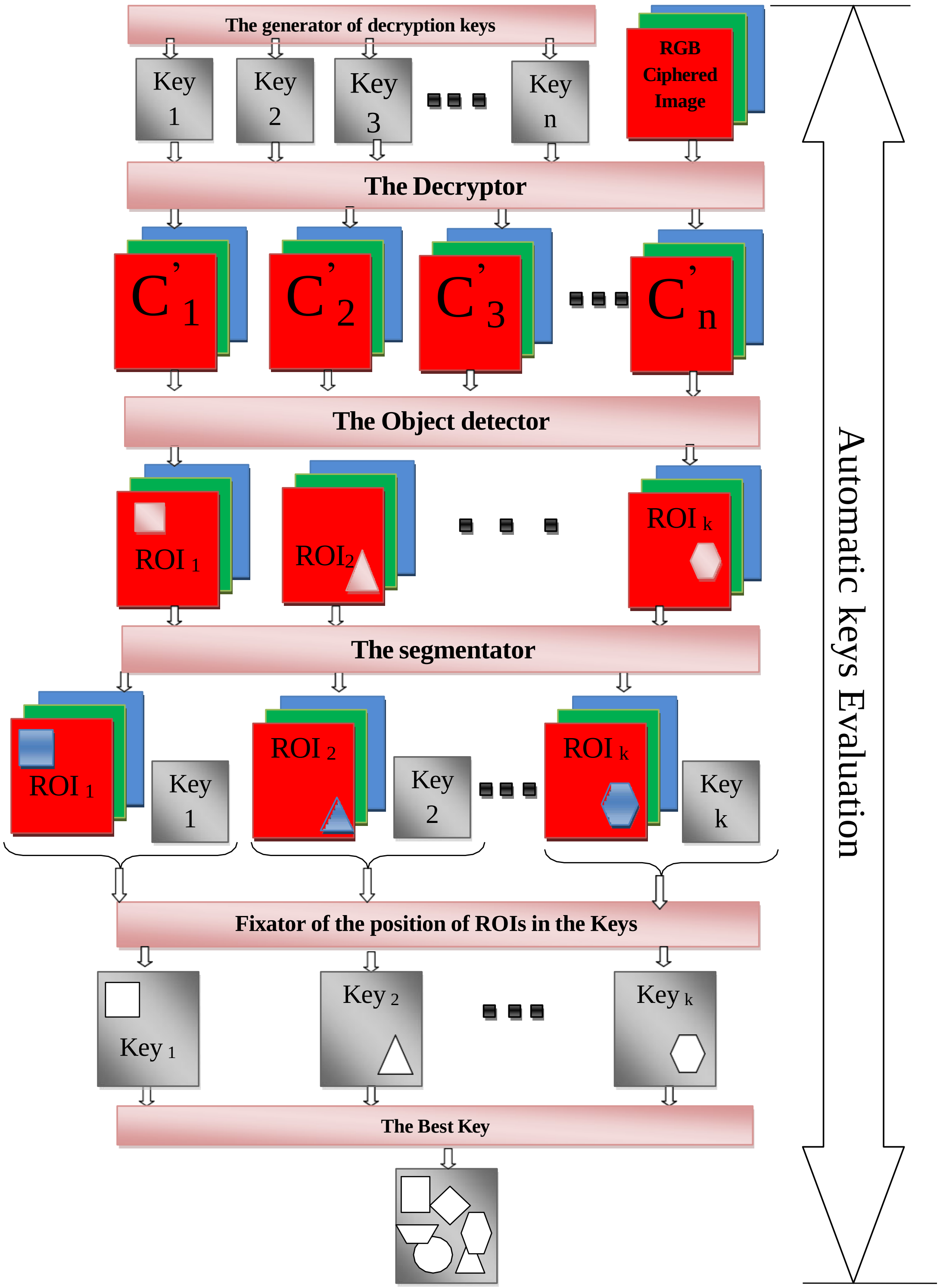

Figure 4.6 – *The platform architecture*

provided by the key $K_i$ is incorrect at first. It immediately takes the value j (the number of detected $ROI_j$)), then accurately says after a specified number of generations that the position $P_{xy}$ provided by the key $K_i$ has become in the position that allows the detection of the area $ROI_j$.
The goal of replacing the state E of a specific location $P_{xy}$ is to designate this position such that it is not taken into account in subsequent generations.

- The fitness function $f$ is properly specified on the reasonable foundation of the fixed position number (E/= 0).

  It is equal to $f = \frac{(Number-of-box-including(E<>0))}{N*M}$.

  where (N*M) is the cipher image size. According to this specific description [284], the obvious fulfillment of the function of fitness allows us to move closer to the target key, generation after generation.

### 4.5.3 Deep learning-based object recognition model (Faster R-cnn)

which will be correctly employed on the $C_i'$ pictures acquired after the practical application of the decryption keys $k_i$ on the encrypted image C, to detect all of the relevant $ROI_j$ areas that occur in these photos in real-time.
Automating the crucial evaluation process overcomes the limitations of human interaction. Indeed, the Faster R-CNN object detection system [285] perceives the image in three (RGB) layers, it uses for that two modules, the first representing a fully convolutional deep network which proposes to him regions by the technique of the selective search, and the second representing a Fast R-CNN detector which will determine the semantics of the encrypted image.[305] In contrast to humans who superficially use the naked eye to evaluate the result of applying a key to the encrypted image .
Additionally, the Faster R-CNN object identification system has previously undergone extensive training, making it capable of accurately and reliably detecting and locating object attributes. as the creation of characteristic maps can provide it with a thorough comprehension in a fair amount of time.
In fact, it is able to identify traits that are invisible to the naked eye, ensuring convergence toward successful outcomes. Utilizing the significant benefits of Faster r-cnn in selective search produced by the sliding windows which suggest regions to the classifier of the faster r-cnn, effectively supports the considered finding of the potential connection between the nearby pixels.

### 4.5.4 Segmentation algorithm using deep learning (Mask R-cnn)

Mask R-cnn [286] is an improved version of Faster R-cnn that uses object segmentation layers to accurately identify the pixels of the important regions that were recognized in the pictures as having a $ROI_j$.
The locations of these visible pixels, as shown in the used $Ki$ keys, correspond to the proper key components. Where $K_{ij}$ [x, y], they will be securely fixed. E will be equal to j, where j is the quantity of discovered ROIs.

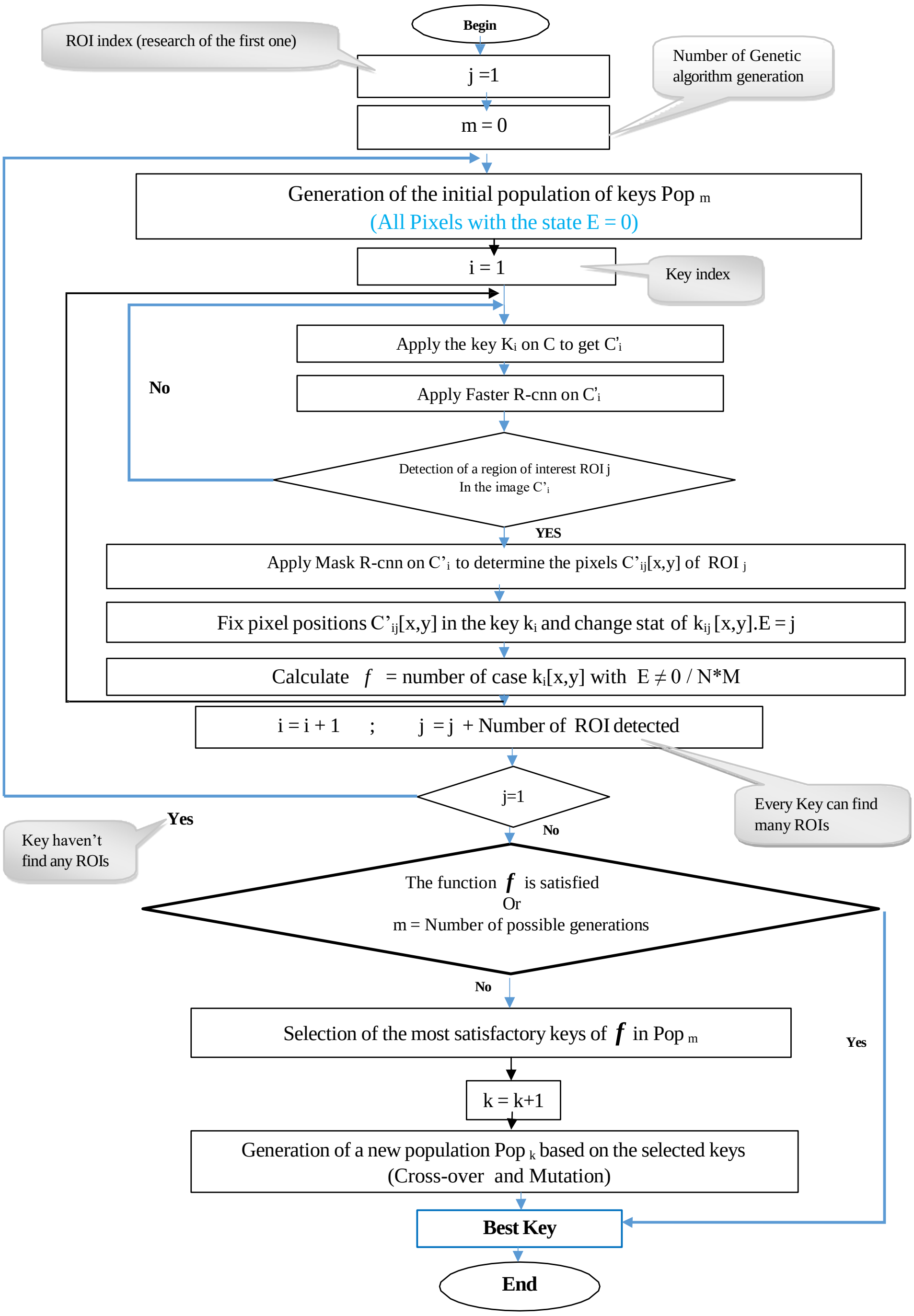

Figure 4.7 – *The activity model.*

As a result, it is feasible to reduce the search space for the locations of other pixels that are yet to be implicated in $ROI_s$ to be masked generation after generation. Indeed, it is evident that all permutation techniques are covered in the $[N * M]!$ with encrypted images of apparent size (N*M)! The search space for the original pixel coordinates is represented by the potential combinations.

In our case the detection of (K) distinct $R_j$ regions of interest, each one of $ROI_j$ contains $Z_j$ pixels, minimize the search space for the positions of the other pixels which remains by $R_j * Z_j$ distinct possibilities, it becomes equal to

$(N * M) - \sum_{j=1}^{K} R_j * Z_j$ ! .

### 4.5.5 The Datasets choice for experiments

The original code of the published Keras version of Faster R-CNN that we have used was typically written by Yann Henon, and he properly used the PASCAL VOC 2007, 2012, and MS COCO datasets. For us [305], we reasonably used Google's Open Images Data set V6.
This data set properly contains a published total of 16M bounding boxes for 600 object classes on 1.9M images, making it the largest existing dataset. The visible and very diverse images often contain complex scenes with several familiar objects (8.3 per image on average). The biggest dataset currently available, this data collection accurately comprises a reported total of 16M bounding boxes for 600 item types on 1.9M photos. The observable and extremely varied visuals frequently feature complex sceneries with several recognizable elements (8.3 per image on average).

### 4.5.6 The platform running process

The initial population $Pop0$ is generated at random in the activity diagram of our platform (figure 2), where a sizable number of keys $K_i$ are created. The platform starts to anticipate people (the keys) from the original population who best meet the fitness function $f$. The framework systematically executes the following procedures on $Pop0$ in order to identify the highest-performing $Pop1$ population that can best meet the fitness function:

– First, each member of the population $Pop_0$, $K_i$ , is appraised objectively. This thorough evaluation willingly goes through the following steps:

- To get the picture $C_i'$ whose pixel coordinates are altered according to the key $K_i$, apply the key $K_i$ to the encrypted image C.
- Deploy the Faster r-cnn to the final picture $C_i'$ to find the $ROI_j$ areas of interest.
- To find the pixels $Cij'$[x, y] of $ROIj$, perform the Mask r-cnn to the picture $C_i'$.

- Fix the positions of the pixels of the $ROI_j$ ($C'_{ij}$ [x, y]) in $K_i$. To do this, change the state of the pixels $K_{ij}$ [x, y] .E from $o$ to $j$.

- Calculate $f = \frac{(Number-of-box-including(E<>0))}{N*M}$.

– The selecting step is then used. Based on their judgment, this step removes the less desirable keys from $Pop_0$ and keeps just the best ones. The keys offered at this moment are distinguished by the ability to identify the majority of critical sections recognized by Faster r-cnn, according to the assessment phases.

– To produce the new $Pop_1$ population, the previously chosen individuals must be crossed correctly in the next step. Therefore, a crossover operator is applied to produce a new descendant from two parents. Although there are various crossover approaches, we will employ the "one-point crossover" in our platform.

To create a new individual, carefully duplicate the important parts of parent 1 and parent 2 using this operator, The crossing point is the precise point at which the parents part ways.

However, we must pay attention to avoid repeating a similar position (we do not replicate the positions already added to the key), and we must avoid forgetting these locations by adding them at the end (do not take them into account and do not touch the positions of cells whose state E is different from zero).

– A mutation procedure is used to diversify the solutions across the generations. This stage entails changing a small portion of the vector (character) in a few members of the new generation at random. This process, which involves, for instance, exchanging two consecutive entries in a person, is carried out with a very low likelihood.
Within the maximum number of potential generations allowed by the software, the stages described above will be reapplied to the $Pop_1$ population. This will provide a population that has individuals (keys) who fully meet the fitness function. If these people cannot be found within the maximum number of generations, the best key found in the most recent generation ($Pop_n$) will be found. The key will have the following characteristics:

- When applied to photos with multiple ROIs, the method will use a maximum of fixed portions (correct portions that reflect ROIs).

- In the event of using the technique on photos with low ROIs, there should be a minimum number of fixed portions. In order to identify the global key in this second situation, it is necessary to apply the strategy across a number of frames and intersect the generated keys.

With the use of two deep learning systems, one for automated identification (faster R-cnn) and another for separation, we proposed a unique cryptanalysis model based on ROI discovery (mask R-cnn). In the short to medium term, we hope to apply our technique to both multiple photos with few objects and images with many objects.

## 4.6 CONCLUSION

In this chapter, in the first and second parts, a series of experiments were carried out to test the resistance of light encryption algorithms to attack by machine learning. All of the experimentation has been done on a short key length, either for predicting the plaintext from the ciphertext or for predicting the secret key from the ciphertext. Despite the good results obtained, they remain restricted in terms of verification and validation because they do not cover all the possible lengths of the private keys that will be used by these algorithms to improve the results obtained, more detailed and in-depth work is mandatory to verify the security and measure the results obtained. An architecture for a platform was suggested in the third section, but it has not yet been implemented since its implementation would need more time and fine-tuning of all the specifics. We intend to complete it very soon and add to the body of knowledge on this subject.

# 5 General Conclusion and Perspectives

Technology innovation in computer hardware and software was indeed extremely fast and accurate from one day to the next, therefore, artificial intelligence methods are usually orchestrated. Day by day, many more high-quality research projects in the specific case of using artificial intelligence have appeared, addressing a variety of significant issues throughout human development, beginning with automatic driving, then cyber-security, medical image processing, artificial vision, the internet of things, the prediction of physical and mental illnesses, and concluding with the simulation of human behaviors.

In the sense of using these techniques to verify and validate the security of cryptographic algorithms, and to measure the resistance of these algorithms to different attacks. Their application has seen huge work and considerable scientific conclusions to detect weaknesses and give more recommendations to the research community in the field of cryptography.

We have seen good results and diverse attacks in existing works, beginning with side-channel attacks, differential attacks, and emulation attacks. Investigative tests have been done on asymmetric and symmetric algorithms with different parameters and multiple constraints, and the results are promising.

Despite the advantages they obtain compared to traditional methods, they cannot replace completely the expertise of humans or the other methods promoted by the sector. It can be an additional arsenal in the toolbox used by experts to carry out their audits and investigation.

In the details of the approaches that exist in the reality train, black box testing by deep learning techniques is completely famous, particularly in the testing of symmetric block cipher cryptographic systems, where their major flaws are articulated in their inability to locate the exact weakness of the studied algorithm despite the good results obtained during the attack phase. This remains an open problem in this area that deserves future work that can offer better answers to this subject, so the answer to this question can offer a big step and a big leap in this context.

Throughout my years of research in this field, I've discovered that a hybrid methods approach can provide several answers to open problems in cryptanalysis by leveraging the benefits of each approach and combining classic and heuristic methods with deep learning models to improve work and results obtained.

Among the strong points, which attract very great attention and provide great motivation to the experts in the field of the exploitation of machine learning techniques to do cryptanalysis, is the aspect of transfer learning.
The fact that an experiment can be advantageous by not redoing some additional tasks opens a great door for the optimization of research techniques in this field, minimizing the time and recourse required for calculations while still improving the results through a phase of inheritance of the experience of a model that is already trained on common characteristics between a set of encryption algorithms.
In the Future, We will investigate deep learning-based cryptanalysis for video and sound encryption and other multimedia encryption systems.
Furthermore, we will use artificial intelligence approaches and tools to address and enhance current traditional cryptanalysis issues. On the other side, because of his rapidity, technological progress in the field of quantum neural networks can also aid in the development and reliability of the tools and models discovered and developed through experimentation.
We are interested in using deep and wide learning models from previous experiments to carry out security audits and assess the safety of lightweight cryptographic algorithms in the hopes of resolving particular field challenges.

# SCIENTIFIC CONTRIBUTIONS

- **Zakaria TOLBA**; Makhlouf Derdour; Mohamed Amine Ferrag; S M Muyeen; Mohamed Benbouzid ,"Automated Deep Learning BLACK-BOX Attack for Multimedia P-BOX Security Assessment", IEEE Access journal,volume 10 , pages 94019-94039,2022.
DOI: 10.1109/access.2022.3204175

- **Zakaria TOLBA**; Makhlouf Derdour; Nour el houda Dehimi ,"Machine learning based cryptanalysis techniques: perspectives, challenges and future directions", 4th International Conference on Pattern Analysis and Intelligent Systems (PAIS), Oum El Bouaghi, Algeria, 2022.
DOI: 10.1109/pais56586.2022.9946889

- **Zakaria TOLBA**; Makhlouf Derdour,"Deep Neural Network Based TensorFlow Model for IoT Lightweight Cipher Attack", International Conference on Artificial Intelligence and its Applications (AIAP), El Oued, Algeria, 2022.
DOI: 10.1007/978-3-030-96311-8_11

- **Zakaria TOLBA**; Makhlouf Derdour,"Deep learning for cryptanalysis attack on IoMT wireless communications via smart eavesdropping", 5th International Conference on Networking and Advanced Systems (ICNAS) , Annaba, Algeria, 2021.
DOI: 10.1109/ICNAS53565.2021.9628924

- **Zakaria TOLBA**; Makhlouf Derdour;Rafik Menassel, "Towards a Novel Cryptanalysis Platform based Regions Of Interest Detection via Deep Learning models", International Conference on Recent Advances in Mathematics and Informatics (ICRAMI), Tebessa, Algeria, 2021.
DOI: 10.1109/icrami52622.2021.9585924

- **Zakaria TOLBA**; Makhlouf Derdour, "TRIVIUM Lightweight Cipher security evaluation using deep learning based cryptanalysis techniques", National Workshop on Deep Learning and Its Applications, Oum El Bouaghi, Algeria, 2022.
URL:http://www.univ-oeb.dz/4th-international-conference-on-pattern-analysis-and-intelligent-systems-pais2022/national-workshop-on-deep-learning-and-its-applications/

# ACRONYMS

| | |
|---|---|
| *ML* | Machine Learning |
| *DL* | Deep Learning |
| *CNN* | Convolution Neural Network |
| *DNN* | Deep Neural Network |
| *RNN* | Recurrent Neural Network |
| *LSTM* | Long Short Term Memory |
| *GAN* | Generative Adversarial Network |
| *DBN* | Deep Believe Network |
| *MLP* | Multi Layer Perceptron |
| *CPS* | Cyber Physical System |
| *IoT* | Internet Of Things |
| *IoMT* | Internet Of Medical Things |
| *DES* | Data Encryption Standard |
| *AES* | Advanced Encryption Standard |
| *RSA* | Rivest Shamir Adleman |
| *AI* | Artificial Intelligence |
| $S-Box$ | Substitution Box |
| $P-Box$ | Permutation Box |
| *SPN* | Substitution Permutation Network |
| *LFSR* | Linear Feedback Shift Register |
| *NLFSR* | Non Linear Feedback Shift Register |
| *ECB* | Electronic Code Book |
| *CBC* | Cipher Code Book |
| *CFB* | Cipher Feedback book |
| *OFB* | Output Feedback book |
| *CTR* | Counter mode |
| *MAC* | Message Authentication Code |
| *AEAD* | Authenticated Encryption Data |
| *IV* | Initial Value |
| *RFID* | Radio Frequency Identification |
| *ROI* | Regions Of Interest |
| *KCA* | Known Ciphertext Attack |
| *KPA* | Known Plaintext Attack |
| *CPA* | Chosen Plaintext Attack |
| *CCA* | Chosen Ciphertext Attack |
| *WW* | World War |
| *KNN* | K-Nearest Neighbors |
| *SVM* | Support Vector Machine |
| *SCA* | Side Channel Analysis |
| *DC* | Differential Cryptanalysis |
| *LC* | Linear Cryptanalysis |
| *COA* | Ciphertext Only Attack |
| *GFS* | Generalized Feistel Structure |
| *MSE* | Mean Squared Error |
| *MAE* | Mean Absolute Error |
| *CONV* | Convolution |
| *MNIST* | Mixed National Institute of Standards and Technology |
| *CML* | Coupled Map Lattice |
| *GCBPM* | Gray Code Based Permutation Method |
| *SWMSN* | Smart Wireless Multimedia Surveillance Network |
| *SDV* | Self Driving Vehicle |
| *DSS* | Drones Surveillance System |
| *GE* | Gate Equivalent |
| *RAM* | Random Access Memory |
| *SGD* | Stochastic Gradient Descent |
| *RELU* | Rectified Exponential Linear Unit |
| *IP* | Internet Protocol |
| *TCP* | Transmission Control Protocol |
| *OSI* | Open Systems Interconnection |